\documentclass[12pt]{article}
\usepackage{amsmath,amssymb,amsthm,amsxtra,overpic,bbm,bm,epsfig,subfigure}
\usepackage{hyperref}
\usepackage{mathrsfs}
\usepackage{enumitem}
\usepackage{graphicx}
\usepackage{color}
\usepackage[table]{xcolor}
\usepackage{comment}
\usepackage{epstopdf}
\usepackage{float}
\usepackage{cite}
\usepackage{slashed,stmaryrd}

\usepackage[title]{appendix}

\allowdisplaybreaks[1]

\def\thefootnote{\fnsymbol{footnote}}

\usepackage{multirow}
\newcommand{\Dlr}{\mbox{$\raisebox{2mm}{\boldmath ${}^\leftrightarrow$}\hspace{-4mm}D^{}_\mu$}}
\newcommand{\Dilr}{\mbox{$\raisebox{2mm}{\boldmath ${}^\leftrightarrow$}\hspace{-4mm}D^I_\mu$}}

\newcommand{\Dl}{\mbox{$\raisebox{2mm}{\boldmath ${}^\leftarrow$}\hspace{-4mm}D^{}_\mu$}}

\newcommand{\Yn}{Y^{}_\nu}
\newcommand{\DYn}{Y^\dagger_\nu}

\newcommand{\TYn}{Y^{\rm T}_\nu}
\newcommand{\Yl}{Y^{}_l}
\newcommand{\DYl}{Y^\dagger_l}
\newcommand{\Yu}{Y^{}_{\rm u}}
\newcommand{\DYu}{Y^\dagger_{\rm u}}
\newcommand{\Yd}{Y^{}_{\rm d}}
\newcommand{\DYd}{Y^\dagger_{\rm d}}

\newcommand{\Lelli}[1][\beta]{\ell^{}_{#1 \rm L}}
\newcommand{\BLelli}[1][\alpha]{\overline{\ell^{}_{#1 \rm L}}}
\newcommand{\REi}[1][\beta]{E^{}_{#1 \rm R}}
\newcommand{\BREi}[1][\alpha]{\overline{E^{}_{#1 \rm R}}}
\newcommand{\LQi}[1][\beta]{Q^{}_{#1 \rm L}}
\newcommand{\BLQi}[1][\alpha]{\overline{Q^{}_{#1 \rm L}}}
\newcommand{\RUi}[1][\beta]{U^{}_{#1 \rm R}}
\newcommand{\BRUi}[1][\alpha]{\overline{U^{}_{#1 \rm R}}}
\newcommand{\RDi}[1][\beta]{D^{}_{#1 \rm R}}
\newcommand{\BRDi}[1][\alpha]{\overline{D^{}_{#1 \rm R}}}

\newcommand{\rmI}{{\rm i}}
\newcommand{\Op}{\mathcal{O}}
\newcommand{\Opr}{\mathcal{R}}
\newcommand{\dC}[2]{ \delta C^{#1}_{#2}}
\newcommand{\dG}[2]{ \delta G^{#1}_{#2}}

\newcommand{\loopfactore}{\frac{1}{\left(4\pi\right)^2 \varepsilon}}
\newcommand{\tr}[1]{{\rm tr}\left( #1 \right)}
\newcommand{\trs}[1]{{\rm tr}\left[ #1 \right]}
\newcommand{\co}{C^{(1)}_{H\ell}}
\newcommand{\ct}{C^{(3)}_{H\ell}}
\newcommand{\bda}[1]{16\pi^2 \mu \frac{{\rm d} #1}{{\rm d}\mu}}

\newcommand{\pl}{P^{}_{\rm L}}
\newcommand{\pr}{P^{}_{\rm R}}

\usepackage{ulem}

\begin{document}

\begin{center}
{\Large\bf Complete One-loop Renormalization-group Equations in the Seesaw Effective Field Theories}
\end{center}

\vspace{0.2cm}

\begin{center}
{\bf  Yilin Wang~$^{a,~b}$}~\footnote{E-mail: wangyilin@ihep.ac.cn} \;, \quad {\bf Di Zhang~$^{a,~b,~c}$}~\footnote{E-mail: di1.zhang@tum.de (corresponding author)}\;, \quad {\bf Shun Zhou~$^{a,~b}$}~\footnote{E-mail: zhoush@ihep.ac.cn}
\\
\vspace{0.2cm}
{\small
$^a$Institute of High Energy Physics, Chinese Academy of Sciences, Beijing 100049, China\\
$^b$School of Physical Sciences, University of Chinese Academy of Sciences, Beijing 100049, China\\
$^c$Physik-Department, Technische Universität München, James-Franck-Straße, 85748 Garching, Germany}
\end{center}

\vspace{1.5cm}

\begin{abstract}
In this paper, we derive the complete set of one-loop renormalization-group equations (RGEs) for the operators up to dimension-six (dim-6) in the seesaw effective field theories (SEFTs). Two kinds of contributions to those RGEs are identified, one from double insertions of the dimension-five (dim-5) Weinberg operator and the other from single insertions of the tree-level dim-6 operators in the SEFTs. A number of new results are presented. First, as the dim-5 Weinberg operator is unique in the standard model effective field theory (SMEFT), its contributions to the RGEs for the SEFTs are equally applicable to the SMEFT. We find the full contributions from the Weinberg operator to one-loop RGEs in the SMEFT, correcting the results existing in previous works, and confirm that those from dim-6 operators are consistent with the results in the literature. Second, in the type-I SEFT, we give the explicit expressions of the RGEs of all the physical parameters involved in the charged- and neutral-current interactions of leptons. Third, the RGEs are numerically solved to illustrate the running behaviors of the non-unitary parameters, mixing angles and CP-violating phases in the non-unitary leptonic flavor mixing matrix. Together with the one-loop matching results of the dim-5 and dim-6 operators and their Wilson coefficients, the present work has established a self-consistent framework up to dim-6 to investigate low-energy phenomena of three types of seesaw models at the one-loop level.
\end{abstract}


\def\thefootnote{\arabic{footnote}}
\setcounter{footnote}{0}
\newpage

\section{Introduction}

As one of the simplest and most natural ways to account for tiny Majorana neutrino masses, the type-I seesaw model extending the standard model (SM) with three right-handed neutrino singlets has been constantly attracting a lot of attention~\cite{Minkowski:1977sc,Yanagida:1979as,Gell-Mann:1979vob,Glashow:1979nm,Mohapatra:1979ia}. Moreover, the type-I seesaw model has provided us with an elegant explanation for the matter-antimatter asymmetry in our Universe via the leptogenesis mechanism~\cite{Fukugita:1986hr}. Typically, the mass scale $M$ of three right-handed neutrinos is much larger than the electroweak scale $\Lambda^{}_{\rm EW}$, i.e., $M \gg \Lambda^{}_{\rm EW} \sim \mathcal{O} \left( 10^2 \right)~{\rm GeV}$, so one can integrate out right-handed neutrinos and construct the low-energy effective field theory (EFT) of the type-I seesaw model with operators of mass dimension higher than four, i.e., the so-called type-I seesaw effective field theory (SEFT), and examine its low-energy phenomena. In this way, the large logarithms originating from the huge difference between the seesaw and electroweak scales in radiative corrections can be easily and systematically resummed through the renormalization-group equations (RGEs) in the EFT. Consequently, the precision studies of the type-I seesaw model at low energies can be substantially simplified~\cite{Weinberg:1980wa}.

With the effective operators up to dimension-six (dim-6) or equivalently to the order of $\mathcal{O} \left(M^{-2}\right)$, the tree-level Lagrangian of the type-I SEFT via integrating out right-handed neutrinos at the tree level has been known for a long time, which contains the unique dimension-five (dim-5) Weinberg operator~\cite{Weinberg:1979sa} and two dim-6 operators $\Op^{(1)}_{H\ell}$ and $\Op^{(3)}_{H\ell}$~\cite{Broncano:2002rw,Broncano:2003fq}. After the spontaneous gauge symmetry breaking, the dim-5 Weinberg operator $\Op^{(5)}$ generates Majorana neutrino masses, while the dim-6 operators $\Op^{(1)}_{H\ell}$ and $\Op^{(3)}_{H\ell}$ modify the couplings of neutrinos with weak gauge bosons and lead to the unitarity violation of the leptonic flavor mixing matrix. Recently, the complete one-loop structure of the type-I SEFT has been obtained by integrating out heavy right-handed neutrinos at the one-loop level~\cite{Zhang:2021jdf} (see also Refs.~\cite{Coy:2018bxr, Zhang:2021tsq, Ohlsson:2022hfl,Coy:2021hyr,Du:2022vso,Crivellin:2022cve}). One can find that in addition to the two dim-6 operators at the tree level, there are other 29 dim-6 operators first appearing at the one-loop level. These one-loop results at the matching scale, together with one-loop RGEs of the Wilson coefficients of dim-5 and dim-6 operators, enable us to carry out full one-loop calculations in the type-I SEFT at the electroweak scale and extract useful information about the type-I seesaw model from low-energy measurements. 
However, to our best knowledge, the complete set of one-loop RGEs for all the SM couplings and Wilson coefficients in the type-I SEFT up to dim-6 are still lacking. Since the type-I SEFT is just a part of the SM effective field theory (SMEFT)~\cite{Buchmuller:1985jz,Grzadkowski:2010es} (see, e.g., Ref.~\cite{Brivio:2017vri} for a recent review), one can directly extract the one-loop RGEs in the type-I SEFT from those in the SMEFT. The one-loop RGE for the Wilson coefficient of the Weinberg operator has been derived in Refs.~\cite{Babu:1993qv,Chankowski:1993tx,Antusch:2001ck}, and the one-loop anomalous dimensions for the Wilson coefficients of the dim-6 operators in the Warsaw basis in the SMEFT have been calculated in Refs.~\cite{Jenkins:2013zja,Jenkins:2013wua,Alonso:2013hga,Alonso:2014zka}.\footnote{The one-loop anomalous dimensions for the Wilson coefficients of dimension-seven and dimension-eight operators in the SMEFT can be found in Refs.~\cite{Liao:2016hru,Liao:2019tep,Chala:2021juk} and Refs.~\cite{Chala:2021pll,AccettulliHuber:2021uoa,DasBakshi:2022mwk,Helset:2022pde,DasBakshi:2023htx}, respectively.} But only the contributions from single insertions of dim-6 operators are included in the latter case. Though the contributions from double insertions of the dim-5 operator to the one-loop RGEs for dim-6 operators have been partially considered in Ref.~\cite{Broncano:2004tz} and then revised in Ref.~\cite{Davidson:2018zuo}, we find that these results are still neither complete nor fully correct. 

In this work, we recalculate the contributions from both double insertions of the dim-5 operator and single insertions of dim-6 operators to the one-loop RGEs in the type-I SEFT and give a set of complete and correct one-loop RGEs in type-I SEFT up to dim-6. Some important observations from our calculations are summarized as follows.
\begin{itemize}
	\item Double insertions of the dim-5 Weinberg operator contribute to the one-loop RGE for the Higgs quartic coupling. Such a contribution is given here for the first time.
	\item The contributions from the double insertions of the dim-5 operator to the one-loop anomalous dimensions of the Wilson coefficients of dim-6 operators $\Op^{}_{H\square}$, $\Op^{}_{HD}$, $\Op^{}_H$, $\Op^{}_{eH}$, $\Op^{}_{uH}$ and $\Op^{}_{dH}$ found in Refs.~\cite{Davidson:2018zuo,Coy:2018bxr} are corrected.
	\item The contributions from single insertions of dim-6 operators (i.e., $\Op^{(1,3)}_{H\ell}$ in the type-I SEFT) are perfectly consistent with those obtained in the SMEFT \cite{Jenkins:2013zja,Jenkins:2013wua,Alonso:2013hga}. This provides a partial crosscheck of the results in Refs. \cite{Jenkins:2013zja,Jenkins:2013wua,Alonso:2013hga}.
\end{itemize}
It is worth pointing out that the contributions from double insertions of the dim-5 operator are exactly the same as those in the SMEFT, so they can be immediately incorporated for completeness into the codes solving the one-loop RGEs in the SMEFT (e.g., {\sf DsixTools}~\cite{Celis:2017hod,Fuentes-Martin:2020zaz}, {\sf Wilson}~\cite{Aebischer:2018bkb} and {\sf RGESolver}~\cite{DiNoi:2022ejg}). Therefore, they are applicable to all other EFTs containing the Weinberg operator, such as the type-II and type-III SEFTs~\cite{Li:2022ipc,Du:2022vso,Coy:2021hyr}. Based on these results and those in Refs. \cite{Jenkins:2013zja,Jenkins:2013wua,Alonso:2013hga}, we also present the complete one-loop RGEs up to dim-6 in the type-II and type-III SEFTs, which can be directly used to study the one-loop renormalization-group (RG)-running effects in those types of SEFTs.

As a practical application of these one-loop RGEs, we take the type-I SEFT for example and further derive the explicit expressions of the one-loop RGEs for the physical parameters involved in the charged- and neutral-current interactions of leptons, like those done for the type-I SEFT with only the Weinberg operator~\cite{Casas:1999tg,Antusch:2003kp,Antusch:2005gp,Mei:2005qp,Xing:2005fw, Ohlsson:2013xva}. Since the leptonic flavor mixing matrix $V$ is non-unitary resulting from the two tree-level dim-6 operators in the type-I SEFT, we adopt the conventional parametrization of the mixing matrix $V \equiv (\mathbbm{1} - \eta)\cdot V^\prime$, with the Hermitian matrix $\eta$ characterizing the unitarity violation and $V^\prime$ being unitary, and derive the RGEs of all the mixing parameters. A few interesting results are obtained. First, we find that the charged-lepton Yukawa coupling matrix $Y^{}_l$, once diagonalized at a given energy scale, does not keep diagonal anymore during RG running because of the extra contributions from dim-6 operators to its RGE. Second, due to the impact of dim-6 operators on the running of $Y^{}_l$, we observe that the running behaviors of the mixing angles in $V^\prime$ could be very different from those in the type-I SEFT with only the dim-5 Weinberg operator. Third, the RGEs of the unitary parameters in $V^\prime$ and the non-unitary parameters in $\eta$ are entangled with each other, so even if the CP-violating phases in $V^\prime$ are vanishing at some energy scale they can be generated radiatively from the non-trivial phases in $\eta$ or vice versa. A similar phenomenon among three CP-violating phases has been discussed in Refs.~\cite{Luo:2005sq,Xing:2006sp,Ohlsson:2012pg,Zhang:2020lsd}. These results demonstrate that it is necessary to implement the one-loop RGEs to link the physical parameters in the full theory (e.g., the parameters residing in the Dirac neutrino Yukawa coupling matrix $Y^{}_\nu$ and right-handed neutrino mass matrix $M^{}_{\rm R}$ in the type-I seesaw model) to the effective parameters in the type-I SEFT, which have received severe constraints from low-energy experimental measurements.

The remaining part of this paper is organized as follows. In Sec.~\ref{sec:deriveRGEs}, we derive the complete set of one-loop RGEs in the type-I SEFT and make some comments on the general features of these RGEs. The explicit expressions of the RGEs of the physical parameters involved in the charged- and neutral-current interactions of leptons are given in Sec.~\ref{sec:explictRGEs}, where the RGEs of the mixing angles and CP-violating phases in the adopted parametrization of the leptonic flavor mixing matrix are analytically derived and numerically solved. The main conclusions are summarized in Sec.~\ref{sec:summary}. Some details are given in a series of appendices. The contributions from tree-level dim-5 and dim-6 operators in the type-I SEFT to the counterterms for the SM couplings and the Wilson coefficients of dim-6 operators in the Warsaw basis are given in Appendix~\ref{app:counterterms}. The complete one-loop RGEs for the Wilson coefficients of dim-6 operators in the type-II and type-III SEFTs are shown in Appendix~\ref{app:SEFTs}. For completeness, the lengthy analytical expressions of the RGEs of the mixing parameters are collected in Appendix~\ref{app:rge}.

\section{Derivation of One-loop RGEs}
\label{sec:deriveRGEs}

To derive the one-loop RGEs in the type-I SEFT, we need only the effective Lagrangian from the tree-level matching of the type-I seesaw model~\cite{Broncano:2002rw,Broncano:2003fq}, i.e.,
\begin{eqnarray}\label{eq:LSEFT}
	\mathcal{L}^{}_{\rm SEFT} = \mathcal{L}^{}_{\rm SM} + \frac{1}{2} \left( C^{\alpha\beta}_5 \Op^{(5)}_{\alpha\beta} + {\rm h.c.} \right) + C^{(1)\alpha\beta}_{H\ell} \Op^{(1)\alpha\beta}_{H\ell} + C^{(3)\alpha\beta}_{H\ell} \Op^{(3)\alpha\beta}_{H\ell} \;,
	\label{eq:LagEFT}
\end{eqnarray}
where $\alpha$ and $\beta$ refer to three flavors of fermions in the SM and the summations over repeated flavor indices are implied. The SM Lagrangian $\mathcal{L}^{}_{\rm SM}$ is given by
\begin{eqnarray}
	\mathcal{L}^{}_{\rm SM} &=& - \frac{1}{4} G^{A}_{\mu\nu} G^{A\mu\nu} - \frac{1}{4} W^I_{\mu\nu} W^{I\mu\nu} - \frac{1}{4} B^{}_{\mu\nu} B^{\mu\nu} + \left( D^{}_\mu H \right)^\dagger \left( D^\mu H \right) - m^2 H^\dagger H - \lambda \left( H^\dagger H \right)^2 
	\nonumber
	\\
	&& + \sum^{}_f \overline{f} \rmI \slashed{D} f - \left[ \BLQi (\Yu)^{}_{\alpha \beta} \tilde{H} \RUi + \BLQi (\Yd)^{}_{\alpha \beta} H \RDi + \BLelli (\Yl)^{}_{\alpha \beta} H \REi + {\rm h.c.} \right]\;,
\end{eqnarray}
where the gauge-fixing and Faddeev-Popov ghost terms stemming from the standard procedure of quantization are suppressed, and the covariant derivative $D^{}_\mu \equiv \partial^{}_\mu - \rmI g^{}_1 Y B^{}_\mu - \rmI g^{}_2 T^I W^I_\mu - \rmI g^{}_s T^A G^A_\mu$ is defined as usual. The effective operators in Eq. \eqref{eq:LSEFT} are found to be
\begin{eqnarray}
	\Op^{(5)}_{\alpha\beta} = \BLelli \widetilde{H} \widetilde{H}^{\rm T} \ell^{\rm c}_{\beta \rm L} \;,\quad \Op^{(1)}_{\alpha\beta} = \left( \BLelli \gamma^\mu \Lelli \right) \left( H^\dagger \rmI \Dlr H \right) \;,\quad \Op^{(3)}_{\alpha\beta} = \left( \BLelli \gamma^\mu \sigma^I \Lelli \right) \left( H^\dagger \rmI \Dilr H \right)
	\label{eq:ope56}
\end{eqnarray}
and their Wilson coefficients at the matching scale $\mu^{}_{\rm M} \sim M \equiv \mathcal{O}(M^{}_{\rm R})$ are
\begin{eqnarray}\label{eq:match_condition}
	C^{}_5 \left( \mu^{}_{\rm M} \right) = \Yn M^{-1}_{\rm R} \TYn  \;,\qquad C^{(1)}_{H\ell} \left( \mu^{}_{\rm M} \right) = - C^{(3)}_{H\ell} \left( \mu^{}_{\rm M} \right) = \frac{1}{4} \Yn M^{-2}_{\rm R} \DYn \;,
\end{eqnarray}
where $\ell^{\rm c}_{\rm L} \equiv {\sf C} \overline{\ell^{}_{\rm L}}^{\rm T}$ with ${\sf C}$ being charge-conjugate matrix, $\Dlr \equiv D^{}_\mu - \Dl$ and $\Dilr \equiv \sigma^I D^{}_\mu - \Dl \sigma^I$ with $\Dl$ acting on the left, $M^{}_{\rm R}$ and $\Yn$ are respectively the mass and Yukawa coupling matrices of the right-handed neutrinos (which have been integrated out to construct the SEFT).

In this work, we implement the background field method (BFM)~\cite{Abbott:1980hw} (see \cite{Abbott:1981ke} for more details and earlier references), dimensional regularization in $d = 4 - 2\varepsilon $ space-time dimensions, and the modified minimal subtraction ($\overline{\rm MS}$) scheme. With the BFM, all fields are split into a background field $\widehat{\phi}$ and a quantum field $\phi$, namely, $\phi \to \phi + \widehat{\phi}$, and the background and quantum gauge fields can be treated in different gauges. In Feynman diagrams, the background fields appear as external legs or internal tree-level propagators while the quantum fields as internal loop propagators. In the case with an unbroken gauge symmetry, it is not necessary to split fermion and Higgs fields into background and quantum fields since they are not involved in the gauge-fixing term with the gauge condition chosen to be the same as that in Refs.~\cite{Abbott:1980hw,Abbott:1981ke} and hence background and quantum fields obey the same Feynman rules. We choose the general $R^{}_\xi$-gauge for the quantum gauge fields and the Feynman gauge for the background gauge fields.

The one-loop RGEs for renormalizable couplings in the SM and that for the Wilson coefficient of the Weinberg operator have been already derived in the previous works~\cite{Machacek:1983tz,Machacek:1983fi,Machacek:1984zw,Luo:2002ey,Babu:1993qv,Chankowski:1993tx,Antusch:2001ck}, and thus we will not repeatedly derive them here. Up to $\mathcal{O}\left( M^{-2} \right)$, dim-6 operators do not contribute to the one-loop RGE for the Wilson coefficient of the Weinberg operator, whereas the dim-5 and dim-6 operators may contribute to those for the SM renormalizable couplings. Therefore, for completeness up to $\mathcal{O} \left( M^{-2} \right)$, we need to take into account the contributions from the dim-5 and dim-6 operators not only to one-loop RGEs for Wilson coefficients of dim-6 operators but also to those for the SM couplings. The one-loop beta functions for the Wilson coefficients of the dim-6 operators can be generally written as
\begin{eqnarray}
	\bda{C^{(6)}_i} = \gamma^{}_{ij} C^{(6)}_j + \widehat{\gamma}^i_{jk} C^{(5)}_j C^{(5)}_k \;,
\end{eqnarray}
where $\gamma^{}_{ij}$ denotes an element of the anomalous dimension matrix resulting from single insertions of dim-6 operators, and $\widehat{\gamma}^i_{jk}$ stands for those induced by double insertions of the Weinberg operator. To find out those anomalous dimensions, one needs to first calculate a set of one-particle-irreducible (1PI) diagrams and obtain the relevant wave-function renormalization constants and counterterms. Fortunately, in the type-I SEFT, the higher-dimensional operators in Eq.~\eqref{eq:LSEFT} do not give additional contributions to the wave-function renormalizations of the SM fields. As a consequence, the SM results can be simply utilized for our purpose. Only those of lepton and Higgs doublets are needed and they are found to be \cite{Antusch:2001vn}
\begin{eqnarray}\label{eq:wavefunction}
	\delta Z^{}_\ell &=& - \frac{1}{4\left( 4\pi\right)^2\varepsilon} \left( 2 \Yl\DYl + \xi^{}_1 g^2_1 + 3\xi^{}_2 g^2_2 \right) \;,
	\nonumber
	\\
	\delta Z^{}_H &=& - \frac{1}{4\left( 4\pi\right)^2\varepsilon} \left[ 4 \tr{\Yl\DYl + 3\Yu\DYu + 3\Yd\DYd} - \left( 3 -\xi^{}_1 \right) g^2_1 - 3\left( 3 - \xi^{}_2 \right) g^2_2 \right] \;,
\end{eqnarray}
where $\xi^{}_{1}$ and $\xi^{}_2$ are gauge-fixing parameters for quantum gauge fields $B^{}_\mu$ and $W^I_\mu$ in the $R^{}_\xi$-gauge, respectively. The cancellation of $\xi$-dependence in the one-loop RGEs offers a crosscheck of the final results.

Now we outline the strategy for the calculations of relevant counterterms. First, we choose a set of 1PI diagrams generated by the tree-level Lagrangian of the type-I SEFT in Eq. \eqref{eq:LSEFT} and covering all classes of dim-6 operators in the Green's basis \cite{Jiang:2018pbd,Gherardi:2020det}. Then, one calculates these diagrams to get all the counterterms for the dim-6 operators in the Green's basis and converts them into those in the Warsaw basis by using the equations of motion (EOMs) of relevant fields. Finally, the anomalous dimensions can be extracted from the resultant counterterms in the Warsaw basis.

\begin{figure}
	\centering
	\includegraphics[width=0.9\textwidth]{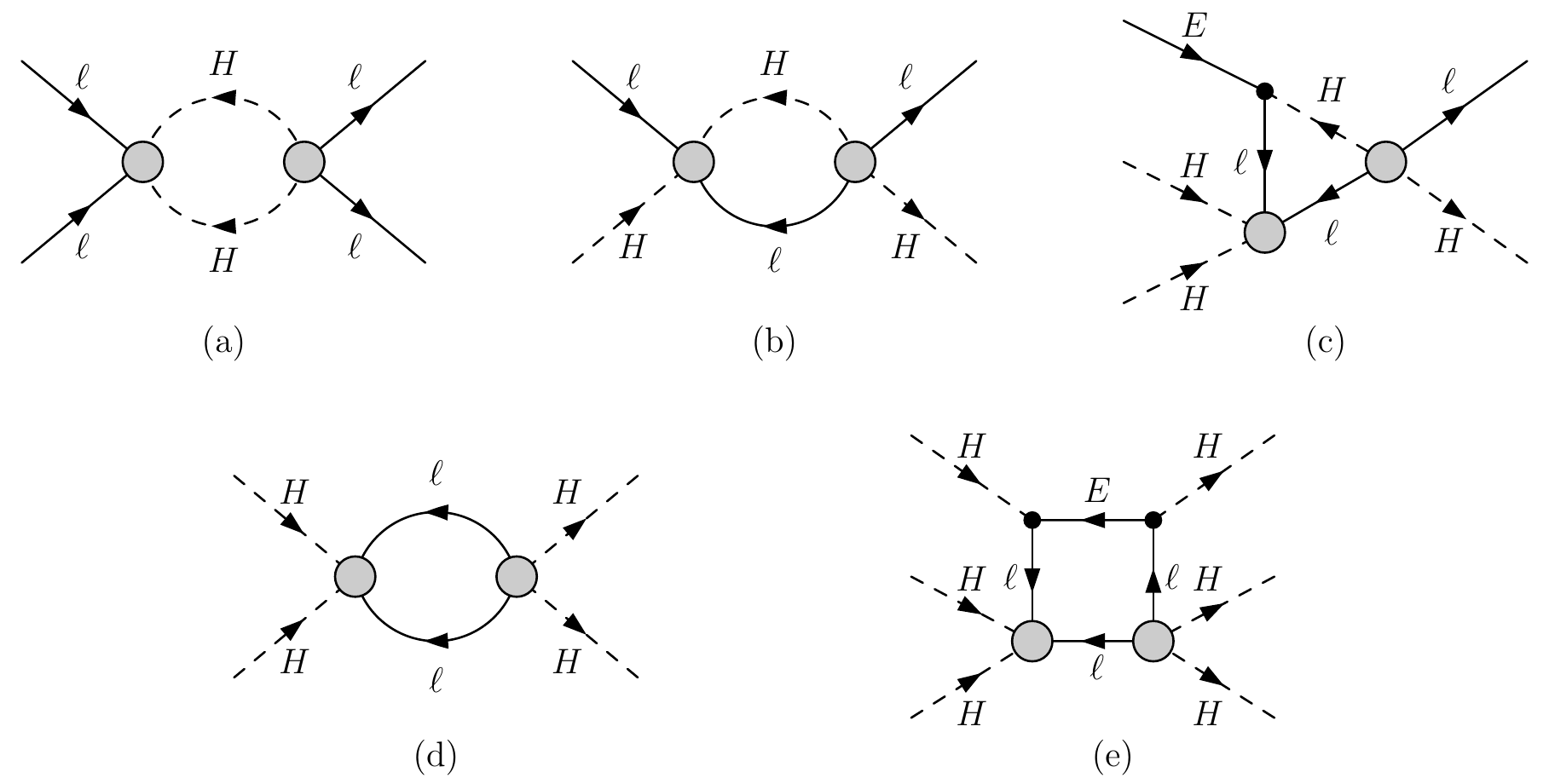}
	\vspace{-0.3cm}
	\caption{Feynman diagrams induced by double insertions of the Weinberg operators, where the grey-filled circles denote the effective vertices of the Weinberg operator. Diagrams (a)-(c) contribute to the lepton-flavor-changing counterterms, whereas Diagrams (d) and (e) lead to the counterterms blind to lepton flavors. For conciseness, the eight diagrams corresponding to crossed external legs of Diagram (e) are not explicitly shown.}
	\label{fig:double-insertions}
\end{figure}
The diagrams involving double insertions of the Weinberg operator are shown in Fig.~\ref{fig:double-insertions} and others involving single insertions of the two dim-6 operators are not shown explicitly in this work due to a large number of such kinds of diagrams. It is worthwhile to stress that Diagram (e) in Fig.~\ref{fig:double-insertions} and its eight crossed diagrams in total do not lead to any ultraviolet (UV) divergences. The calculations of these Feynman diagrams are straightforward and can be easily performed by making use of the Mathematica packages {\sf FeynRules}~\cite{Christensen:2008py,Alloul:2013bka}, {\sf FeynArts} ~\cite{Hahn:2000kx} and {\sf FeynCalc}~\cite{Shtabovenko:2016sxi,Shtabovenko:2020gxv}, together with {\sf FeynHelper}~\cite{Shtabovenko:2016whf} connecting {\sf FeynCalc} to {\sf Package-X}~\cite{Patel:2015tea,Patel:2016fam}. Therefore, we do not present all the details of loop calculations but just list the final results of the counterterms in the Warsaw basis in Appendix~\ref{app:counterterms}. The 19 dim-6 operators (barring flavor structures and Hermitian conjugates) in the Warsaw basis needed to eliminate extra UV divergences at the one-loop level in the type-I SEFT are listed in Table~\ref{tab:operators}.
\begin{table}[!t]
	\centering 
	\renewcommand\arraystretch{1.55}
	\resizebox{\textwidth}{!}{ %
	\begin{tabular}{c|c|c|c|c|c}
		\hline\hline
		\multicolumn{2}{c}{ $ H^{6}$~and~$ H^{4}D^{2}$} & \multicolumn{2}{|c|}{$\psi^{2}H^{3}$} & \multicolumn{2}{c}{$\rm (\overline{L}L)(\overline{L}L)$}
		\tabularnewline
		\hline
		\cellcolor{gray!20}{$\Op^{}_H$} & \cellcolor{gray!20}{$\left( H^\dagger H \right)^3 $} & \cellcolor{gray!20}{$\Op^{\alpha\beta}_{eH}$} & \cellcolor{gray!20}{$\left( \BLelli H \REi \right) \left( H^\dagger H \right)$} & \cellcolor{gray!20}{$\Op^{\alpha\beta\gamma\lambda}_{\ell\ell}$} & \cellcolor{gray!20}{$\left( \BLelli \gamma^\mu \Lelli \right) \left( \BLelli[\gamma] \gamma^{}_\mu \Lelli[\lambda] \right) $}
		\tabularnewline
		\cellcolor{gray!20}{$\Op^{}_{H\square}$} & \cellcolor{gray!20}{$\left(H^\dagger H \right) \square \left( H^\dagger H \right) $} & \cellcolor{gray!20}{$\Op^{\alpha\beta}_{uH}$} & \cellcolor{gray!20}{$\left( \BLQi \widetilde{H} \RUi \right) \left( H^\dagger H \right)$ } & $\Op^{(1)\alpha\beta\gamma\lambda}_{\ell q}$ & $\left( \BLelli \gamma^\mu \Lelli \right) \left( \BLQi[\gamma] \gamma^{}_\mu \LQi[\lambda] \right) $
		\tabularnewline
		\cellcolor{gray!20}{$\Op^{}_{HD}$} & \cellcolor{gray!20}{$\left(H^\dagger D^\mu H \right)^\ast  \left( H^\dagger D^{}_\mu H \right) $} & \cellcolor{gray!20}{$\Op^{\alpha\beta}_{dH}$} & \cellcolor{gray!20}{$\left( \BLQi H \RDi \right) \left( H^\dagger H \right)$}  & $\Op^{(3)\alpha\beta\gamma\lambda}_{\ell q}$ & $\left( \BLelli \gamma^\mu \sigma^I \Lelli \right) \left( \BLQi[\gamma] \gamma^{}_\mu \sigma^I \LQi[\lambda] \right) $
		\tabularnewline
		\hline
		\multicolumn{4}{c|}{$\psi^2H^2D$} & \multicolumn{2}{c}{$\rm (\overline{L} L)(\overline{R}R)$} 
		\tabularnewline
		\hline
		\cellcolor{gray!20}{$\Op^{(1)\alpha\beta}_{H\ell}$} & \cellcolor{gray!20}{$\left( \BLelli \gamma^\mu \Lelli \right) \left( H^\dagger \rmI \Dlr H \right)$} & $\Op^{(3)\alpha\beta}_{Hq}$& $\left( \BLQi \gamma^\mu \sigma^I \LQi \right) \left( H^\dagger \rmI \Dilr H \right) $ & $\Op^{\alpha\beta\gamma\lambda}_{\ell e}$ & $ \left( \BLelli \gamma^\mu \Lelli \right) \left( \BREi[\gamma] \gamma^{}_\mu \REi[\lambda] \right)$
		\tabularnewline
		\cellcolor{gray!20}{$\Op^{(3)\alpha\beta}_{H\ell}$} & \cellcolor{gray!20}{$\left( \BLelli \gamma^\mu \sigma^I \Lelli \right) \left( H^\dagger \rmI \Dilr H \right) $} & $\Op^{\alpha\beta}_{Hu}$ & $\left( \BRUi \gamma^\mu \RUi \right) \left( H^\dagger \rmI \Dlr H \right)$ & $\Op^{\alpha\beta\gamma\lambda}_{\ell u}$ & $ \left( \BLelli \gamma^\mu \Lelli \right) \left( \BRUi[\gamma] \gamma^{}_\mu \RUi[\lambda] \right)$
		\tabularnewline
		$\Op^{\alpha\beta}_{He}$ & $\left( \BREi \gamma^\mu \REi \right) \left( H^\dagger \rmI \Dlr H \right)$ & $\Op^{\alpha\beta}_{Hd}$ & $\left( \BRDi \gamma^\mu \RDi \right) \left( H^\dagger \rmI \Dlr H \right)$ & $\Op^{\alpha\beta\gamma\lambda}_{\ell d}$ & $ \left( \BLelli \gamma^\mu \Lelli \right) \left( \BRDi[\gamma] \gamma^{}_\mu \RDi[\lambda] \right)$
		\tabularnewline
		$\Op^{(1)\alpha\beta}_{Hq}$ & $\left( \BLQi \gamma^\mu \LQi \right) \left( H^\dagger \rmI \Dlr H \right)$ & & & &
		\tabularnewline
		\hline\hline
	\end{tabular}
}
	\caption{The dim-6 operators in the Warsaw basis needed to eliminate all UV divergences at the one-loop level in the type-I SEFT. Moreover, the counterterms given by the 9 operators in the gray shaded region acquire contributions from double insertions of the Weinberg operator in addition to those from single insertions of the two tree-level dim-6 operators in the type-I SEFT.}
	\label{tab:operators} 
\end{table}

As an illustrative example and also as a crosscheck, we explicitly calculate Diagram (d) in Fig.~\ref{fig:double-insertions} by hand. With the type-I SEFT Lagrangian given in Eq.~\eqref{eq:LSEFT} and the Feynman rules for the fermion-number-violating interactions~\cite{Denner:1992me,Denner:1992vza}, one can easily write down the amplitude corresponding to Diagram (d), i.e., 
\begin{eqnarray}\label{eq:amp}
	\rmI \mathcal{M}^{}_{\rm (d)} &=& - \frac{1}{2} \mu^{4\varepsilon} \int \frac{{\rm d}^d k}{\left( 2\pi \right)^d} \trs{ \rmI C^{\alpha\beta \ast}_5 \left( \epsilon^{fa} \epsilon^{eb} + \epsilon^{fb}\epsilon^{ea} \right) \pl \frac{\rmI}{\slashed{k}} \rmI C^{\alpha\beta}_5 \left( \epsilon^{ed} \epsilon^{fc} + \epsilon^{ec} \epsilon^{fd} \right) \pr \frac{\rmI}{\slashed{k} + \slashed{p}^{}_1 + \slashed{p}^{}_2}} 
	\nonumber
	\\
	&=& -2 \left( \delta^{ad}\delta^{bc} + \delta^{ac}\delta^{bd} \right) \tr{C^{}_5 C^\dagger_5} \mu^{4\varepsilon} \int \frac{{\rm d}^d k}{\left( 2\pi \right)^d} \frac{k\cdot\left( k + p^{}_1 + p^{}_2 \right)}{k^2 \left( k+p^{}_1 + p^{}_2 \right)^2 }
	\nonumber
	\\
	&=& - \frac{2\rmI \mu^{2\varepsilon}}{\left( 4\pi \right)^2} \left( \delta^{ad}\delta^{bc} + \delta^{ac}\delta^{bd} \right) \tr{C^{}_5 C^\dagger_5} \left[ {\mathbf A}^{}_0 \left(0\right) - \frac{1}{2} \left( p^{}_1 + p^{}_2 \right)^2 {\mathbf B}^{}_0 \left( p^{}_1+p^{}_2 , 0, 0 \right)  \right]
	\nonumber
	\\
	&=& \frac{\rmI \mu^{2\varepsilon}}{\left( 4\pi \right)^2\varepsilon} \left( \delta^{ad}\delta^{bc} + \delta^{ac}\delta^{bd} \right) \left( p^{}_1 + p^{}_2 \right)^2 \tr{C^{}_5 C^\dagger_5}  + {\rm UV~finite} \;,
\end{eqnarray}
where the minus sign and the factor $1/2$ in the first line originate from the symmetric fermion loop, $\mu$ denotes an arbitrary parameter of mass-dimension one to keep the mass dimensions of all the couplings (including the Wilson coefficients of higher-dimensional operators) in $d$ dimensions the same as those in four dimensions, $a,b, \cdots, f=1,2$ are weak isospin indices, and $\epsilon$ is the two-dimensional antisymmetric tensor. In the third line of Eq. \eqref{eq:amp}, ${\mathbf A}^{}_0 \left( 0 \right)$ and ${\mathbf B}^{}_0\left( p^{}_1 + p^{}_2, 0, 0 \right)$ stand for the Passarino-Veltman (PV) scalar integrals~\cite{tHooft:1978jhc,Passarino:1978jh} and we have followed the notations in Ref.~\cite{Denner:1991kt}. Noticing the UV divergences of the PV scalar integrals, i.e., $(d-4){\mathbf A}^{}_0 \left( 0 \right) = 0$ and $(d-4){\mathbf B}^{}_0 \left( p^{}_1 + p^{}_2, 0, 0  \right) = -2$~\cite{Denner:1991kt}, one can easily achieve the final result in the last line of Eq.~\eqref{eq:amp}. To eliminate the UV divergence in Eq.~\eqref{eq:amp}, one has to take account of the contributions from the relevant counterterms. Only the four dim-6 operators in  the $H^4 D^2$-class in the Green's basis may contribute,\footnote{Note that the quartic term $\left( H^\dagger H \right)^2$ in the SM does not contribute since its counterterm is independent of external momenta.} i.e.,
\begin{eqnarray}\label{eq:h4d2}
	&&\Op^{}_{H\square} = \left( H^\dagger H \right) \square \left( H^\dagger H \right) \;,\qquad \Op^{}_{HD} = \left( H^\dagger D^\mu H \right)^\ast \left( H^\dagger D^{}_\mu H \right) \;,
	\nonumber
	\\
	&&\Opr^{\prime}_{HD} = \left( H^\dagger H \right) \left( D^\mu H\right)^\dagger \left( D^{}_\mu H \right) \;,\qquad \Opr^{\prime\prime}_{HD} = \left( H^\dagger H \right) D^\mu \left( H^\dagger \rmI \Dlr H \right) \;,
\end{eqnarray}
where $\Op^{\dots}_{\dots}$ and $\Opr^{\dots}_{\dots}$ denote independent operators in the Warsaw basis and redundant operators in the Green's basis, respectively. Then the following condition
\begin{eqnarray}\label{eq:condition}
	&&\frac{\rmI}{\left( 4\pi \right)^2\varepsilon} \left( \delta^{ad}\delta^{bc} + \delta^{ac}\delta^{bd} \right) \left( p^{}_1 + p^{}_2 \right)^2 \tr{C^{}_5 C^\dagger_5} - 2\rmI \left[ \delta^{ac} \delta^{bd} \left( q^{}_1 - p^{}_1 \right)^2 + \delta^{ad} \delta^{bc} \left( q^{}_1 - p^{}_2 \right)^2 \right] \dG{}{H\square}
	\nonumber
	\\
	&& + \rmI \left( q^{}_1 \cdot p^{}_2 + q^{}_2 \cdot p^{}_1 \right) \left( \delta^{ac}\delta^{bd} \dG{}{HD} + \delta^{ad} \delta^{bc} \dG{\prime}{HD} \right) + \rmI \left(q^{}_1 \cdot p^{}_1 + q^{}_2 \cdot p^{}_2 \right)  \left( \delta^{ad}\delta^{bc} \dG{}{HD} + \delta^{ac} \delta^{bd} \dG{\prime}{HD} \right)
	\nonumber
	\\
	&& - \left( \delta^{ad}\delta^{bc} + \delta^{ac}\delta^{bd} \right) \left( q^2_1 + q^2_2 - p^2_1 - p^2_2 \right) \dG{\prime\prime}{HD} = 0
\end{eqnarray}
with $p^{}_1 + p^{}_2 = q^{}_1 + q^{}_2$ and $\dG{\dots}{\dots}$ being coefficients in the Green's basis must be satisfied to guarantee the cancellation of UV divergence. With the help of Eq. \eqref{eq:condition}, one arrives at
\begin{eqnarray}\label{eq:wcs-h4d2}
	\dG{}{HD} = \dG{\prime}{HD} = - \loopfactore \tr{C^{}_5 C^\dagger_5} \;, \quad \dG{}{H\square} = \dG{\prime\prime}{HD} = 0 \;.
\end{eqnarray}
Applying the EOMs of the Higgs doublet and with the help of Eqs.~\eqref{eq:h4d2} and \eqref{eq:wcs-h4d2}, one achieves the final results in the Warsaw basis, namely,
\begin{eqnarray}\label{eq:counterterms-5}
	&&\delta \lambda = \frac{m^2 }{\left( 4\pi \right)^2\varepsilon} \tr{C^{}_5 C^\dagger_5} \;,
	\nonumber
	\\
	&&\dC{}{H} = - \frac{ 2\lambda }{\left( 4\pi \right)^2\varepsilon} \tr{C^{}_5 C^\dagger_5} \;,~~ \dC{}{H\square} = - \frac{1}{2\left( 4\pi \right)^2\varepsilon} \tr{C^{}_5 C^\dagger_5} \;, ~~\dC{}{HD} = - \frac{1}{\left( 4\pi \right)^2\varepsilon} \tr{C^{}_5 C^\dagger_5} \;,\quad
	\nonumber
	\\
	&& \dC{}{eH} = \frac{-Y^{}_l }{ 2\left( 4\pi \right)^2\varepsilon} \tr{C^{}_5 C^\dagger_5} \;, ~~\dC{}{uH} =  \frac{-Y^{}_{\rm u} }{ 2\left( 4\pi \right)^2\varepsilon} \tr{C^{}_5 C^\dagger_5} \;, ~~\dC{}{dH} =  \frac{-Y^{}_{\rm d} }{ 2\left( 4\pi \right)^2\varepsilon} \tr{C^{}_5 C^\dagger_5} \;,
\end{eqnarray}
in which $\dC{}{\dots}$ refers to the coefficients in the Warsaw basis and the explicit forms of the corresponding operators are shown in Table~\ref{tab:operators}. From the counterterms in Eq. \eqref{eq:counterterms-5}, it is straightforward to extract the contributions from double insertions of the Weinberg operator as in Diagram (d) in Fig.~\ref{fig:double-insertions} to the anomalous dimensions of relevant dim-6 operators. Two comments on the results in Eq.~\eqref{eq:counterterms-5} are in order.
\begin{itemize}
	\item Though the lepton-flavor-blind contributions from Diagram (d) in Fig.~\ref{fig:double-insertions} have been taken into account in Ref.~\cite{Davidson:2018zuo}, it has not been mentioned that the non-vanishing counterterm $\delta \lambda$ could contribute to the one-loop RGE of the Higgs quartic coupling $\lambda$. Hence our explicit result given here is new.
	\item Even after considering different notations for the Higgs quartic coupling in Ref.~\cite{Davidson:2018zuo} and in this work, our results in the last two lines of Eq.~\eqref{eq:counterterms-5} have an additional factor of $1/2$. We have crosschecked our results by manual calculations and by the aforementioned Mathematica packages. In addition, one can also confirm our results by calculating the Feynman diagram with four external Higgs legs and one external gauge-boson leg (or two external gauge-boson legs).
\end{itemize}

Following the above procedure, one can calculate all relevant diagrams and then obtain all the counterterms corresponding to both dim-6 operators in Table \ref{tab:operators} and the renormalizable couplings, which are collected in Appendix~\ref{app:counterterms}. Together with the one-loop RGEs for the renormalizable couplings in the SM and that for the Wilson coefficient of the Weinberg operator \cite{Machacek:1983tz,Machacek:1983fi,Machacek:1984zw,Luo:2002ey,Babu:1993qv,Chankowski:1993tx,Antusch:2001ck}, the complete set of one-loop RGEs for the renormalizable couplings and the Wilson coefficients of higher-dimensional operators in the type-I SEFT can be found with the help of the counterterms in Appendix \ref{app:counterterms} and the formulas in Ref.~\cite{Antusch:2001ck}. The final results are summarized as below.
\subsubsection*{Renormalizable terms}
\begin{itemize}
	\item {\bf The Yukawa couplings}
\begin{eqnarray}
	\bda{\Yl} &=& \left[- \frac{15}{4} g^2_1 - \frac{9}{4} g^2_2 + T + \frac{3}{2} \Yl\DYl  - 2 m^2 \left( \co + 3\ct \right)  \right] \Yl \;,
	\nonumber
	\\
	\bda{\Yu} &=& \left[ -\frac{17}{12} g^2_1 - \frac{9}{4} g^2_2 - 8 g^2_s + T + \frac{3}{2} \left( \Yu\DYu - \Yd\DYd \right) \right] \Yu \;,
	\nonumber
	\\
	\bda{\Yd} &=& \left[ -\frac{5}{12} g^2_1 - \frac{9}{4} g^2_2 - 8 g^2_s + T -  \frac{3}{2} \left( \Yu\DYu - \Yd\DYd \right) \right] \Yd \;,
 \label{eq:YukawaRGE}
\end{eqnarray}
where $T \equiv \tr{\Yl \DYl + 3\Yu\DYu + 3\Yd\DYd }$ has been defined for later convenience.

\item {\bf The gauge couplings}
\begin{eqnarray}
	\bda{g^{}_1} &=& \frac{41}{6} g^3_1 \;,
	\nonumber
	\\
	\bda{g^{}_2} &=& - \frac{19}{6} g^3_2 \;,
	\nonumber
	\\
	\bda{g^{}_s} &=& -7 g^3_s \;.
 \label{eq:gaugeRGE}
\end{eqnarray}

\item {\bf The Higgs quadratic and quartic couplings}
\begin{eqnarray}
	\bda{m^2} &=&  \left( - \frac{3}{2} g^2_1 - \frac{9}{2} g^2_2 + 12\lambda + 2T \right) m^2 \;,
	\nonumber
	\\
	\bda{\lambda} &=& 24 \lambda^2 - 3\lambda \left( g^2_1 + 3g^2_2 \right) + \frac{3}{8} \left( g^2_1 + g^2_2 \right)^2 + \frac{3}{4} g^4_2 + 4\lambda T - 2{\rm tr} \left[ \left( \Yl \DYl \right)^2  \right.
	\nonumber
	\\
	&& + 3\left( \Yu \DYu \right)^2 + \left. 3 \left( \Yd \DYd \right)^2 \right] +  m^2   \tr{ 2 C^{}_5 C^\dagger_5 - \frac{8}{3} g^2_2 \ct+ 8\ct \Yl\DYl} \;. \quad
 \label{eq:HiggsRGE}
\end{eqnarray}
\end{itemize}

As can be seen from above, only the beta-functions for the charged-lepton Yukawa coupling matrix $\Yl$ and the Higgs quartic coupling $\lambda$ receive the contributions from the Wilson coefficients of higher-dimensional operators. The result of the RGE of $\lambda$ in Eq.~(\ref{eq:HiggsRGE}) is new and given in the complete form.

\subsubsection*{The Weinberg operator}

\begin{eqnarray}
	\bda{C^{}_5} &=& \left( -3g^2_2 + 4\lambda +2T \right) C^{}_5 - \frac{3}{2} \left[ \Yl\DYl C^{}_5 + C^{}_5 \left( \Yl\DYl  \right)^{\rm T} \right] \;.
 \label{eq:C5RGE}
\end{eqnarray}
Note that the one-loop RGE for $C^{}_5$ does not receive any contributions from double insertions of the Weinberg operator itself or single insertions of dim-6 operators up to $\mathcal{O} \left( M^{-2} \right)$. Therefore, the one-loop RGE in Eq.~(\ref{eq:C5RGE}) remains unchanged when compared to that up to $\mathcal{O} \left( M^{-1} \right)$ in the type-I SEFT.

\subsubsection*{Dimension-six operators}

\begin{itemize}
	\item {\bf $\bm{ H^6}$ and $\bm{ H^4 D^2}$}
\begin{eqnarray}
	\bda {C^{}_{H\square}} &=& - 2\, \trs{ \frac{1}{2} C^{}_5 C^\dagger_5 + \frac{1}{3} g^2_1 \co - g^2_2 \ct + (\co+3\ct)\Yl\DYl } \;,
	\nonumber
	\\
	\bda{C^{}_{HD}} &=& - 2\, \tr{ C^{}_5 C^\dagger_5 + \frac{4}{3} g^2_1 \co +  4\co \Yl\DYl } \;,
	\nonumber
	\\
	\bda{C^{}_{H}} &=& 4\, \tr{ -\lambda C^{}_5 C^\dagger_5 + \frac{4}{3} \lambda g^2_2 \ct - 4\lambda \ct \Yl\DYl } \;.
\end{eqnarray}

\item $\bm {\psi^2 H^3}$
\begin{eqnarray}
	\bda{C^{}_{eH}} &=& 2 \left[ \frac{3}{4}  C^{}_5 C^\dagger_5  \Yl + \tr{ - \frac{1}{2} C^{}_5 C^\dagger_5 + \frac{2}{3} g^2_2 \ct - 2\ct \Yl \DYl} \Yl + \co \Yl\DYl\Yl  \right.
	\nonumber
	\\
	&& + \left. \left( 2\lambda - 3g^2_1 \right) \co \Yl + 3 \left( 2\lambda - g^2_1 \right) \ct \Yl \right] \;,
	\nonumber
	\\
	\bda{C^{}_{uH}} &=& \tr{ - C^{}_5 C^\dagger_5 + \frac{4}{3} g^2_2 \ct -  4\ct \Yl\DYl} \Yu \;,
	\nonumber
	\\
	\bda{C^{}_{dH}} &=& \tr{ - C^{}_5 C^\dagger_5 + \frac{4}{3} g^2_2 \ct - 4\ct \Yl\DYl} \Yd \;.
\end{eqnarray}

\item $\bm{\psi^2H^2 D}$
\begin{eqnarray} \label{eq:psi2h2d}
	\bda{C^{(1)}_{H\ell}}&=& -\frac{3}{2} C^{}_5 C^\dagger_5 + \frac{2}{3} g^2_1  \tr{\co} \mathbbm{1} + \left( \frac{1}{3} g^2_1 + 2 T \right) \co 
	\nonumber
	\\
	&& + \frac{1}{2} \left[ \left( 4\co + 9\ct \right) \Yl\DYl + \Yl\DYl \left( 4\co + 9\ct \right) \right] \;,
	\nonumber
	\label{eq:cHl1}
	\\
	\bda{C^{(3)}_{H\ell}}&=& C^{}_5 C^\dagger_5 + \frac{2}{3} g^2_2  \tr{\ct} \mathbbm{1} + \left( -\frac{17}{3}  g^2_2 + 2 T \right) \ct 
	\nonumber
	\\
	&& + \frac{1}{2} \left[ \left( 3\co + 2\ct \right) \Yl\DYl + \Yl\DYl \left( 3\co + 2\ct \right) \right] \;,
	\nonumber
	\\
	\bda{C^{}_{He}} &=&  \frac{4}{3} g^2_1 \tr{\co} \mathbbm{1} - 2\DYl\co\Yl  \;,
	\nonumber
	\\
	\bda{C^{(1)}_{Hq}} &=& - \frac{2}{9} g^2_1 \tr{\co} \mathbbm{1} \;,
	\nonumber
	\\
	\bda{C^{(3)}_{Hq}} &=& \frac{2}{3} g^2_2 \tr{\ct} \mathbbm{1} \;,
	\nonumber
	\\
	\bda{C^{}_{Hu}} &=& - \frac{8}{9} g^2_1 \tr{\co} \mathbbm{1} \;,
	\nonumber
	\\
	\bda{C^{}_{Hd}} &=& \frac{4}{9} g^2_1 \tr{\co} \mathbbm{1} \;,
\end{eqnarray}
where the wave-function renormalizations for the lepton and Higgs doublets in Eq.~\eqref{eq:wavefunction} are exploited to derive the RGEs of $C^{(1)}_{H\ell}$ and $C^{(3)}_{H\ell}$. As is expected, the gauge-fixing parameters $\xi^{}_1$ and $\xi^{}_2$ have been canceled out and thus do not appear in the RGEs of $C^{(1)}_{H\ell}$ and $C^{(3)}_{H\ell}$. 

\item $\mathbf {\left( \overline{L} L \right) \left( \overline{L} L \right) }$
\begin{eqnarray}
	\bda{C^{\alpha\beta\gamma\lambda}_{\ell\ell}} &=& -\frac{1}{2} C^{\alpha\gamma}_5 C^{\beta\lambda \ast}_5 -  \frac{1}{3} g^2_1 \delta^{\gamma\lambda} C^{(1)\alpha\beta}_{H\ell} + \frac{1}{3} g^2_2 \left( 2\delta^{\gamma\beta} C^{(3)\alpha\lambda}_{H\ell} - \delta^{\gamma\lambda} C^{(3) \alpha\beta}_{H\ell} \right) 
	\nonumber
	\\
	&& - \frac{1}{2} \left( \Yl \DYl \right)^{}_{\alpha\beta} \left( \co - \ct \right)^{\gamma\lambda} - \frac{1}{2} \left( \co - \ct \right)^{\alpha\beta} \left( \Yl\DYl \right)^{}_{\gamma\lambda} 
	\nonumber
	\\
	&& -  \left( \Yl\DYl \right)^{}_{\alpha\lambda} C^{(3)\gamma\beta}_{H\ell} - C^{(3)\alpha\lambda}_{H\ell} \left( \Yl \DYl \right)^{}_{\gamma\beta} \;,
	\nonumber
	\\
	\bda{C^{(1)\alpha\beta\gamma\lambda}_{\ell q}} &=&   \frac{1}{9} g^2_1 \delta^{\gamma\lambda} C^{(1)\alpha\beta}_{H\ell} +  C^{(1)\alpha\beta}_{H\ell} \left( \Yu\DYu - \Yd\DYd \right)^{}_{\gamma\lambda} \;,
	\nonumber
	\\
	\bda{C^{(3)\alpha\beta\gamma\lambda}_{\ell q}} &=&  \frac{1}{3} g^2_2 \delta^{\gamma\lambda} C^{(3)\alpha\beta}_{H\ell} -  C^{(3)\alpha\beta}_{H\ell} \left( \Yu\DYu + \Yd\DYd \right)^{}_{\gamma\lambda} \;.
\end{eqnarray}

\item $\mathbf {\left( \overline{L} L \right) \left( \overline{R} R \right) }$
\begin{eqnarray}
	\bda{C^{\alpha\beta\gamma\lambda}_{\ell e}} &=& - \frac{2}{3} g^2_1 \delta^{\gamma\lambda} C^{(1)\alpha\beta}_{H\ell} + 2C^{(1)\alpha\beta}_{H\ell} \left( \DYl\Yl \right)^{}_{\gamma\lambda} \;,
	\nonumber
	\\
	\bda{C^{\alpha\beta\lambda\gamma}_{\ell u}} &=& \frac{4}{9} g^2_1 \delta^{\gamma\lambda} C^{(1)\alpha\beta}_{H\ell} - 2C^{(1)\alpha\beta}_{H\ell} \left( \DYu\Yu \right)^{}_{\gamma\lambda} \;,
	\nonumber
	\\
	\bda{C^{\alpha\beta\lambda\gamma}_{\ell d}} &=& - \frac{2}{9} g^2_1 \delta^{\gamma\lambda} C^{(1)\alpha\beta}_{H\ell}  + 2C^{(1)\alpha\beta}_{H\ell} \left( \DYd \Yd \right)^{}_{\gamma\lambda} \;.
 \label{eq:LLRRRGE}
\end{eqnarray}
\end{itemize}

Thus far, we have presented the complete set of one-loop RGEs in the type-I SEFT. Some remarks on the above results are helpful.
\begin{itemize}
	\item Only one-loop RGEs for the Wilson coefficients of the 9 dim-6 operators shown in the grey-shaded region in Table~\ref{tab:operators} and those for the Higgs quartic coupling acquire the contributions from double insertions of the unique dim-5 operator. Moreover, these results are generic and thus valid not only for the type-I SEFT but also for the SMEFT in general. For this reason, we separately list these results in Appendix~\ref{subapp:con-5}. Together with those given in Refs.~\cite{Jenkins:2013zja,Jenkins:2013wua,Alonso:2013hga,Alonso:2014zka}, they constitute the complete set of one-loop RGEs for the Wilson coefficients of dim-6 operators in the SMEFT. Therefore they can readily be incorporated into the codes solving the one-loop RGEs in the SMEFT, such as {\sf DsixTools}~\cite{Celis:2017hod,Fuentes-Martin:2020zaz}, {\sf Wilson}~\cite{Aebischer:2018bkb} and {\sf RGESolver}~\cite{DiNoi:2022ejg}.
	
	\item Single insertions of the two dim-6 operators, i.e., $\Op^{(1)}_{H\ell}$ and $\Op^{(3)}_{H\ell}$, contribute to all the 19 dim-6 operators in Table~\ref{tab:operators}, to the Higgs quartic coupling $\lambda$, and to the charged-lepton Yukawa coupling matrix $Y^{}_l$. These observations are perfectly consistent with those in the latest versions of Refs.~\cite{Jenkins:2013zja,Jenkins:2013wua,Alonso:2013hga} when the contributions from single insertions of other dim-6 operators are switched off. In this sense, we have accomplished a partial crosscheck of the previous results in the literature. It is worth pointing out that there exists a more convenient operator basis~\cite{Broncano:2004tz} for the type-I SEFT at the tree level, namely, $\left\{\Op^{(-)}_{H\ell}, \Op^{(+)}_{H\ell} \right\} = \left\{ \left( \Op^{(1)}_{H\ell} - \Op^{(3)}_{H\ell} \right)/2, \left( \Op^{(1)}_{H\ell} + \Op^{(3)}_{H\ell} \right)/2 \right\}$ with $\left\{ C^{(-)}_{H\ell}, C^{(+)}_{H\ell} \right\} = \left\{ C^{(1)}_{H\ell} - C^{(3)}_{H\ell}, C^{(1)}_{H\ell} + C^{(3)}_{H\ell} \right\}$. At the matching scale $\mu^{}_{\rm M}$, $ C^{(-)}_{H\ell} \left( \mu^{}_{\rm M} \right) = Y^{}_\nu M^{-2}_{\rm R} Y^\dagger_\nu /2$ and $C^{(+)}_{H\ell} \left( \mu^{}_{\rm M} \right) = 0$ hold. Therefore, only one effective operator $\Op^{(-)}_{H\ell}$ remains in this basis at the tree level. To convert all the above results into those in this particular basis, one can substitute $C^{(1)}_{H\ell}$ and $C^{(3)}_{H\ell}$ in Eqs.~\eqref{eq:YukawaRGE}-\eqref{eq:LLRRRGE} for $\left( C^{(-)}_{H\ell} + C^{(+)}_{H\ell} \right)/2$ and $\left( -C^{(-)}_{H\ell} + C^{(+)}_{H\ell} \right)/2$, respectively, and take $C^{(+)}_{H\ell}$ appearing on the right-hand sides of those equations to be zero. Nevertheless, we retain the Warsaw basis with $\Op^{(1)}_{H\ell}$ and $\Op^{(3)}_{H\ell}$ throughout this work to compare our results with those in previous works~\cite{Coy:2018bxr,Davidson:2018zuo,Jenkins:2013zja,Jenkins:2013wua,Alonso:2013hga} and keep results in a slightly more general form, which can be (partially) applied to other cases, such as the type-III SEFT.
	
	\item In the type-I SEFT, only 19 dim-6 operators can be generated via the RGEs at the one-loop level, which should be compared with 31 dim-6 operators arising from the one-loop matching \cite{Zhang:2021jdf}. The one-loop Wilson coefficients of those 19 dim-6 operators at the matching scale~\cite{Zhang:2021jdf} all contain divergent terms, which consist of both infrared (IR) and UV divergences from the hard-momentum part of loop integrals in the UV model. Those IR divergences are exactly the UV divergences (with a relative minus sign) in the EFT~\cite{Bilenky:1993bt,Dittmaier:2021fls}. This is the reason why only the 19 dim-6 operators can be induced by the one-loop RGEs.\footnote{There is an exception in the type-II SEFT, where the dim-6 operator $\Op^{}_{Hud}$ can be induced by one-loop RGEs as shown in Eq.~\eqref{eq:type-II-rge} but does not result from the one-loop matching. The reason is that the terms proportional to $\left( C^{}_{HD} - 2C^{}_{H\square} \right)$ are vanishing at the matching scale due to the tree-level matching condition $C^{}_{HD} = 2C^{}_{H\square}$. However, $C^{}_{HD} $ and $C^{}_{H\square}$ obey different RGEs as shown in Eq.~\eqref{eq:type-II-rge1}. As a result, though the running of $C^{}_{Hud}$ is determined by $\left( C^{}_{HD} - 2C^{}_{H\square}\right)$, $C^{}_{Hud}$ will obtain non-vanishing values below the matching scale.}
	
	\item As can be seen from Appendix \ref{app:counterterms}, the counterterms of the dim-6 operators $\Op^{}_{eB}$ and $\Op^{}_{eW}$ do not exist in the type-I SEFT. Therefore, radiative decays of charged leptons in the type-I SEFT, i.e, $l^{}_{\beta} \to l^{}_\alpha + \gamma$ [for $(\alpha,\beta) = (e,\mu),(e,\tau),(\mu,\tau)$], do not involve any divergences, as all divergences are canceled out among themselves. This is indeed the case since the one-loop Wilson coefficients of $\Op^{}_{eB}$ and $\Op^{}_{eW}$ are finite~\cite{Zhang:2021jdf}. The divergences in the contributions from active neutrinos turn out to be canceled with each other even though the leptonic flavor mixing matrix is not unitary~\cite{Xing:2020ivm}.
\end{itemize}
The above results have provided us with a self-consistent framework to examine the low-energy phenomena of the type-I seesaw model up to the order of ${\cal O}(M^{-2})$ at the one-loop level. Meanwhile, the one-loop RGEs in Appendix~\ref{app:SEFTs} can be implemented for the same purpose as for the type-II and type-III seesaw models. In the next section, we shall focus on the type-I SEFT and derive the explicit forms of the one-loop RGEs for the physical parameters involved in the charged- and neutral-current weak interactions of leptons. Similar studies can be carried out also for the type-II and type-III SEFTs.

\section{RG Running of Non-unitary Parameters}
\label{sec:explictRGEs}

The complete set of one-loop RGEs given in Eqs.~(\ref{eq:YukawaRGE})-(\ref{eq:LLRRRGE}) provide a link between the low-energy observables and the parameters associated with the type-I seesaw model at the matching scale. In general, it is necessary to take into account RG-running effects when one attempts to extract the UV model parameters from low-energy observables or begins with the UV model to predict low-energy observables. Therefore, as an application of one-loop RGEs, we are going to discuss the RG-running effects on the non-unitary leptonic flavor mixing in this section. The upper bounds on the non-unitary parameters obtained from low-energy observables will definitely be affected without considering the RG running. Since we focus only on the observables related to the leptonic flavor mixing, the ordinary parametrization of the non-unitary flavor mixing matrix will be adopted even above the electroweak scale. Though the parameters in this parametrization are not observable at high-energy scales, this specific parametrization can help us single out physical parameters from relevant coupling matrices. Moreover, one can discuss the overall RG-running effects on those physical parameters and examine their RG-running behaviors in an analytical way with such a parametrization. Certainly, the RGE corrections to physical parameters can also be studied in various interesting processes, such as the lepton-flavor-violating decays of $Z$-gauge boson and the Higgs boson $Z/H \to l^\pm_\alpha + l^\mp_\beta$ (with $\alpha \neq \beta = e, \mu, \tau$), the $\mu$-$e$ conversion in heavy nuclei and the cosmological baryon asymmetry via leptogenesis~\cite{Xing:2020ghj,Xing:2020erm}, which however are beyond the scope of the present work. In the following, we concentrate on the non-unitary leptonic flavor mixing and show how to derive and employ the analytical expressions of the one-loop RGEs for the relevant parameters.

\subsection{The Low-energy Lagrangian}
\label{subsec:Lag}

As is well known, after the spontaneous gauge symmetry breaking, the dim-5 Weinberg operator in Eq.~(\ref{eq:LSEFT}) accounts for tiny Majorana neutrino masses~\cite{Weinberg:1979sa}, while two dim-6 operators therein modify the weak interactions of leptons~\cite{Broncano:2002rw,Broncano:2003fq}. More explicitly, in the latter case, we have the extra contributions to the weak interactions, i.e.,
\begin{eqnarray}
	C_{Hl}^{(1)\alpha \beta} \mathcal{O}_{Hl}^{(1)\alpha \beta} &\rightarrow& -\frac{g_2^{}}{2 c_{\rm W}^{}} C_{Hl}^{(1)\alpha \beta} v^2_{} \left(\overline{\nu_{\alpha{\rm L}}^{}} \gamma^\mu_{} \nu_{\beta{\rm L}}^{} + \overline{l_{\alpha{\rm L}}^{}} \gamma^\mu_{} l_{\beta{\rm L}}^{}\right) Z_\mu^{} \;, \label{eq:OHl1}\\
	C_{Hl}^{(3)\alpha \beta} \mathcal{O}_{Hl}^{(3)\alpha \beta} &\rightarrow& + \frac{g_2^{}}{\sqrt{2}} C_{Hl}^{(3)\alpha \beta} v^2_{} \left( \overline{\nu_{\alpha{\rm L}}^{}} \gamma^\mu_{} l_{\beta{\rm L}}^{} W_\mu^+ + \overline{l_{\alpha{\rm L}}^{}} \gamma^\mu_{} \nu_{\beta{\rm L}}^{} W_\mu^- \right) \nonumber\\
	&& + \frac{g_2^{}}{2 c_{\rm W}^{}} C_{Hl}^{(3)\alpha \beta} v^2_{} \left(\overline{\nu_{\alpha{\rm L}}^{}} \gamma^\mu_{} \nu_{\beta{\rm L}}^{} - \overline{l_{\alpha{\rm L}}^{}} \gamma^\mu_{} l_{\beta{\rm L}}^{}\right) Z_\mu^{} \;, \label{eq:OHl3}
\end{eqnarray}
where $c^{}_{\rm W} \equiv \cos \theta^{}_{\rm W}$ has been defined with $\theta_{\rm W}^{}$ being the Weinberg angle and $v \approx 246~{\rm GeV}$ denotes the vacuum expectation value (vev) of the SM Higgs doublet $H$. As indicated in Eqs.~(\ref{eq:OHl1}) and (\ref{eq:OHl3}), $\mathcal{O}_{Hl}^{(1)}$ changes the neutral-current interactions of leptons, whereas $\mathcal{O}_{Hl}^{(3)}$ modifies both charged- and neutral-current interactions. Then the effective tree-level Lagrangian relevant for lepton masses and leptonic weak interactions in the type-I SEFT becomes
\begin{eqnarray}
	\mathcal{L}_{\rm SEFT}^{l} &=&  - \left[ \overline{l_{\alpha{\rm L}}^{}} \left(M_l^{}\right)_{\alpha\beta}^{} l_{\beta {\rm R}} + \frac{1}{2} \overline{\nu_{\alpha{\rm L}}^{}} \left(M_\nu^{}\right)_{\alpha\beta}^{} \nu_{\beta {\rm L}}^{\rm c} + {\rm h.c.} \right] \nonumber\\
	&~& +\left[ \frac{g_2^{}}{\sqrt{2}} \, \overline{l_{\alpha{\rm L}}^{}} \gamma^\mu_{} \left(\delta_{\alpha \beta}^{} - \tilde{\eta}_{\alpha \beta}^{} \right) \nu_{\beta {\rm L}}^{} W_\mu^- + {\rm h.c.} \right]+ \frac{g_2^{}}{2 c_{\rm W}^{}} \, \overline{\nu_{\alpha{\rm L}}^{}} \gamma^\mu_{} \left(\delta_{\alpha \beta}^{} - \tilde{\eta}_{\alpha \beta}^\prime \right) \nu_{\beta {\rm L}}^{} Z_\mu^{} \nonumber\\
	&~& -\frac{g_2^{}}{2 c_{\rm W}^{}} \, \overline{l_{\alpha{\rm L}}^{}} \gamma^\mu_{} \left[\left(1 - 2 s^2_{\rm W}\right) \delta_{\alpha \beta}^{} + \left( \tilde{\eta}^\prime - 2 \tilde{\eta} \right)_{\alpha \beta}^{} \right] l_{\beta {\rm L}}^{} Z_\mu^{} + \frac{g_2^{}}{c_{\rm W}^{}} s^2_{\rm W} \overline{l_{\alpha {\rm R}}^{}} \gamma^\mu_{} l_{\alpha {\rm R}}^{} Z_\mu^{}\;,
	\label{eq:effLagflavor}
\end{eqnarray}
where the charged-lepton mass matrix $M_l^{} = Y_l^{} \, v/\sqrt{2}$ and the neutrino mass matrix $M_\nu^{} = -C_5^{} \, v^2_{} /2$ are given in the first line, and the charged- and neutral-current interactions of leptons are shown in the last two lines with $s^{}_{\rm W} \equiv \sin \theta^{}_{\rm W}$ and
\begin{eqnarray}
	\tilde{\eta} \equiv - C_{Hl}^{(3)} \, v^2_{} \;, \quad \tilde{\eta}^\prime \equiv \left(C_{Hl}^{(1)} - C_{Hl}^{(3)}\right) v^2_{} \;.
	\label{eq:eta}
\end{eqnarray}
Notice that the charged- and neutral-current interactions of leptons are not flavor-diagonal anymore due to the presence of two dim-6 operators. After diagonalizing the lepton mass matrices via $U^\dagger_l M^{}_l U^{\prime}_l = \widehat{M}^{}_l \equiv {\rm diag}\{m^{}_e, m^{}_\mu, m^{}_\tau\}$ and $U^\dagger_\nu M^{}_\nu U^*_\nu = \widehat{M}^{}_\nu \equiv {\rm diag}\{m^{}_1, m^{}_2, m^{}_3\}$, with $m^{}_\alpha$ (for $\alpha = e, \mu, \tau$) and $m^{}_i$ (for $i = 1, 2, 3$) being respectively the charged-lepton and neutrino masses, one can introduce the $3\times 3$ unitary matrix $V^\prime \equiv U^\dagger_l U^{}_\nu$, which is just the leptonic flavor mixing matrix, i.e., the Pontecorvo-Maki-Nakagawa-Sakata (PMNS) matrix~\cite{Pontecorvo:1957cp,Maki:1962mu}, in the limit of vanishing Wilson coefficients of two dim-6 operators. For later convenience, we further explicitly write $V^\prime \equiv P \cdot U \cdot Q$, where $P \equiv {\rm diag}\{e^{{\rm i}\phi^{}_e}, e^{{\rm i}\phi^{}_\mu}, e^{{\rm i}\phi^{}_\tau}\}$ and $Q \equiv {\rm diag}\{e^{{\rm i}\rho}, e^{{\rm i}\sigma}, 1\}$ are two diagonal phase matrices and $U$ is a unitary matrix parametrized in the standard way~\cite{ParticleDataGroup:2022pth}
\begin{eqnarray}
    U = \begin{pmatrix}
        c_{12}^{} c_{13}^{} \quad&\quad s_{12}^{} c_{13}^{} \quad&\quad s_{13}^{} {\rm e}^{-{\rm i} \delta} \\
        - s_{12}^{} c_{23}^{} - c_{12}^{} s_{13}^{} s_{23}^{} {\rm e}^{{\rm i} \delta} \quad&\quad +c_{12}^{} c_{23}^{} - s_{12}^{} s_{13}^{} s_{23}^{} {\rm e}^{{\rm i} \delta} \quad&\quad c_{13}^{} s_{23}^{} \\
        +s_{12}^{} s_{23}^{} - c_{12}^{} s_{13}^{} c_{23}^{} {\rm e}^{{\rm i} \delta} \quad&\quad -c_{12}^{} s_{23}^{} - s_{12}^{} s_{13}^{} c_{23}^{} {\rm e}^{{\rm i} \delta} \quad&\quad c_{13}^{} c_{23}^{}
    \end{pmatrix}\;,
    \label{eq:Upara}
\end{eqnarray}
where $s^{}_{ij} \equiv \sin \theta^{}_{ij}$ and $c^{}_{ij} \equiv \cos \theta^{}_{ij}$ with $\theta^{}_{ij}$ for $ij = 12, 13, 23$ being three mixing angles, while $\delta$ and $\{\rho, \sigma\}$ are the Dirac-type and Majorana-type CP violating phases, respectively. It should be noticed that the mixing angles $\{\theta^{}_{12}, \theta^{}_{13}, \theta^{}_{23}\}$ and CP-violating phases $\{\rho, \sigma, \delta\}$ are not exactly the same as in the case of a unitary leptonic flavor mixing matrix, since the flavor mixing matrix appearing in the charged-current interaction of leptons is not unitary as will be shown shortly. But, for simplicity, we adopt the same notations for those parameters as they coincide with the ordinary ones in the unitarity limit.

Converting into the mass basis of charged leptons and neutrinos, we can rewrite the effective Lagrangian in Eq.~(\ref{eq:effLagflavor}) as below
\begin{eqnarray}
	\mathcal{L}_{\rm SEFT}^{l} &=& - \left( \overline{l_{{\rm L}}^{}} \, \widehat{M}_l^{} \, l_{ {\rm R}} + \frac{1}{2} \, \overline{\nu_{{\rm L}}^{}} \, \widehat{M}_\nu^{} \, \nu_{ {\rm L}}^{\rm c} + {\rm h.c.} \right) \nonumber\\
	&~& + \left( \frac{g_2^{}}{\sqrt{2}} \, \overline{l_{{\rm L}}^{}} \gamma^\mu_{} V \nu_{{\rm L}}^{} W_\mu^- + {\rm h.c.} \right) + \frac{g_2^{}}{2 c_{\rm W}^{}} \, \overline{\nu_{{\rm L}}^{}} \gamma^\mu_{} N^\dagger N \nu_{{\rm L}}^{} Z_\mu^{} \nonumber\\
	&~& -\frac{g_2^{}}{2 c_{\rm W}^{}} \, \overline{l_{{\rm L}}^{}} \gamma^\mu_{} \left[ \left(1 - 2 s_{\rm W}^2\right) + \left( \eta^\prime - 2 \eta \right) \right] l_{ {\rm L}}^{} Z_\mu^{}  + \frac{g_2^{} }{c_{\rm W}^{}} s_{\rm W}^2 \overline{l_{{\rm R}}^{}} \gamma^\mu_{} l_{{\rm R}}^{} Z_\mu^{}\;,
	\label{eq:effLagmass}
\end{eqnarray}
where the leptonic flavor mixing matrix in the charged-current interaction of leptons has been defined as $V \equiv (\mathbbm{1} - \eta) \cdot U \cdot Q$, and the coupling matrix $N \equiv (\mathbbm{1} - \eta^\prime/2) \cdot U \cdot Q$ has been introduced in the neutral-current interaction of neutrinos but the higher-order term of ${\cal O}(|\eta^\prime|^2)$ should be neglected. In addition, we have also redefined the phases of the mass eigenfields of charged-leptons to absorb the phase matrix $P$ and the Hermitian matrices $\eta$ and $\eta^\prime$ are given by
\begin{eqnarray}
    \eta &\equiv& P^\dagger U^\dagger_l 
    \tilde{\eta} \, U^{}_l P = \begin{pmatrix}
    \eta_{ee}^{} & \left|\eta_{e\mu}^{}\right| {\rm e}^{+{\rm i} \phi_{e\mu}^{}} & \left|\eta_{e\tau}^{}\right| {\rm e}^{+{\rm i} \phi_{e\tau}^{}} \\
     \left|\eta_{e\mu}^{}\right| {\rm e}^{-{\rm i} \phi_{e\mu}^{}} & \eta_{\mu\mu}^{} & \left|\eta_{\mu\tau}^{}\right| {\rm e}^{+{\rm i} \phi_{\mu\tau}^{}} \\
     \left|\eta_{e\tau}^{}\right| {\rm e}^{-{\rm i} \phi_{e\tau}^{}} & \left|\eta_{\mu\tau}^{}\right| {\rm e}^{-{\rm i} \phi_{\mu\tau}^{}} & \eta_{\tau\tau}^{}
    \end{pmatrix} \;, \label{eq:etapara}\\
    \eta^\prime &\equiv& P^\dagger U^\dagger_l 
    \tilde{\eta}^\prime U^{}_l P = \begin{pmatrix}
    \eta_{ee}^{\prime} & \left|\eta_{e\mu}^{\prime}\right| {\rm e}^{+{\rm i} \phi_{e\mu}^{\prime}} & \left|\eta_{e\tau}^{\prime}\right| {\rm e}^{+{\rm i} \phi_{e\tau}^{\prime}} \\
     \left|\eta_{e\mu}^{\prime}\right| {\rm e}^{-{\rm i} \phi_{e\mu}^{\prime}} & \eta_{\mu\mu}^{\prime} & \left|\eta_{\mu\tau}^{\prime}\right| {\rm e}^{+{\rm i} \phi_{\mu\tau}^{\prime}} \\
     \left|\eta_{e\tau}^{\prime}\right| {\rm e}^{-{\rm i} \phi_{e\tau}^{\prime}} & \left|\eta_{\mu\tau}^{\prime}\right| {\rm e}^{-{\rm i} \phi_{\mu\tau}^{\prime}} & \eta_{\tau\tau}^{\prime}
    \end{pmatrix} \;. \label{eq:etaprimepara}
\end{eqnarray}

As in Refs.~~\cite{Antusch:2006vwa,Antusch:2014woa}, the leptonic flavor mixing matrix $V$ is non-unitary and parametrized in terms of the Hermitian matrix $\eta$ and the mixing parameters in the unitary matrix $U$ and the phase matrix $Q$. It is worth stressing that the neutrino charged- and neutral-current interactions in the type-I SEFT are governed by the matrices $V$ and $N$, respectively, whose deviations from those in the SM are determined by $\eta$ and $\eta^\prime$. However, in the scheme of minimal unitarity violation (MUV) considered in Ref.~\cite{Antusch:2006vwa}, the condition $\eta^\prime = 2\eta$ valid only at the matching scale, as implied by Eqs.~(\ref{eq:match_condition}) and (\ref{eq:eta}), is imposed such that both $V$ and $N$ depend only on $\eta$. Moreover, as indicated in the last line of Eq.~(\ref{eq:effLagmass}), the flavor-changing-neutral-current (FCNC) interaction of charged leptons disappears in this case. As one can observe from Eq.~(\ref{eq:effLagmass}), this is not the case in general. Even if $\eta^\prime = 2\eta$ is assumed at some energy scale, the running effects according to their RGEs governed by Eq.~(\ref{eq:cHl1}) will break such an identity.

\subsection{Analytical Results}
\label{subsec:analytical}

Apart from the charged-lepton and neutrino masses, the physical parameters relevant for low-energy phenomena are contained in the two Hermitian matrices $\eta$ and $\eta^\prime$, the mixing matrix $U$ and the phase matrix $Q$. In this subsection, we start with the one-loop RGEs in the type-I SEFT and derive the analytical expressions for the RGEs of the physical parameters. Then, adopting the specific parametrizations of $V^\prime$, $\eta$ and $\eta^\prime$ given in the previous subsection, we further find out the RGE of each individual parameter. 

However, one should notice that the non-vanishing values of $C^{}_H$, $C^{}_{He}$ and $C^{}_{eH}$ induced by the RG running can affect the vev of Higgs field, the neutral-current interactions of charged leptons and the charged-lepton masses, respectively. Those effects should be included in Eq.~\eqref{eq:effLagmass} for consistently studying the RG-running behaviors of all relevant physical parameters. Unfortunately, in this case, it is almost impossible to derive the analytical expressions of the RGEs for most of the physical parameters and only numerical calculations can be carried out to examine their running behaviors. This will hinder us from a clear understanding of non-unitary effects on the running of all physical parameters. In order to illustrate the running behaviors of the relevant parameters analytically, we simply ignore the non-vanishing values of $C^{}_H$, $C^{}_{He}$ and $C^{}_{eH}$ from the RG running, especially that of the last one. Therefore, we only need to take into account the RGEs for $Y^{}_l$, $C^{}_5$, $C^{(1)}_{H\ell}$ and $C^{(3)}_{H\ell}$, from which the differential RGEs of all physical parameters can be derived. But one should keep in mind that a contribution of $\mathcal{O} \left[ v^2/\left( 16 \pi^2 M^2\right) \right]$ from $C^{}_H$, $C^{}_{He}$ and $C^{}_{eH}$ to some physical parameters has been ignored. As one will see later, the differential RGEs for all lepton mixing parameters under the standard parametrization of the PMNS matrix are obtained as a by-product for the first time if all effects of the dim-6 operators are switched off.

According to the RGEs of $Y_l^{}$ in Eq.~(\ref{eq:YukawaRGE}), $C_5^{}$ in Eq.~(\ref{eq:C5RGE}), $C_{Hl}^{(1)}$ and $C_{Hl}^{(3)}$ in Eq.~(\ref{eq:cHl1}), as well as the definitions of $\eta$ and $\eta^\prime$ in Eqs.~\eqref{eq:eta}, (\ref{eq:etapara}) and (\ref{eq:etaprimepara}), one can find
\begin{eqnarray}
    \dot{Y_l^{}} &=& \left[ \alpha_l^{} + C_l^l \left( Y_l^{} Y_l^\dagger \right) - 2 \, \frac{m^2_{}}{v^2_{}} U_l^{} P \left(\eta^\prime_{} - 4 \eta \right) P^\dagger U_l^\dagger\right] Y_l^{} \;, \label{eq:RGEYlnew} \\
    \dot{\kappa} &=& \alpha_\kappa^{} \kappa + C_\kappa^{} \left[ \left( Y_l^{} Y_l^\dagger \right) \kappa + \kappa  \left( Y_l^{} Y_l^\dagger \right)^{\rm T}_{} \right] \;, \label{eq:RGEkappa} \\
	\dot{\eta}_{\alpha\beta}^{} &=& {\rm i} \left(\dot{\phi}_\beta^{} - \dot{\phi}_\alpha^{} \right) \eta_{\alpha\beta}^{} + \sum_{\gamma }^{} \left( \dot{U}_l^\dagger U_l^{} \right)_{\alpha\gamma}^{}  \eta_{\gamma\beta}^{} \, {\rm e}^{{\rm i} \left( \phi_\gamma^{} - \phi_\alpha^{} \right)}_{} + \sum_{\gamma}^{} \left(U_l^\dagger  \dot{U}_l^{} \right)_{\gamma\beta}^{} \eta_{\alpha\gamma}^{} \, {\rm e}^{{\rm i}\left( \phi_\beta^{} - \phi_\gamma^{}\right)}_{}  \nonumber \quad \\
	&&- \sum_i^{} \kappa_i^2 v^2_{} U_{\alpha i}^{} U_{\beta i}^\ast + \frac{2}{3} g_2^2 {\rm tr} \left(\eta\right) \delta_{\alpha\beta}^{} - \left(\frac{17}{3} g_2^2 - 2 T\right) \eta_{\alpha\beta}^{}+ \frac{1}{2} \left( y_\alpha^2 + y_\beta^2 \right) \left(5 \eta_{\alpha\beta}^{} - 3 \eta_{\alpha\beta}^\prime\right) ,  \label{eq:RGEeta} \\
	\dot{\eta}_{\alpha\beta}^\prime &=&  {\rm i} \left(\dot{\phi}_\beta^{} - \dot{\phi}_\alpha^{} \right) \eta_{\alpha\beta}^\prime + \sum_{\gamma }^{} \left( \dot{U}_l^\dagger U_l^{} \right)_{\alpha\gamma}^{}  \eta_{\gamma\beta}^{\prime} \, {\rm e}^{{\rm i} \left( \phi_\gamma^{} - \phi_\alpha^{} \right)}_{} + \sum_{\gamma}^{} \left(U_l^\dagger  \dot{U}_l^{} \right)_{\gamma\beta}^{} \eta_{\alpha\gamma}^{\prime} \, {\rm e}^{{\rm i}\left( \phi_\beta^{} - \phi_\gamma^{}\right)}_{} \nonumber\\
	&&+ \frac{2}{3} \left( g_2^2- g_1^2 \right) {\rm tr} \left(\eta\right) \delta_{\alpha\beta}^{} + \frac{2}{3} g_1^2 {\rm tr} \left(\eta^\prime_{}\right) \delta_{\alpha\beta}^{} - \frac{1}{3} \left( 17 g_2^2 + g_1^2 \right) \eta_{\alpha\beta}^{} +\left(\frac{1}{3} g_1^2 + 2 \, T\right) \eta_{\alpha\beta}^\prime \nonumber\\
	&&- \frac{5}{2} \sum_i^{} \kappa_i^2 v^2_{} U_{\alpha i}^{} U_{\beta i}^\ast + \frac{1}{2} \left( y_\alpha^2 + y_\beta^2 \right) \left( \eta_{\alpha\beta}^\prime - 8 \eta_{\alpha\beta}^{}\right) \;, \label{eq:RGEetaprime}
\end{eqnarray}
where $\alpha, \beta, \gamma = e, \mu, \tau$ are lepton flavor indices, $C_l^l = 3/2$, $C_\kappa^{} = -3/2$ and $\alpha_l^{} = - (15/4) \, g_1^2 - (9/4) \,g_2^2 + T$, $\alpha_\kappa^{} = - 3 g_2^2 + 4 \lambda + 2 \, T$ have been defined, $\kappa \equiv C^{}_5$ is introduced for convenience, and the notation $\dot{X} \equiv 16 \pi^2 \mu \,( {\rm d} X/{\rm d} \mu)$ has been used for $X = Y^{}_l$, $\kappa$, $\eta$ and $\eta^\prime$. In addition, the charged-lepton Yukawa coupling matrix is diagonalized as $U^\dagger_l Y^{}_l U^\prime_l = {\rm diag}\{y^{}_e, y^{}_\mu, y^{}_\tau\} \equiv \widehat{Y}^{}_l$, while $U^\dagger_\nu \kappa U^*_\nu = {\rm diag}\{\kappa^{}_1, \kappa^{}_2, \kappa^{}_3\} \equiv \widehat{\kappa}$. Some helpful comments on Eqs.~(\ref{eq:RGEYlnew})-(\ref{eq:RGEetaprime}) are in order. 
\begin{itemize}
     \item Although three phases $\{\phi^{}_e, \phi^{}_\mu, \phi^{}_\tau\}$ in the diagonal matrix $P$ can be absorbed by redefining the charged-lepton fields at a given energy scale and thus are unphysical, they do appear in the RGEs of $Y^{}_l$, $\eta$ and $\eta^\prime$. The reason is simply that one has to diagonalize the running matrices $Y^{}_l$ and $\kappa$ at each energy scale to obtain the unitary matrix $V^\prime$ and then specify the phases to be absorbed by the charged-lepton fields. Therefore, the unphysical phases $\phi^{}_\alpha$ (for $\alpha = e, \mu, \tau$) themselves are running with respect to the energy scale. As we shall see soon, however, the RGEs of the physical parameters in $V^\prime$, $\eta$ and $\eta^\prime$ are independent of these unphysical phases.
     
     \item It should be noted that the RGEs in Eqs.~(\ref{eq:RGEYlnew})-(\ref{eq:RGEetaprime}) are reduced to the familiar ones in the type-I SEFT with only the dim-5 operator when $\eta$ and $\eta^\prime$ are set to zero. With the contributions from $\eta$ and $\eta^\prime$, the evolution of $Y^{}_l$ becomes qualitatively different. Even if $Y^{}_l$ is diagonalized at one energy scale, it turns out to be non-diagonal at another. As a consequence, the running behaviors of the mixing parameters in $V^\prime = U^\dagger_l U^{}_\nu$ will be affected via the unitary matrix $U^{}_l$ arising from the charged-lepton sector. 
     
     \item There are nine CP-violating phases in total, namely, $\{\phi^{}_{\alpha \beta}, \phi^\prime_{\alpha \beta}\}$ for $\alpha = e\mu, e\tau, \mu\tau$ from $\eta$ and $\eta^\prime$ and $\{\rho, \sigma, \delta\}$ from $V^\prime$, and their evolution is entangled with each other. Therefore, as long as CP violation exists in the UV model, even if one of the CP-violating phases is vanishing at some energy scale, it will be radiatively generated through other non-trivial phases. Certainly, when CP symmetry is preserved at the high-energy scale, it remains to hold at low energies.
\end{itemize}

By means of Eqs.~\eqref{eq:RGEYlnew}-\eqref{eq:RGEetaprime}, one can derive the analytical expressions for the RGEs of all physical parameters involved in the non-unitary lepton flavor mixing. As explained in Appendix~\ref{app:rge}, we first follow the standard procedure of diagonalizing the matrices $Y_l^{}$ and $\kappa$, and then extract the RGEs of their eigenvalues and the associated unitary matrices $U^{}_l$ and $U^{}_\nu$. The RGEs of the eigenvalues of $Y_l^{}$ and $\kappa$ are
\begin{eqnarray} \label{eq:rge-lep}
	\dot{y}_\alpha = \left[ \alpha_l^{} + C_l^l \, y_\alpha^2 - 2 \, \frac{m^2_{}}{v^2_{}} \left(\eta^\prime_{} - 4 \eta \right)_{\alpha \alpha}^{} \right] y_\alpha \;, 
\end{eqnarray}
and 
\begin{eqnarray} \label{eq:rge-neu}
	\dot{\kappa}_i^{} = \left( \alpha_\kappa^{} + 2 C_\kappa^{} {\rm Re} {\cal S}_{ii}^{} \right) \kappa_i^{} \;, 
\end{eqnarray}
with ${\cal S} \equiv V^{\prime \dagger}_{} \widehat{Y}_l^2 V^\prime$ and $V^\prime \equiv U^\dagger_l U^{}_\nu$. The RGEs of two unitary mixing matrices $U_l^{}$ and $U_\nu^{}$ are given in Eqs.~\eqref{eq:Uloffd} and \eqref{eq:Unuoffd}. Consequently, the RGEs in Eqs.~\eqref{eq:RGEYlnew}-\eqref{eq:RGEetaprime} have been converted into those for the eigenvalues of $Y^{}_l$ and $\kappa$ given in Eqs.~\eqref{eq:rge-lep} and \eqref{eq:rge-neu}, respectively, and those for the unitary matrix $V^\prime$ and the Hermitian matrices $\left\{\eta, \eta^{\prime} \right\}$, namely,
\begin{eqnarray}
\dot{V}_{\alpha i}^{\prime} &=& \sum_\beta \left( \dot{U}_l^\dagger U_l^{} \right)_{\alpha\beta}^{} V_{\beta i}^{\prime} + \sum_j V_{\alpha j}^{\prime} \left( U_\nu^\dagger \dot{U}_\nu^{} \right)_{ji}^{} 
\nonumber
\\
&=& \sum_{\beta \neq \alpha} 2 \, \frac{m^2_{}}{v^2_{}} y_{\alpha\beta}^{} \left(\eta^\prime - 4\eta\right)_{\alpha \beta}^{} {\rm e}^{{\rm i} \left(\phi_\alpha^{}-\phi_\beta^{} \right)}_{} V_{\beta i}^{\prime} + \sum_{j\neq i} V_{\alpha j}^{\prime} \frac{ C_\kappa^{} }{\kappa_i^2 - \kappa_j^2} \left[ \left(\kappa_i^2 + \kappa_j^2\right) {\cal S}_{ji}^{} + 2 \kappa_i^{} \kappa_j^{} {\cal S}_{ji}^\ast \right] \;,
\label{eq:pmnsrge}
 \\
 \dot{\eta}_{\alpha\beta}^{} &=& {\rm i} \left(\dot{\phi}_\beta^{} - \dot{\phi}_\alpha^{} \right) \eta_{\alpha\beta}^{} + \sum_{\gamma \neq \alpha}^{} 2 \, \frac{m^2_{}}{v^2_{}}  y^{}_{\alpha\gamma} \left(\eta^\prime_{} - 4 \eta\right)_{\alpha \gamma}^{} \eta_{\gamma\beta}^{} + \sum_{\varrho \neq \beta}^{} 2 \, \frac{m^2_{}}{v^2_{}} y^{}_{\beta \varrho} \, \eta_{\alpha\varrho}^{} \left(\eta^\prime_{} - 4 \eta\right)_{\varrho\beta}^{} 
 \nonumber
 \\
&-& \sum_i^{} \kappa_i^2 v^2_{} U_{\alpha i}^{} U_{\beta i}^\ast + \frac{2}{3} g_2^2 {\rm tr} \left(\eta\right) \delta_{\alpha\beta}^{} + \left(-\frac{17}{3} g_2^2 + 2 T\right) \eta_{\alpha\beta}^{}+ \frac{1}{2} \left( y_\alpha^2 + y_\beta^2 \right) \left(5 \eta_{\alpha\beta}^{} - 3 \eta_{\alpha\beta}^\prime\right) \;,
\label{eq:rge-eta}
\\
\dot{\eta}_{\alpha\beta}^\prime &=&  {\rm i} \left(\dot{\phi}_\beta^{} - \dot{\phi}_\alpha^{} \right) \eta_{\alpha\beta}^\prime + \sum_{\gamma \neq \alpha}^{} 2 \, \frac{m^2_{}}{v^2_{}} y^{}_{\alpha\gamma} \left(\eta^\prime_{} - 4 \eta\right)_{\alpha \gamma}^{} \eta_{\gamma\beta}^\prime + \sum_{\varrho \neq \beta}^{} 2 \, \frac{m^2_{}}{v^2_{}} y^{}_{\beta\varrho} \, \eta_{\alpha\varrho}^\prime \left(\eta^\prime_{} - 4 \eta\right)_{\varrho\beta}^{} 
\nonumber
\\
&+& \frac{2}{3} \left( g_2^2- g_1^2 \right) {\rm tr} \left(\eta\right) \delta_{\alpha\beta}^{} + \frac{2}{3} g_1^2 {\rm tr} \left(\eta^\prime_{}\right) \delta_{\alpha\beta}^{} - \frac{1}{3} \left( 17 g_2^2 + g_1^2 \right) \eta_{\alpha\beta}^{} +\left(\frac{1}{3} g_1^2 + 2 \, T\right) \eta_{\alpha\beta}^\prime 
\nonumber
\\
&-& \frac{5}{2} \sum_i^{} \kappa_i^2 v^2_{} U_{\alpha i}^{} U_{\beta i}^\ast + \frac{1}{2} \left( y_\alpha^2 + y_\beta^2 \right) \left( \eta_{\alpha\beta}^\prime - 8 \eta_{\alpha\beta}^{}\right) \;,
\label{eq:rge-etap}
\end{eqnarray}
where $(\dot{U}^\dagger_\nu U^{}_\nu)^{}_{i i} =0 $ (for $i=1,2,3$) and ${\rm Re} [ (\dot{U}^\dagger_l U^{}_l)^{}_{\alpha \alpha} ] = 0$ (for $\alpha = e, \mu, \tau$) have been taken into account, and terms involving ${\rm Im} [ (\dot{U}^\dagger_l U^{}_l)^{}_{\alpha \alpha} ]$ have been discarded in each equation. Fortunately, one can throw them away safely though ${\rm Im} [ (\dot{U}^\dagger_l U^{}_l)^{}_{\alpha \alpha} ]$ is unknown at all. Because the terms involving ${\rm Im} [ (\dot{U}^\dagger_l U^{}_l)^{}_{\alpha \alpha} ]$ in Eqs.~\eqref{eq:pmnsrge}-\eqref{eq:rge-etap} contribute to the beta function in the form of $\dot{X}= {\rm i} \mathbbm{R} X$ with $\mathbbm{R}$ being real, they make contributions only to the unphysical phases in the end and do not affect the results for any physical parameters.

To derive the RGEs for all physical parameters involved in leptonic weak interactions, we can substitute $V^\prime = P\cdot U \cdot Q$ with the previously specified parametrizations of the matrices $P$, $Q$ and $U$ into Eq.~(\ref{eq:pmnsrge}) and work out results for mixing angles and phases in $V^\prime$ first. Then, with the help of the parameterizations shown in Eqs.~\eqref{eq:etapara} and \eqref{eq:etaprimepara}, results for the moduli and phases of elements in $\eta$ and $\eta^\prime$ can be easily achieved via Eqs.~\eqref{eq:rge-eta} and \eqref{eq:rge-etap}. The calculational details and the final results can be found in Appendix~\ref{app:rge}. Our results for three mixing angles and three CP-violating phases in Eqs.~\eqref{eq:theta12}-\eqref{eq:delta} will be reduced to those in the case where all non-unitary effects are switched off and the standard parametrization of the PMNS matrix is adopted. Although the analytical expressions of the one-loop RGEs of leptonic flavor mixing parameters in the standard parametrization~\cite{Casas:1999tg,Antusch:2003kp,Antusch:2005gp,Mei:2005qp} and those in the Fritzsch-Xing parametrization~\cite{Xing:2005fw} have been derived in the unitarity limit, the strong hierarchy among charged-lepton Yukawa couplings $y_\tau \gg y_e^{}, y_\mu^{}$ has been implemented to derive approximate results. In this sense, as a by-product, we have obtained for the first time the RGEs of the mixing parameters in the standard parametrization without any approximations.

\subsection{Numerical Results}

In the type-I SEFT, once the model parameters (i.e., $Y^{}_\nu$, $M^{}_{\rm R}$ and those already present in the SM) are given at the matching scale, one can determine the Wilson coefficients of the dim-5 and dim-6 operators and then solve the set of RGEs in Eqs.~\eqref{eq:YukawaRGE}-\eqref{eq:C5RGE} together with the first two in Eq.~\eqref{eq:psi2h2d} [or the RGEs in Eqs.~\eqref{eq:rge-lep}, \eqref{eq:rge-neu} and \eqref{eq:theta12}-\eqref{eq:phimutaup} instead of those for $Y^{}_l$, $C^{}_5$ and $C^{(1,3)}_{H\ell}$] numerically to obtain the relevant physical parameters at the electroweak scale. To examine the strength of running effects on relevant parameters and look into their running behaviors, we consider the following two different scenarios.
\begin{itemize}
    \item {\bf SEFT}: This scenario refers to the type-I SEFT, in which the matching condition $\eta^\prime = 2\eta$ should be satisfied at the matching scale $\mu^{}_{\rm M}$, as implied by Eq.~(\ref{eq:match_condition}). Instead of arbitrarily choosing the model parameters at the matching scale, we first take their low-energy values measured or constrained by experiments as initial conditions and solve the RGEs numerically to estimate the values at the matching scale. Then, the values of $\eta^\prime$ and $\eta$ are adjusted to fulfill the identity $\eta^\prime = 2\eta$ at $\mu^{}_{\rm M}$ whereas others are kept the same or slightly modified. With all these properly chosen values at $\mu^{}_{\rm M}$ as initial conditions, we again solve the RGEs to obtain those at the low-energy scale. In this way, the obtained values should automatically fall into the experimentally-allowed regions.
    
    \item {\bf EFT}: In this scenario, we regard the effective Lagrangian in Eq.~(\ref{eq:LSEFT}) as a general EFT, so the Wilson coefficients of the dim-5 and dim-6 operators are free parameters. Given the same values of relevant parameters at the low-energy scale as in the previous scenario, we numerically solve the same set of RGEs and evaluate the values at an arbitrary high-energy scale. Such an analysis will be instructive to see how much the effective parameters depend on the energy scale, in particular for the underlying UV model at a superhigh-energy scale.
\end{itemize}

With the above setup, we now specify the input values and explain the benchmark values of some free parameters assumed for illustration. First of all, we choose a benchmark value of the low-energy scale $\mu^{}_{\rm B} = 200$ GeV and summarize the corresponding values of all involved parameters at this energy scale in Table~\ref{table: initial values}. Some explanations for the input values are necessary.
\begin{table}[t!]
	\centering
	\begin{tabular}{c|c||c|c||c|c||c|c}
		\hline \hline
		$m_u^{}/{\rm MeV}$ & 1.2504 & $m_e^{}/{\rm MeV}$ & 0.5239 & $g_1^{}$ & 0.3589 & $\eta_{ee}^{}$, $\eta_{ee}^\prime/2$ & $1.25 \times 10^{-3}_{}$\\
		\hline
		$m_d^{}/{\rm MeV}$ & 2.7176 & $m_\mu^{}/{\rm GeV}$ & 0.1104 & $g_2^{}$ & 0.6468 & $\eta_{\mu\mu}^{}$, $\eta_{\mu\mu}^\prime/2$ & $2.21 \times 10^{-4}_{}$\\
		\hline
		$m_s^{}/{\rm MeV}$ & 54.120 & $m_\tau^{}/{\rm GeV}$ & 1.8748 & $g_s^{}$ & 1.1525 & $\eta_{\tau\tau}^{}$, $\eta_{\tau\tau}^\prime/2$ & $2.81 \times 10^{-3}_{}$\\
		\hline
		$m_c^{}/{\rm GeV}$ & 0.6299 & $m_1^{}/{\rm eV}$ & 0.05 & 
        $\lambda$ & 0.1235 & $|\eta_{e\mu}^{}|$, $|\eta_{e\mu}^\prime|/2$ & $1.20 \times 10^{-5}_{}$\\
		\hline
		$m_b^{}/{\rm GeV}$ & 2.8731 & $m_2^{}/{\rm eV}$ & 0.05074 & $m^2_{}/{\rm GeV}^2_{}$ & $-8672.61$ & $|\eta_{e\tau}^{}|$, $|\eta_{e\tau}^\prime|/2$ & $1.35 \times 10^{-3}_{}$ \\
		\hline
		$m_t^{}/{\rm GeV}$ & 173.075 & $m_3^{}/{\rm eV}$ & 0.07079 & $\delta_{}^{\rm q}$ & 1.144 & $|\eta_{\mu\tau}^{}|$, $|\eta_{\mu\tau}^\prime|/2$ & $6.13 \times 10^{-4}_{}$\\
		\hline
	    $\sin \theta_{12}^{\rm q}$ & 0.2250 & $\sin \theta_{12}^{}$ & 0.5505 & $\delta$ & 3.438 & $\phi_{e\mu}^{}$, $\phi_{e\mu}^\prime$& $\pi/3$ \\
		\hline
		$\sin \theta_{23}^{\rm q}$ & 0.04182 & $\sin \theta_{23}^{}$ & 0.7563 & $\rho$ & $\pi/6$  &$\phi_{e\tau}^{}$, $\phi_{e\tau}^\prime$& $\pi/3$ \\
		\hline
		$\sin \theta_{13}^{\rm q}$ & 0.00369 & $\sin \theta_{13}^{}$ & 0.1484 & $\sigma$ & $\pi/4$ &$\phi_{\mu\tau}^{}$, $\phi_{\mu\tau}^\prime$ & $\pi/3$ \\
		\hline
		\hline
	\end{tabular}
	\vspace{0.2cm}
	\caption{Summary of the input values of all the relevant parameters at the benchmark energy scale $\mu^{}_{\rm B} = 200~{\rm GeV}$. See the main text for further explanations.}
	\label{table: initial values}
\end{table}

\begin{enumerate}
    \item The quark and lepton Yukawa couplings, gauge couplings $\{g_{1}^{}, g^{}_2, g^{}_s \}$ and the Higgs couplings $\{\lambda, m^2_{}\}$ are taken from Ref.~\cite{Alam:2022cdv}, where all these SM parameters are evaluated in the $\overline{\rm MS}$ scheme at the energy scale $\mu^{}_{\rm B} = 200~{\rm GeV}$. The vev of the Higgs field is given at the true vacuum $v = \sqrt{-m^2_{}/\lambda}$ at $\mu^{}_{\rm B}$. In our calculations, $v$ is fixed as a normalization constant for the parameters of mass-dimension one while $m^2$ and $\lambda$ are subject to the RG running.
    
    \item The quark flavor mixing angles $\{\theta^{\rm q}_{12}, \theta^{\rm q}_{13}, \theta^{\rm q}_{23}\}$ and the CP-violating phase $\delta^{\rm q}$ in the Cabibbo-Kobayashi-Maskawa (CKM) matrix are taken from Ref.~\cite{ParticleDataGroup:2022pth}. The leptonic flavor mixing angles $\{\theta^{}_{12}, \theta^{}_{13}, \theta^{}_{23}\}$, the Dirac CP-violating phase $\delta$ and neutrino mass-squared differences $\Delta m^2_{21} \equiv m^2_2 - m^2_1$ and $\Delta m^2_{31} \equiv m^2_3 - m^2_1$ in the case of normal mass ordering are quoted from the global-fit analysis of neutrino oscillation experiments in NuFIT 5.2~\cite{Esteban:2020cvm}. For illustration, we assume the normal neutrino mass ordering $m^{}_1 < m^{}_2 < m^{}_3$ and take the lightest neutrino mass to be $m^{}_1 = 0.05~{\rm eV}$. Since the Majorana CP-violating phases $\{\rho, \sigma\}$ are completely unknown for the moment, we simply set them to be $\rho = \pi/6$ and $\sigma = \pi/4$.
    
    \item The upper bounds on the non-unitary parameters in $\eta$ from the global-fit analysis in Ref.~\cite{Fernandez-Martinez:2016lgt} are chosen.\footnote{In Ref.~\cite{Fernandez-Martinez:2016lgt}, the global constraints are obtained in the full theory, where both the charged- and neutral-current interactions of leptons depend only on $\eta$ because of the condition $\eta^\prime = 2\eta$. A similar issue exists for those obtained in the MUV scheme~\cite{Antusch:2006vwa,Antusch:2014woa}. However, from the EFT perspective, the relation $\eta^\prime=2\eta$ holds only at the matching scale. The different RGEs of $\eta$ and $\eta^\prime$ lead to the breaking of such a relation at low energies, where the physical processes are implemented to constrain $\eta$. Even so, we still adopt these constraints as inputs for illustration.} The parameters in $\eta^\prime$ at low energies are in principle different from those in $\eta$, but their magnitudes may be comparable. For simplicity, we take $|\eta^{\prime}_{\alpha \beta}| = 2|\eta^{}_{\alpha \beta}|$ for $\alpha, \beta = e, \mu, \tau$ at $\mu^{}_{\rm B}$. Furthermore, as the CP-violating phases $\{\phi^{}_{e\mu}, \phi^{}_{e\tau}, \phi^{}_{\mu \tau}\}$ and $\{\phi^{\prime}_{e\mu}, \phi^{\prime}_{e\tau}, \phi^{\prime}_{\mu \tau}\}$ are not constrained experimentally, we assume all of them to be $\pi/3$ just for an illustrative purpose.
\end{enumerate}    

\begin{figure}
	\centering
	\;\includegraphics[scale=0.62]{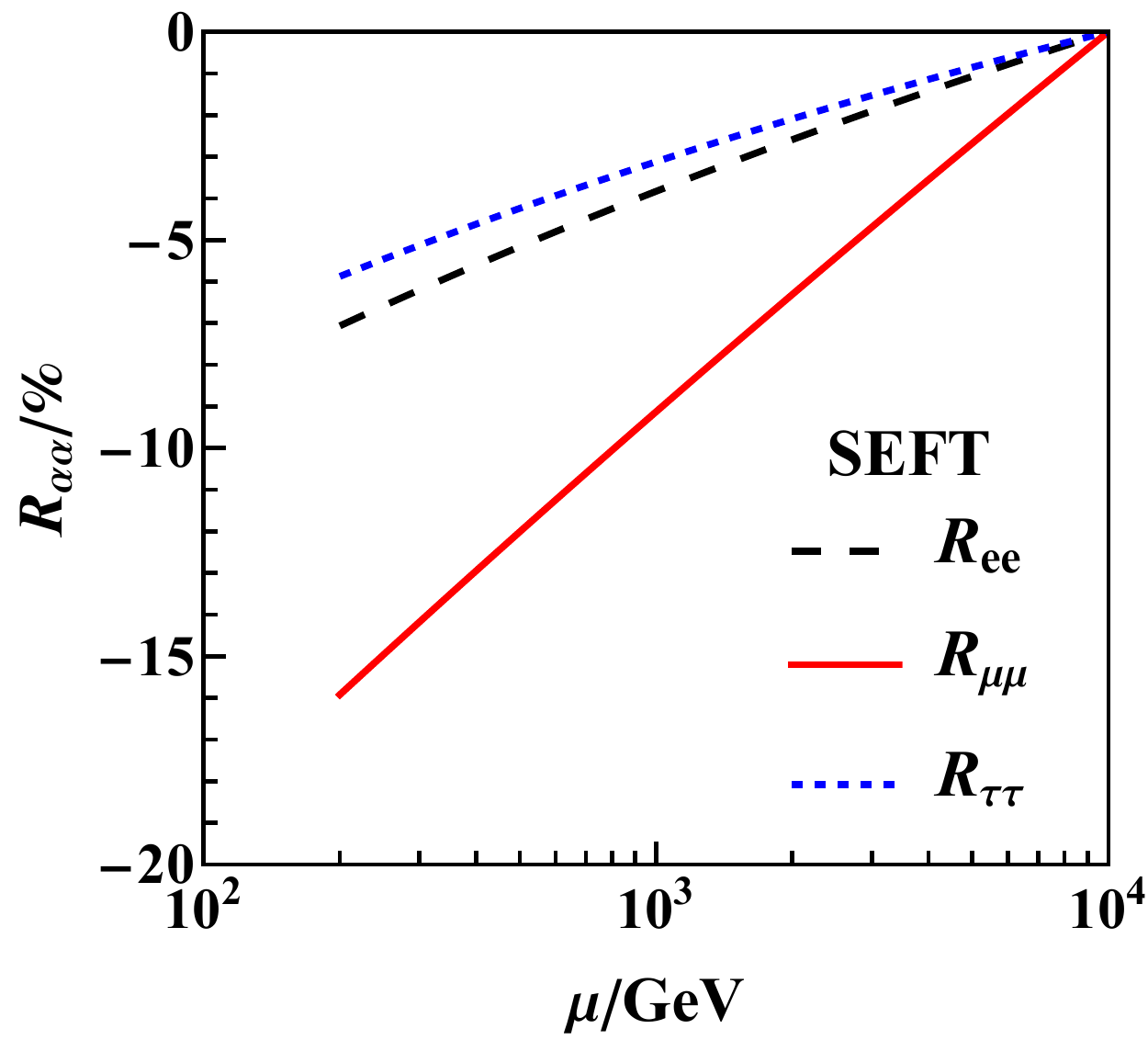}\quad
	\includegraphics[scale=0.62]{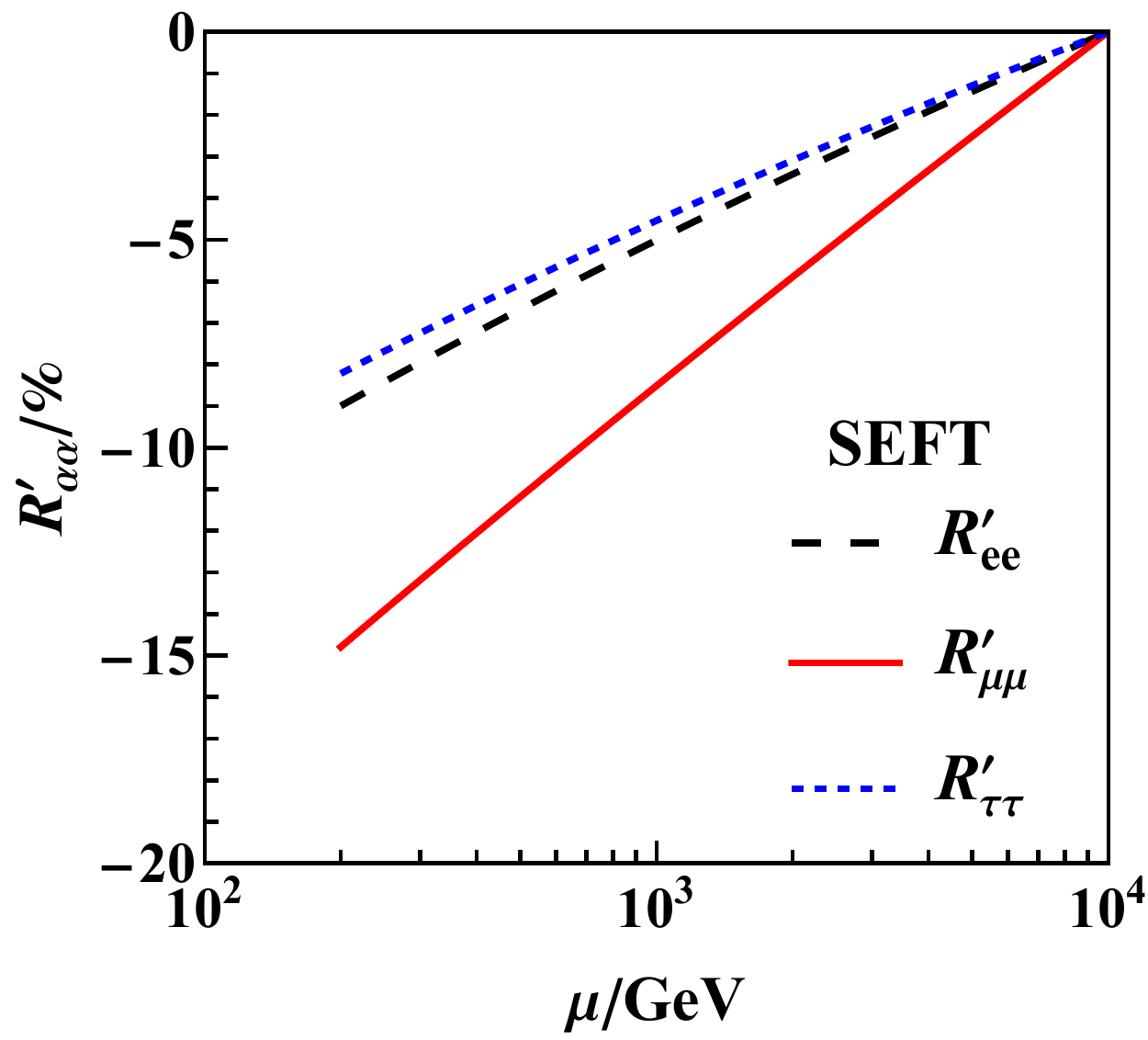}\\
	\includegraphics[scale=0.63]{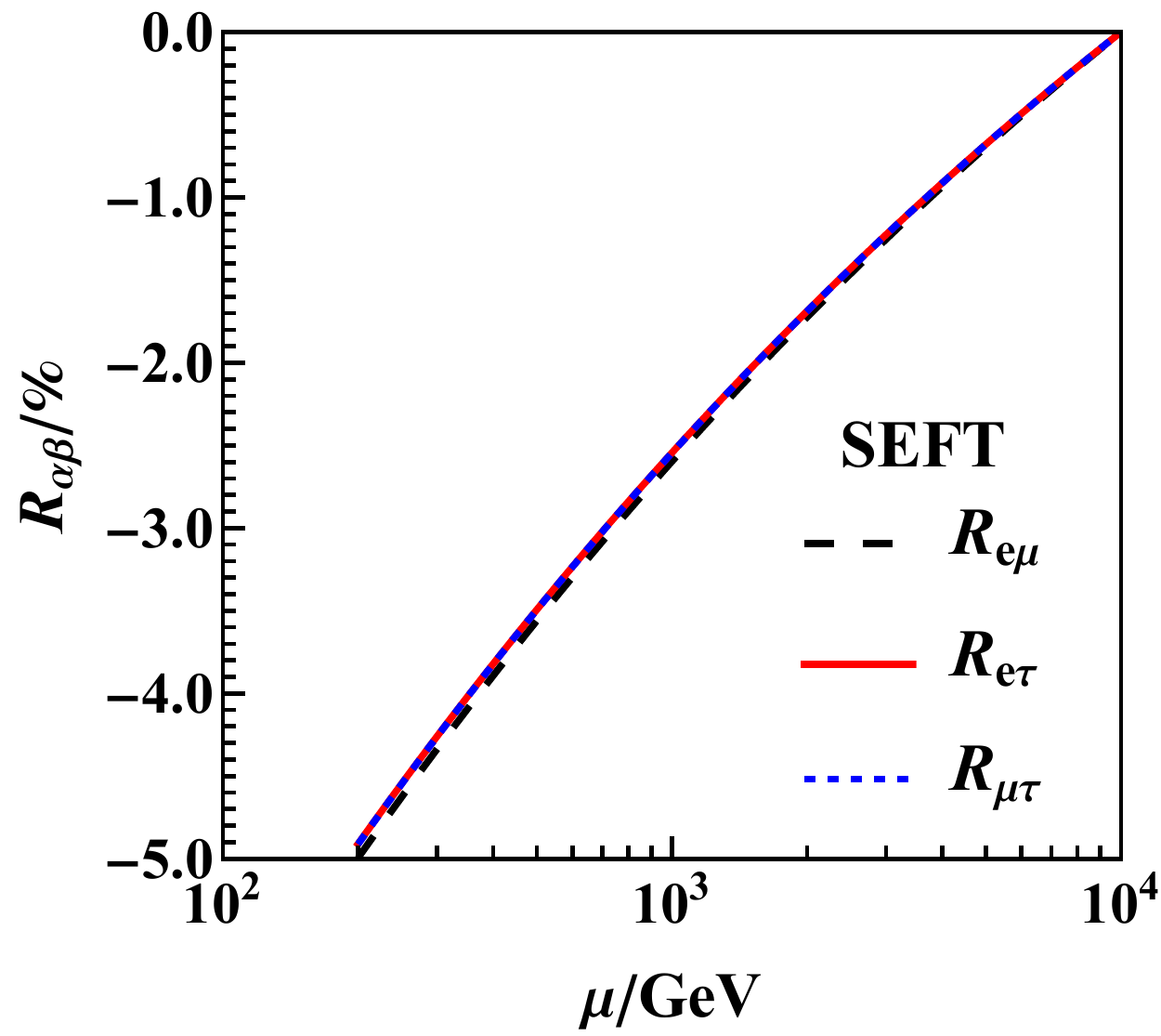}\;
    \includegraphics[scale=0.63]{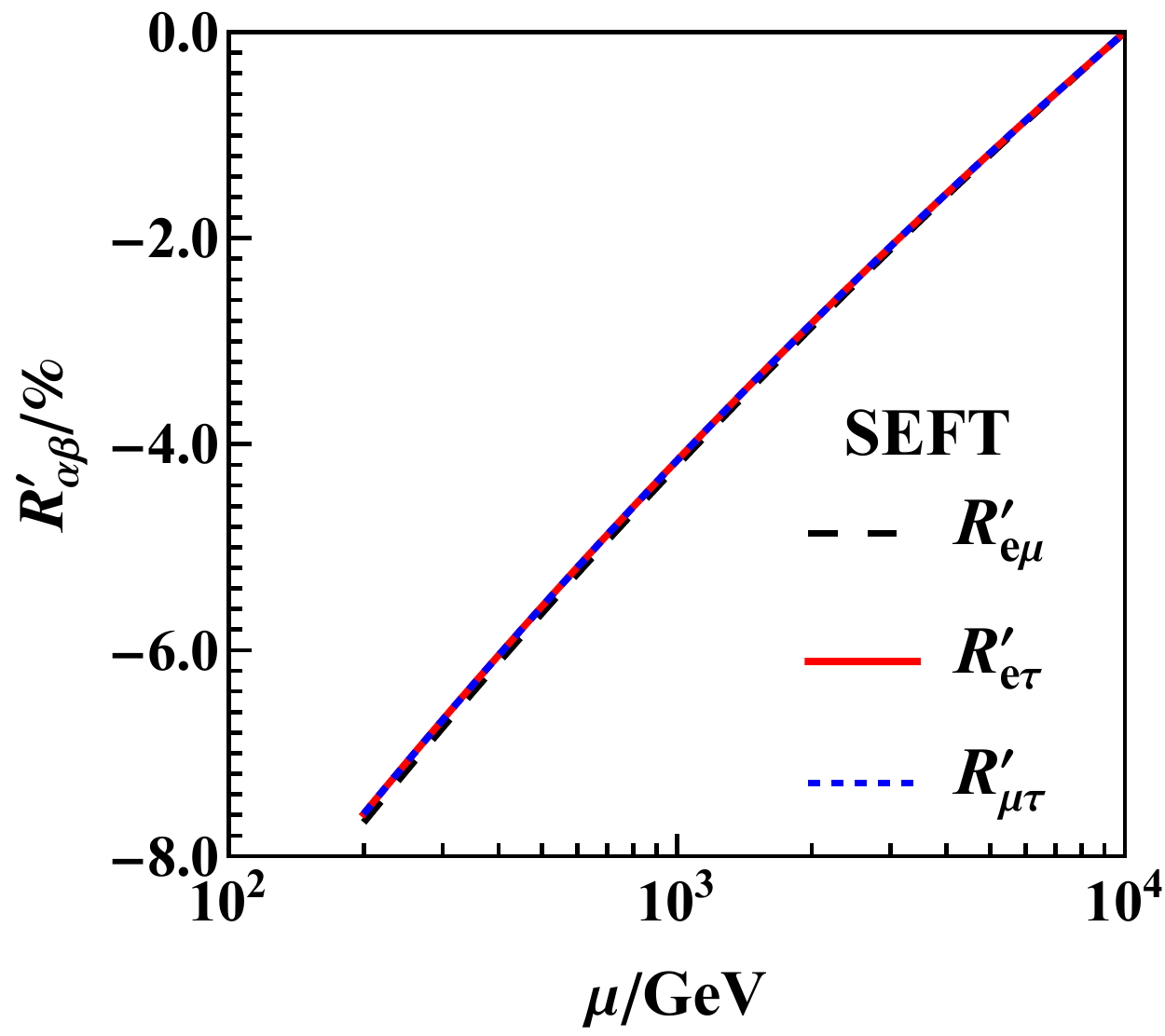}\\
	\;\;\includegraphics[scale=0.625]{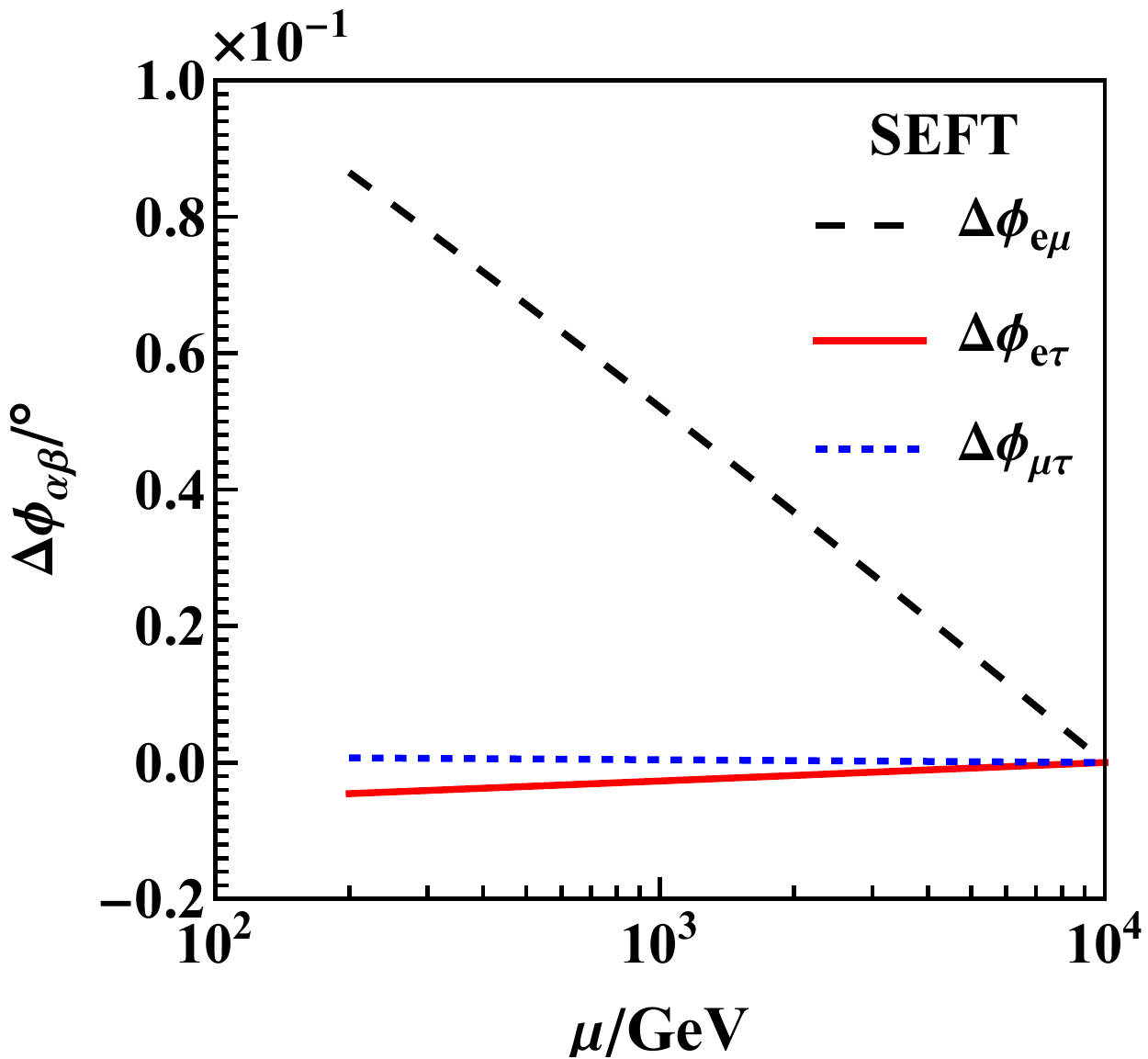}\quad
	\includegraphics[scale=0.625]{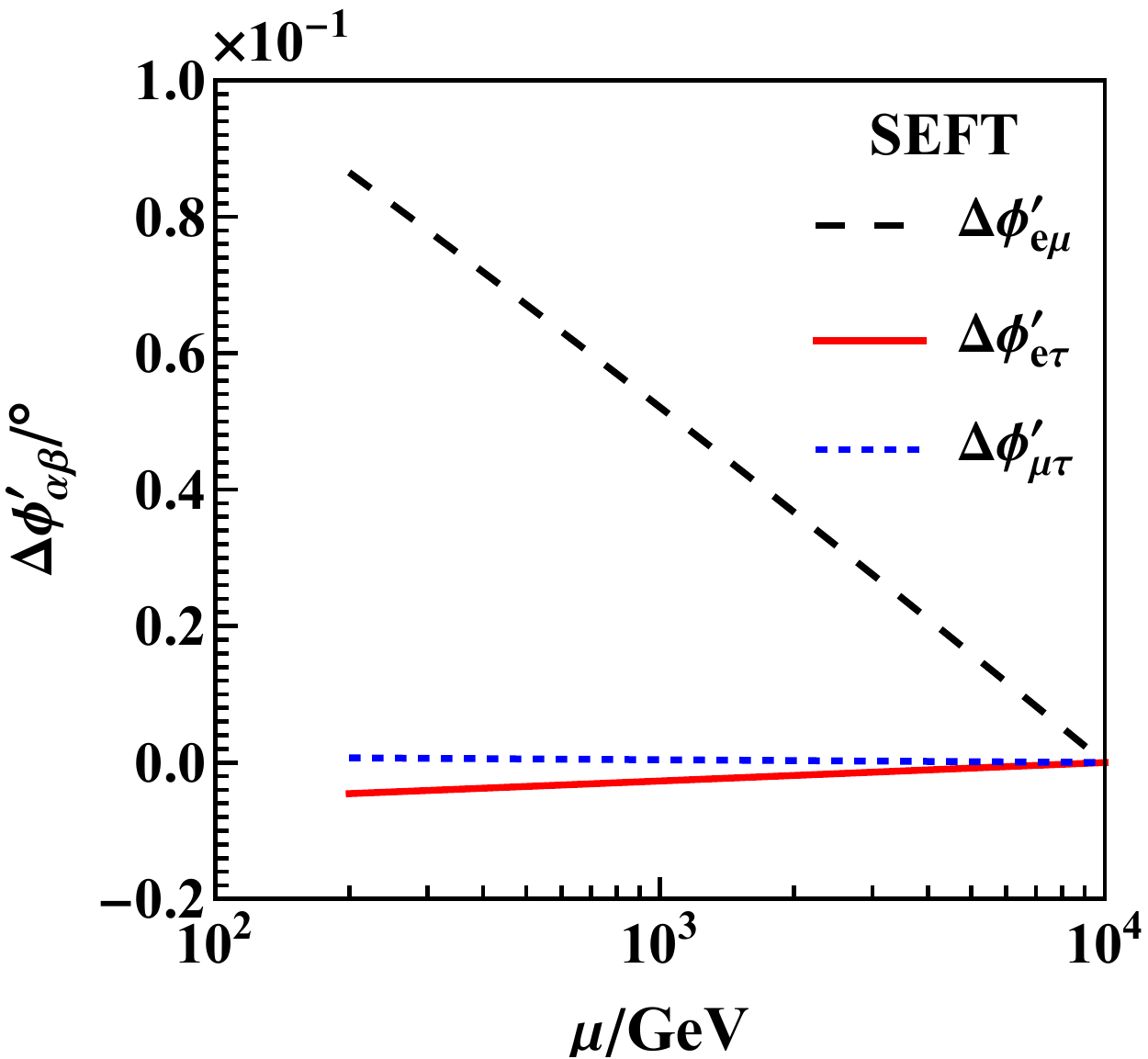}
	\vspace{-0.3cm}
	\caption{Running behaviors of non-unitary parameters in $\eta$ (left panel) and $\eta^\prime$ (right panel) in the type-I SEFT with the matching scale $\mu^{}_{\rm M} = 10^4~{\rm GeV}$. The running of diagonal elements $\eta_{\alpha\alpha}^{(\prime)}$ and the moduli of non-diagonal elements $|\eta_{\alpha\beta}^{(\prime)}|$ are shown as $R_{\alpha\beta}^{(\prime)} \equiv [|\eta_{\alpha\beta}^{(\prime)}(\mu)|  - |\eta_{\alpha\beta}^{(\prime)}(\mu^{}_{\rm M})| ] / |\eta_{\alpha\beta}^{(\prime)}(\mu^{}_{\rm M})| \times 100\%$ for $\alpha, \beta=e,\mu,\tau$, while the non-unitary phases $\phi_{\alpha\beta}^{(\prime)}$ as $\Delta \phi_{\alpha\beta}^{(\prime)} \equiv \phi_{\alpha\beta}^{(\prime)} (\mu) - \phi_{\alpha\beta}^{(\prime)} (\mu^{}_{\rm M})$.} 
	\label{fig:SEFT-eta}
\end{figure}

\begin{figure}
	\centering
	\includegraphics[scale=0.62]{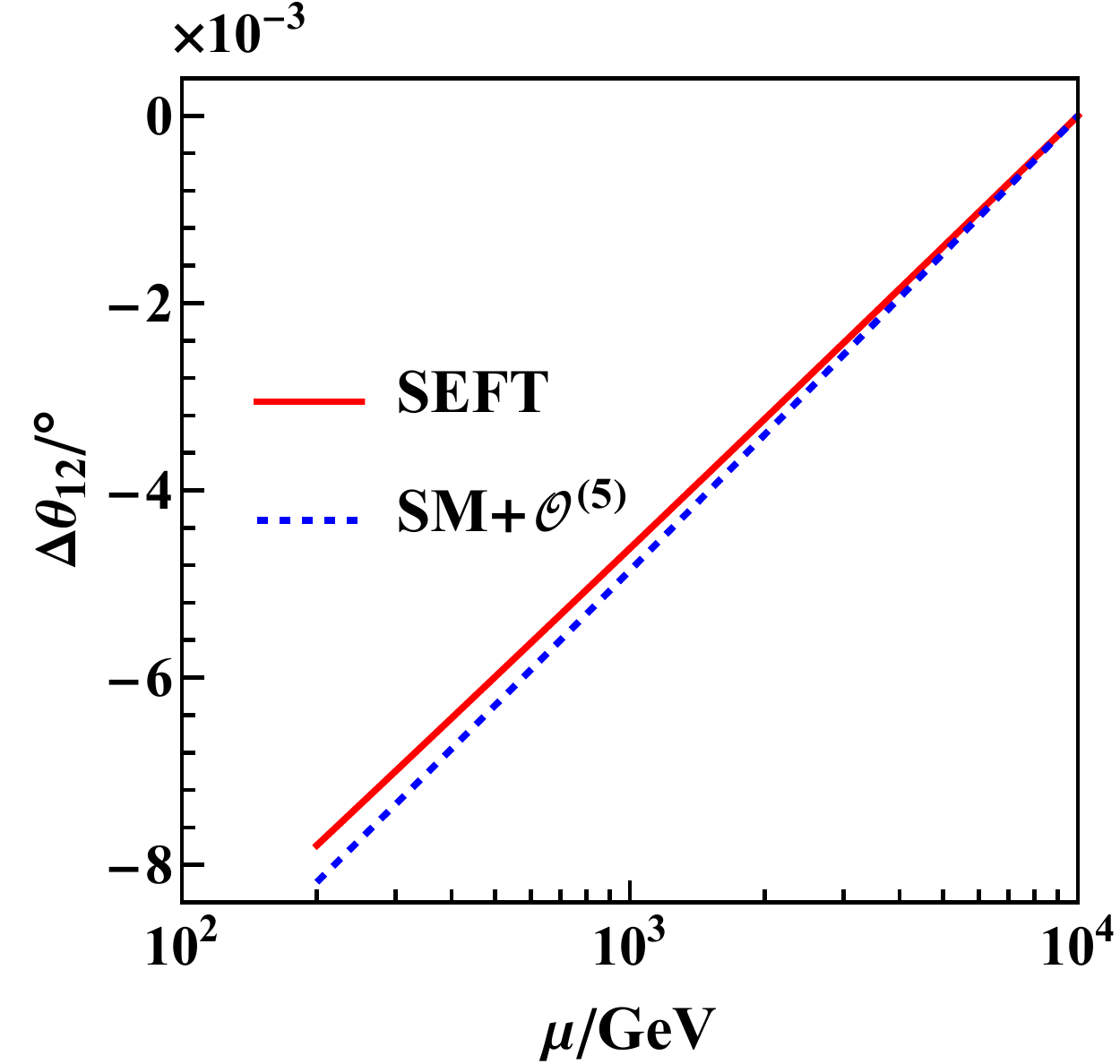}\quad
	\includegraphics[scale=0.62]{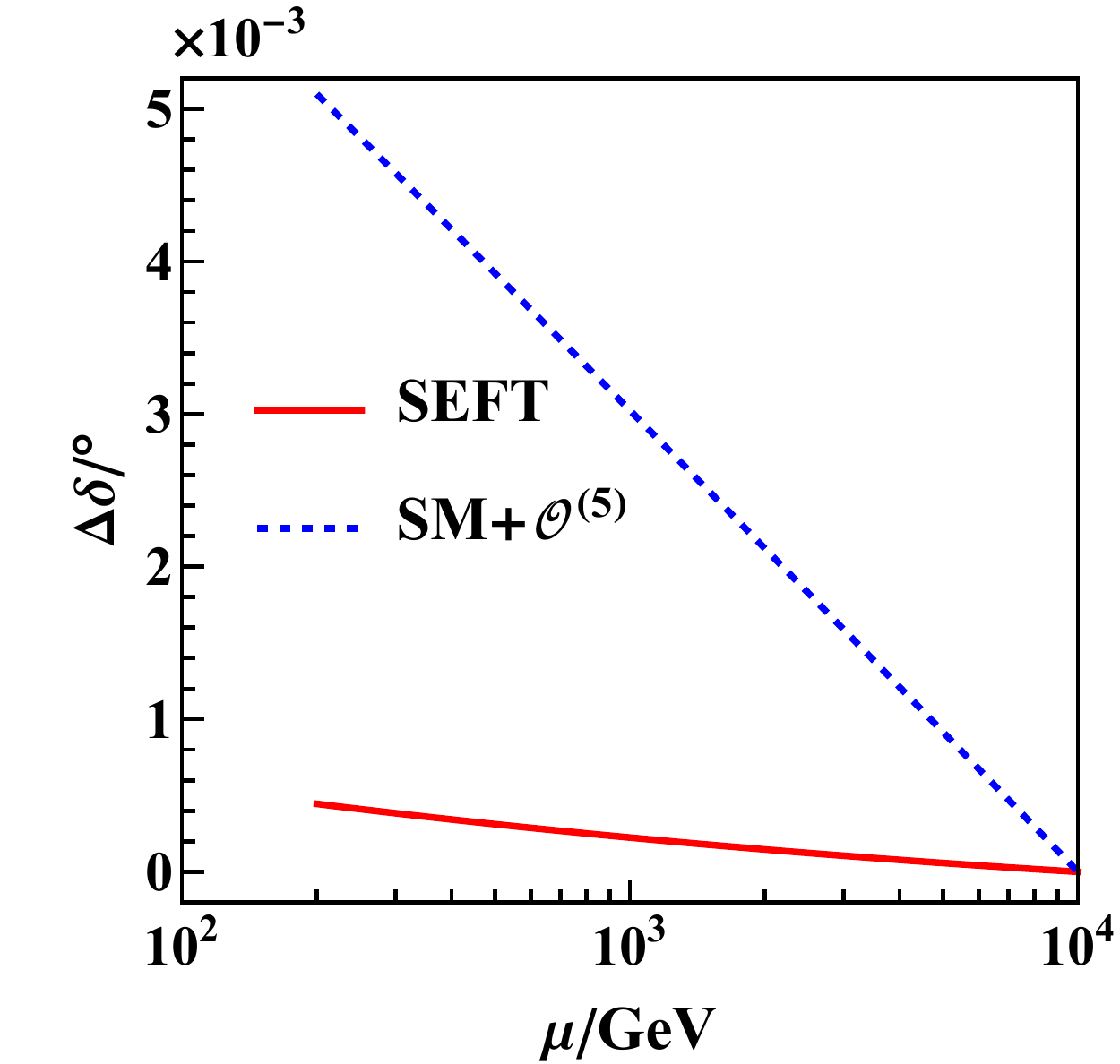}\\
	\includegraphics[scale=0.62]{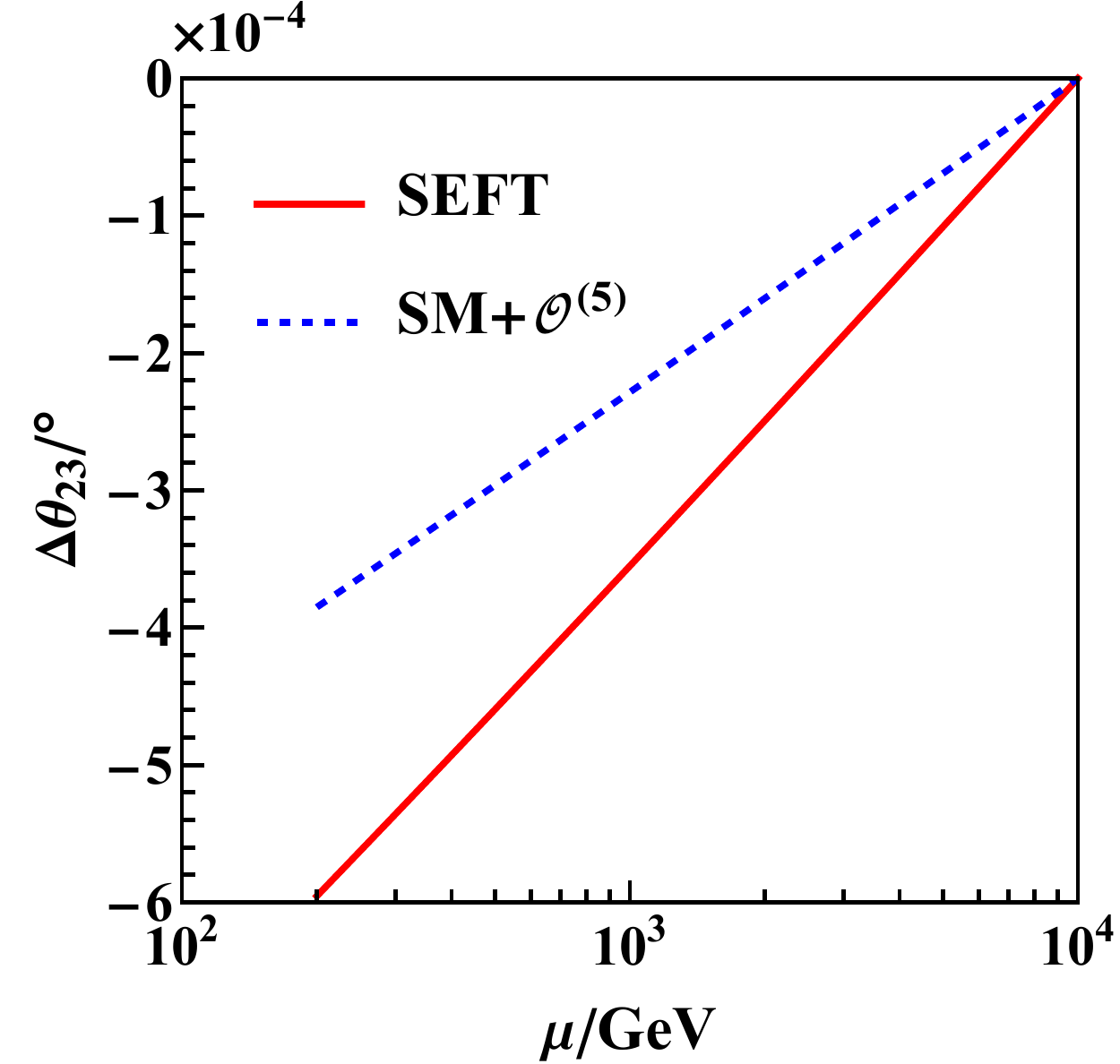}\quad
    \includegraphics[scale=0.62]{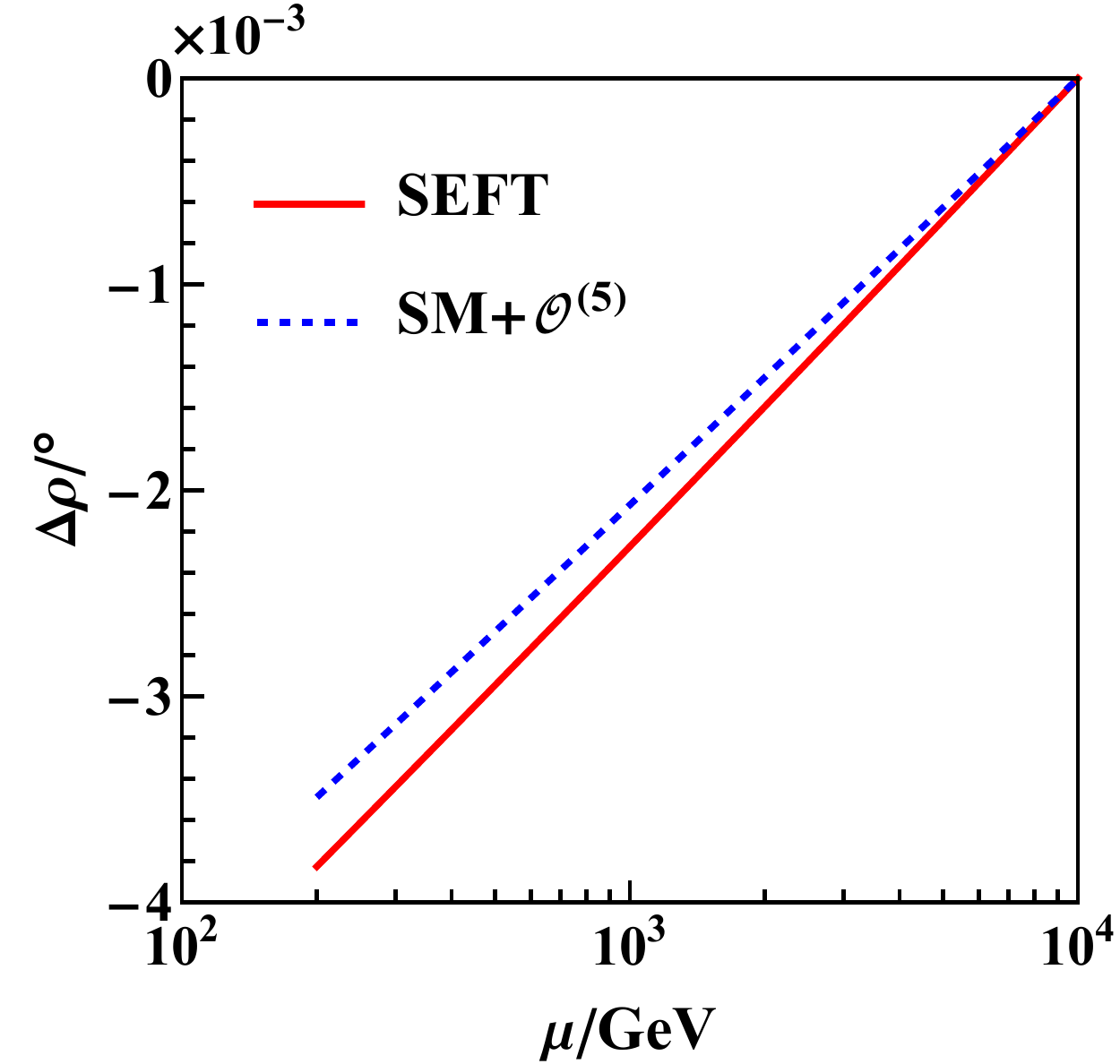}\\
	\includegraphics[scale=0.62]{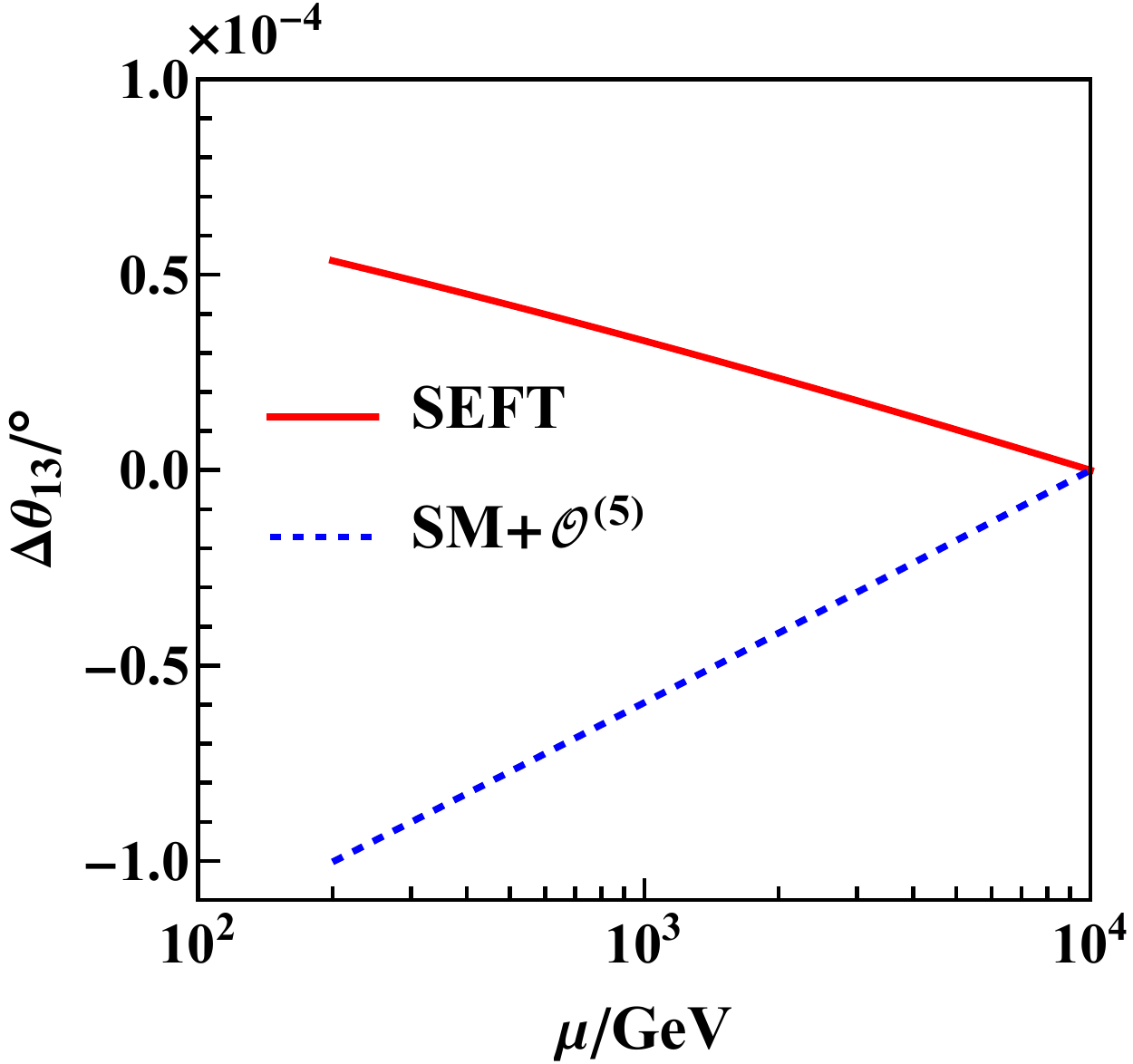}\quad
	\includegraphics[scale=0.62]{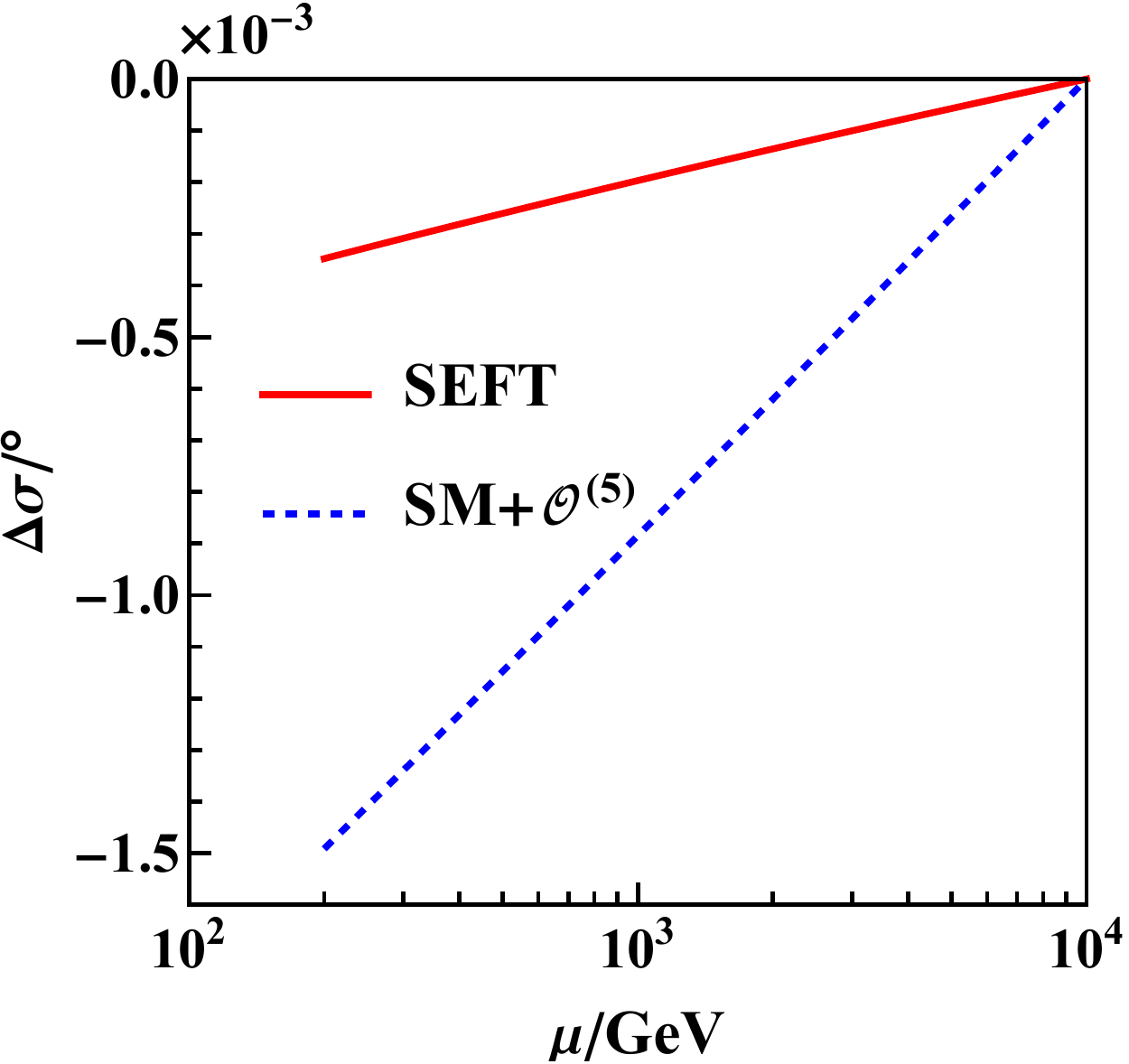}
	\vspace{-0.3cm}
	\caption{Running behaviors of three mixing angles $\{\theta_{12}^{},\theta_{23}^{},\theta_{13}^{}\}$ (left panel) and Dirac and Majorana CP-violating phases $\{\delta,\rho,\sigma\}$ (right panel) in the type-I SEFT with the matching scale $\mu^{}_{\rm M} = 10^4~{\rm GeV}$. The values of $\Delta P \equiv P(\mu) - P(\mu^{}_{\rm M})$ for all parameters in the SEFT are shown as red lines, while those in the case with only the dim-5 operator as blue dashed lines. }
	\label{fig:SEFT-mixing}
\end{figure}

\begin{figure}
	\centering
	\includegraphics[scale=0.61]{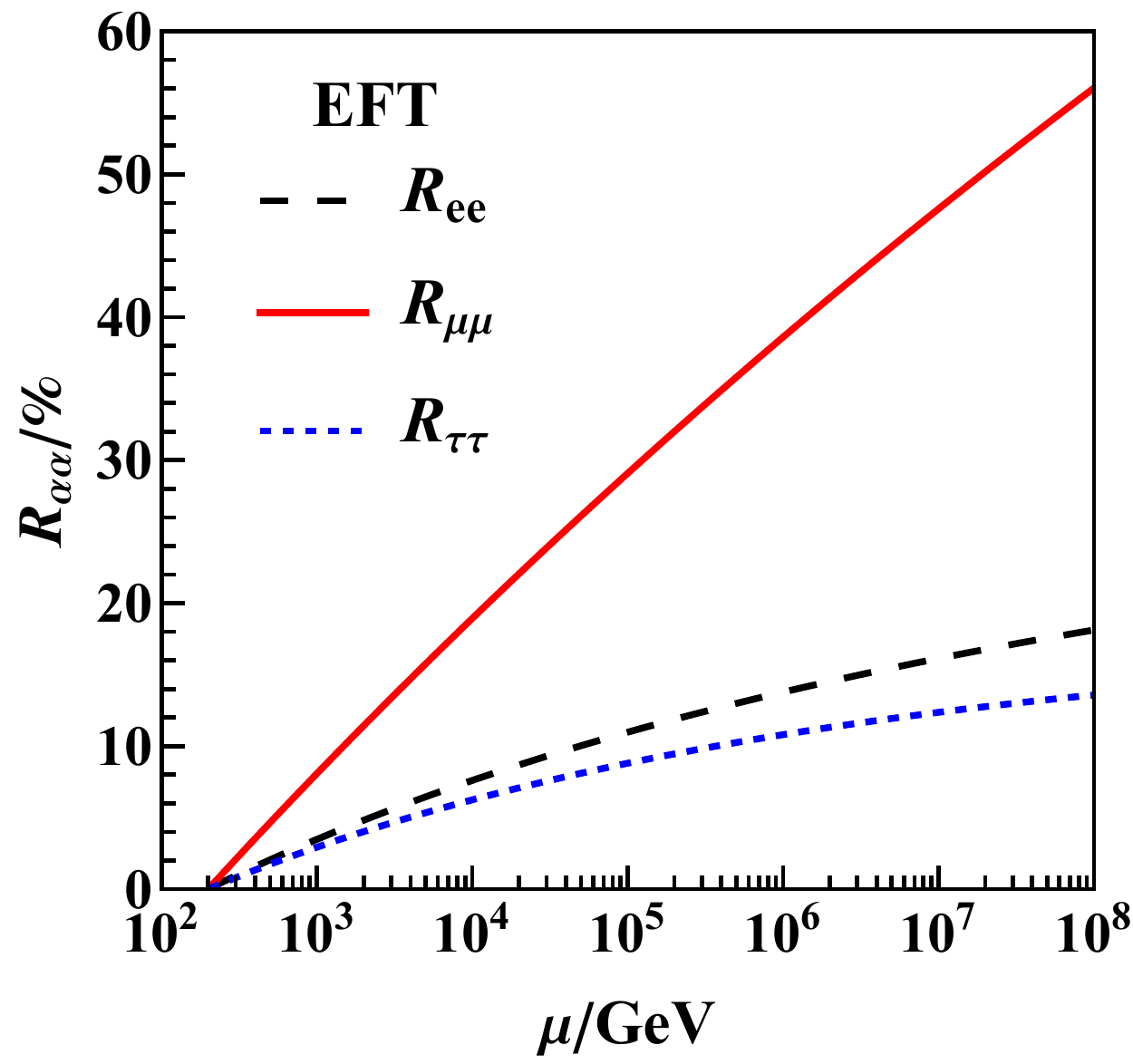}\quad\;
	\includegraphics[scale=0.61]{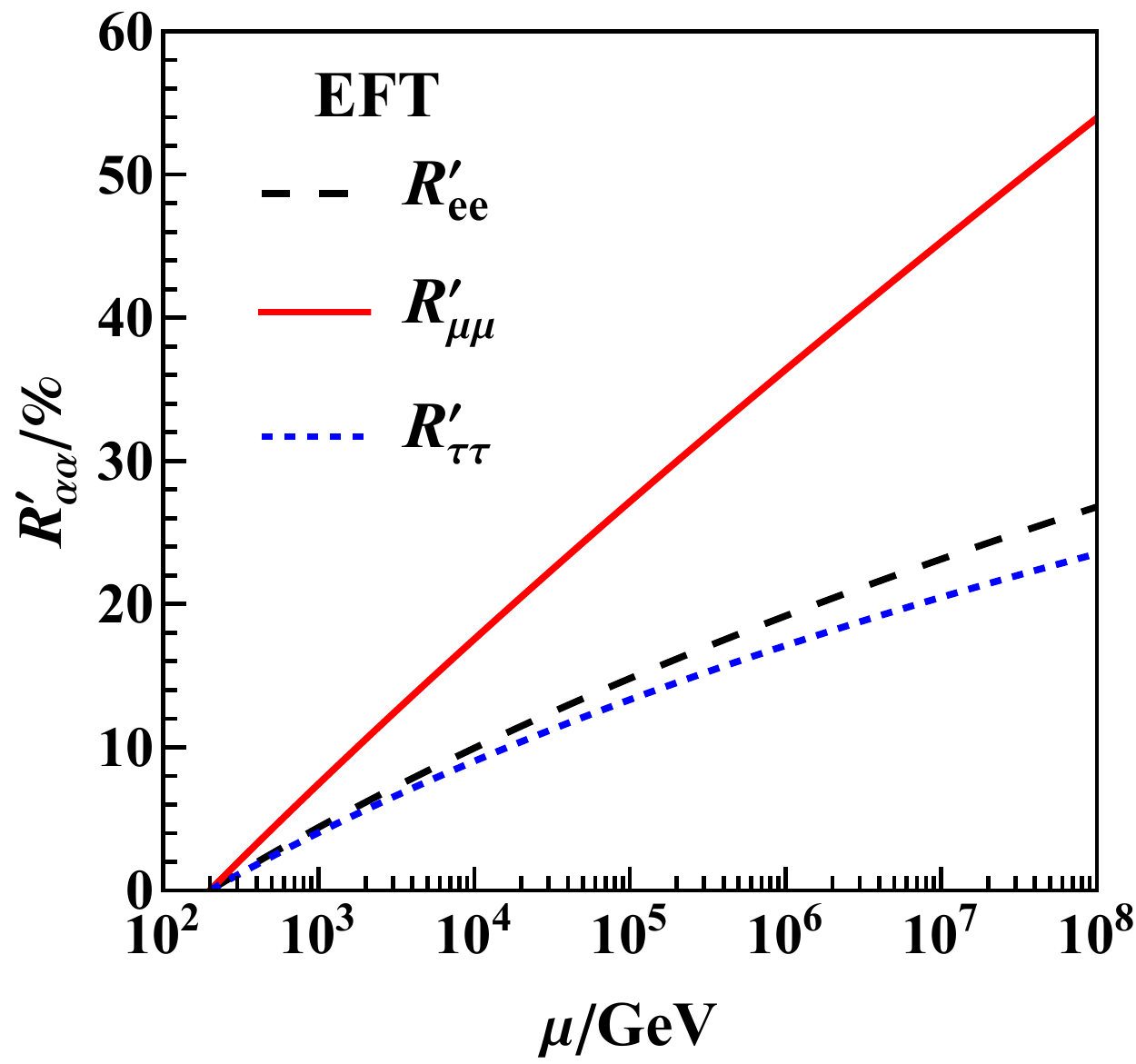}\\
	\includegraphics[scale=0.61]{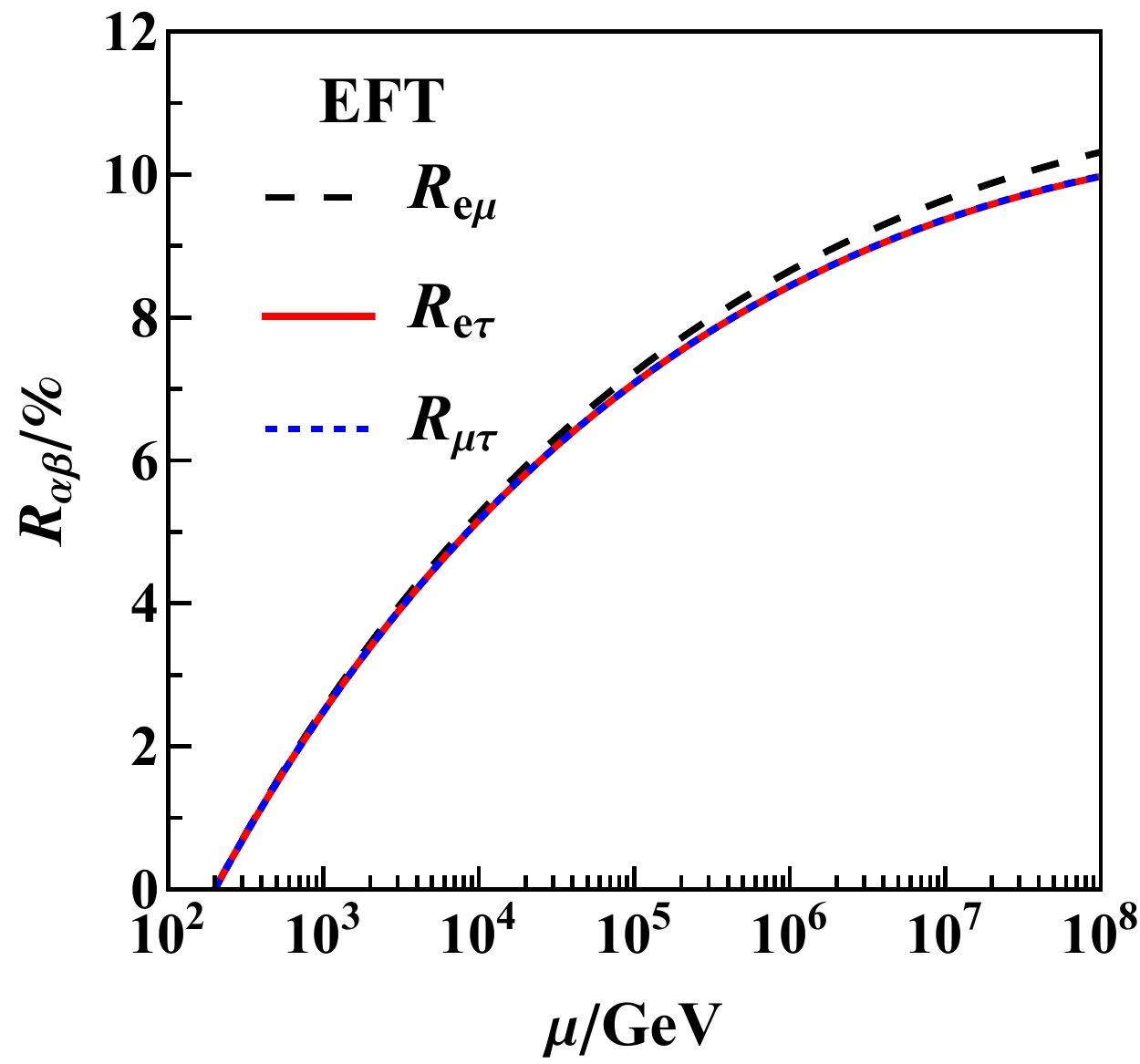}\quad\;
    \includegraphics[scale=0.61]{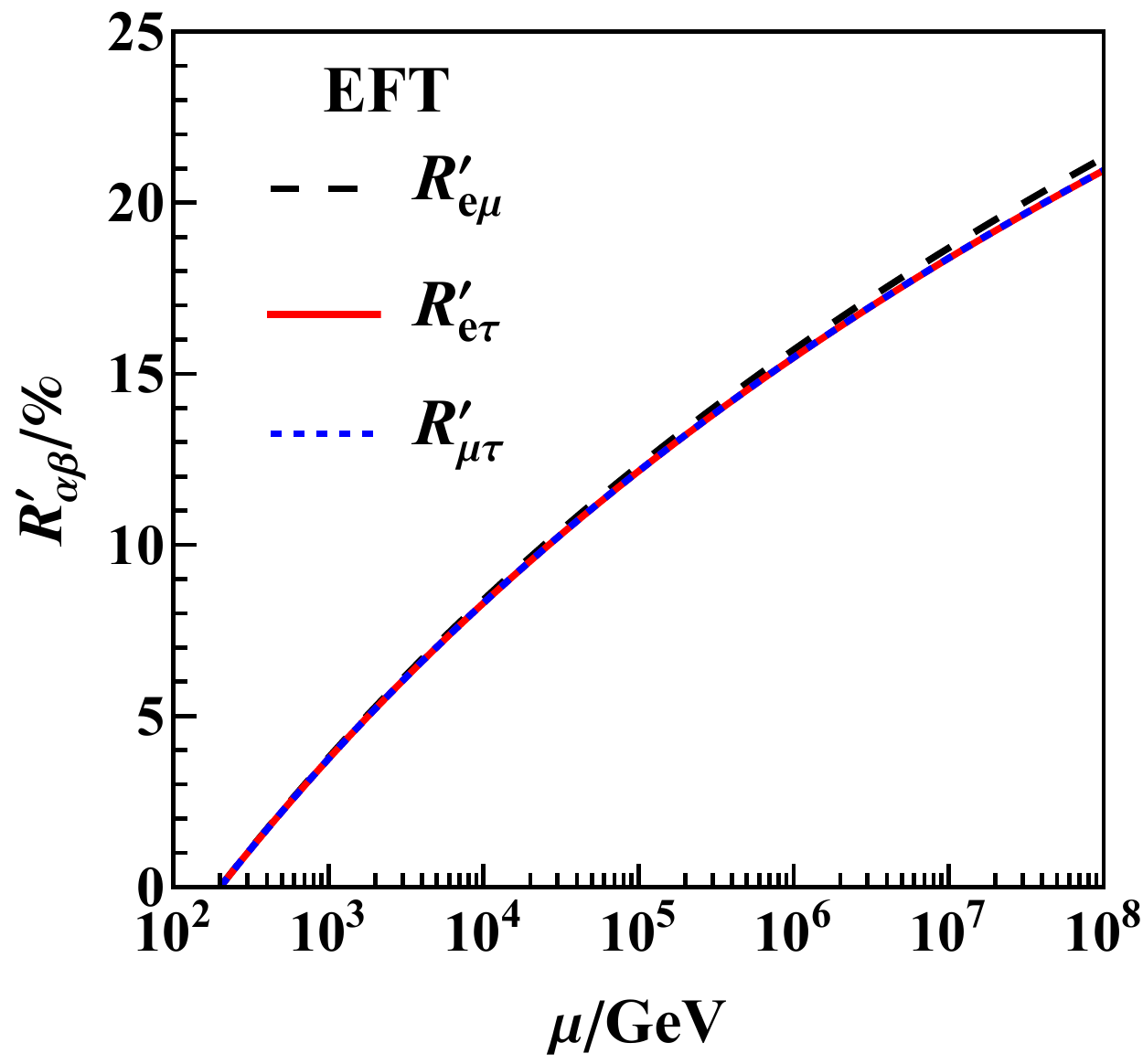}\\
	\includegraphics[scale=0.645]{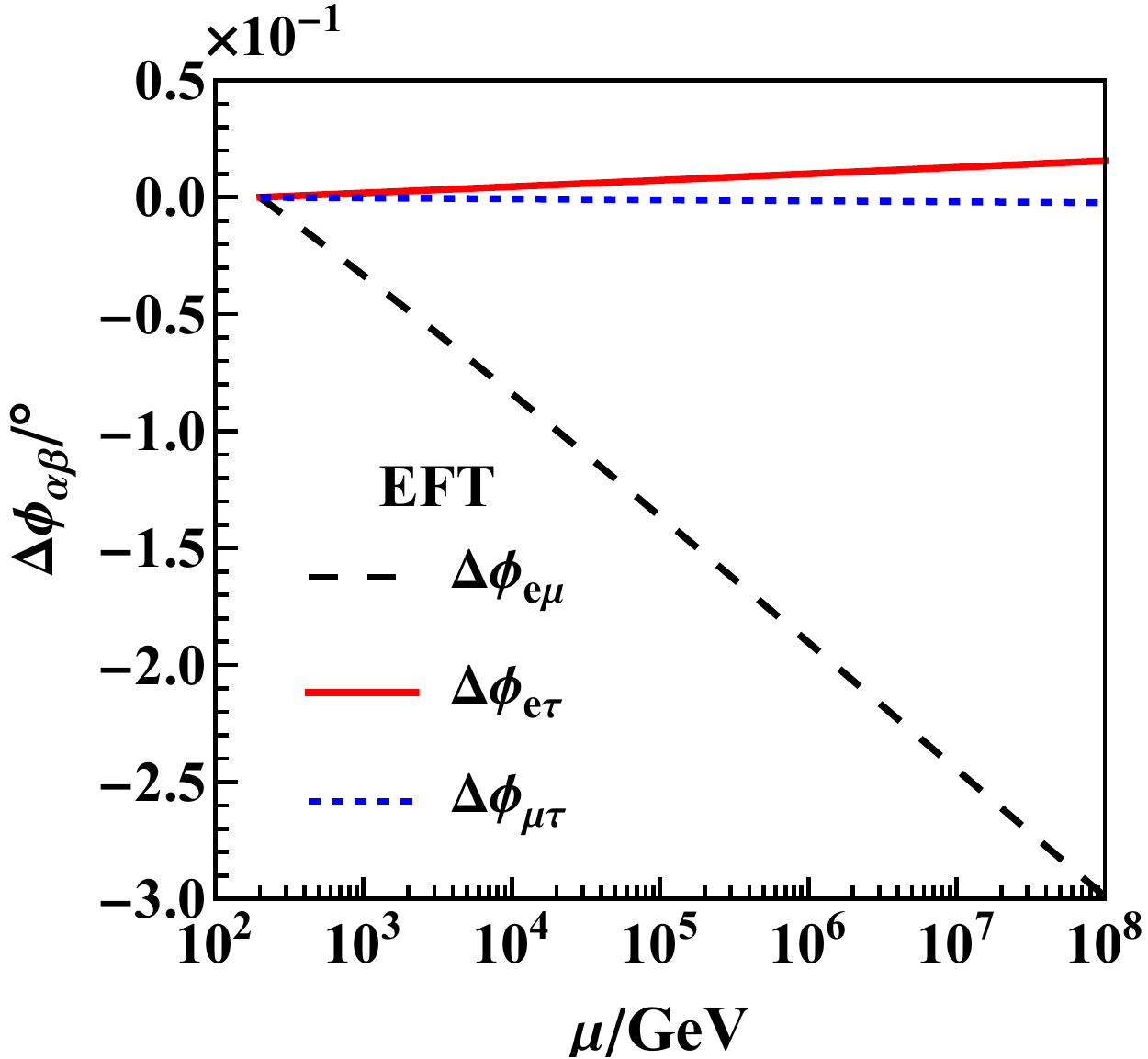}
	\includegraphics[scale=0.645]{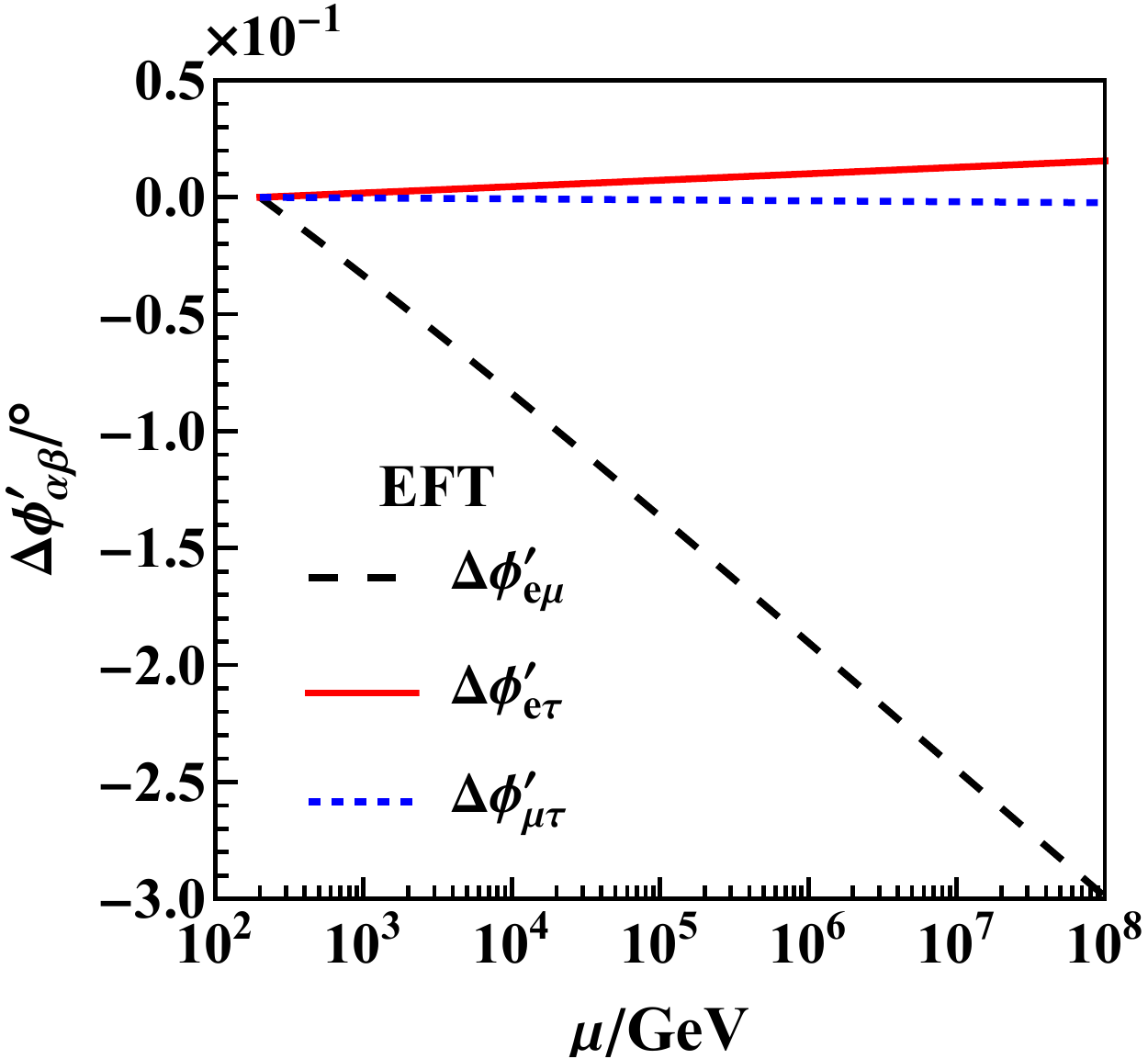}
	\vspace{-0.3cm}
	\caption{Running behaviors of non-unitary parameters in $\eta$ (left panel) and $\eta^\prime$ (right panel) in the general EFT from $\mu_{\rm B}^{} = 200~{\rm GeV}$ to the cutoff scale $\mu^{}_\Lambda = 10^8~{\rm GeV}$. The running of diagonal elements $\eta_{\alpha\alpha}^{(\prime)}$ and the moduli of non-diagonal elements $|\eta_{\alpha\beta}^{(\prime)}|$ are shown as $R_{\alpha\beta}^{(\prime)} \equiv [|\eta_{\alpha\beta}^{(\prime)}(\mu)|  - |\eta_{\alpha\beta}^{(\prime)}(\mu^{}_{\rm B})|] / |\eta_{\alpha\beta}^{(\prime)}(\mu^{}_{\rm B})|\times 100\%$ for $\alpha, \beta=e,\mu,\tau$, while the non-unitary phases $\phi_{\alpha\beta}^{(\prime)}$ as $\Delta \phi_{\alpha\beta}^{(\prime)} \equiv \phi_{\alpha\beta}^{(\prime)} (\mu) - \phi_{\alpha\beta}^{(\prime)} (\mu^{}_{\rm B})$.} 
	\label{fig:eta}
\end{figure}

\begin{figure}
	\centering
	\includegraphics[scale=0.618]{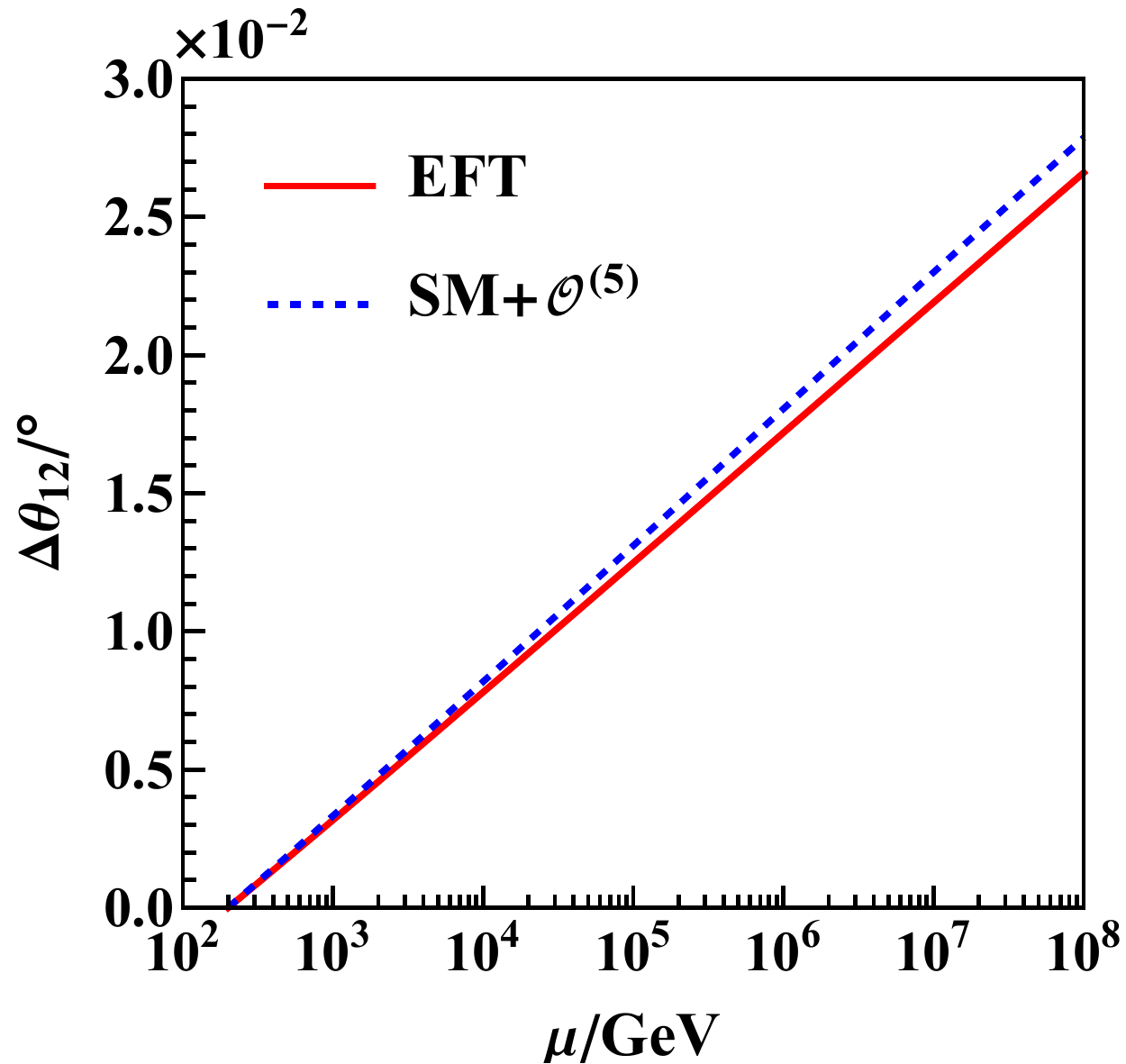}\quad
	\includegraphics[scale=0.62]{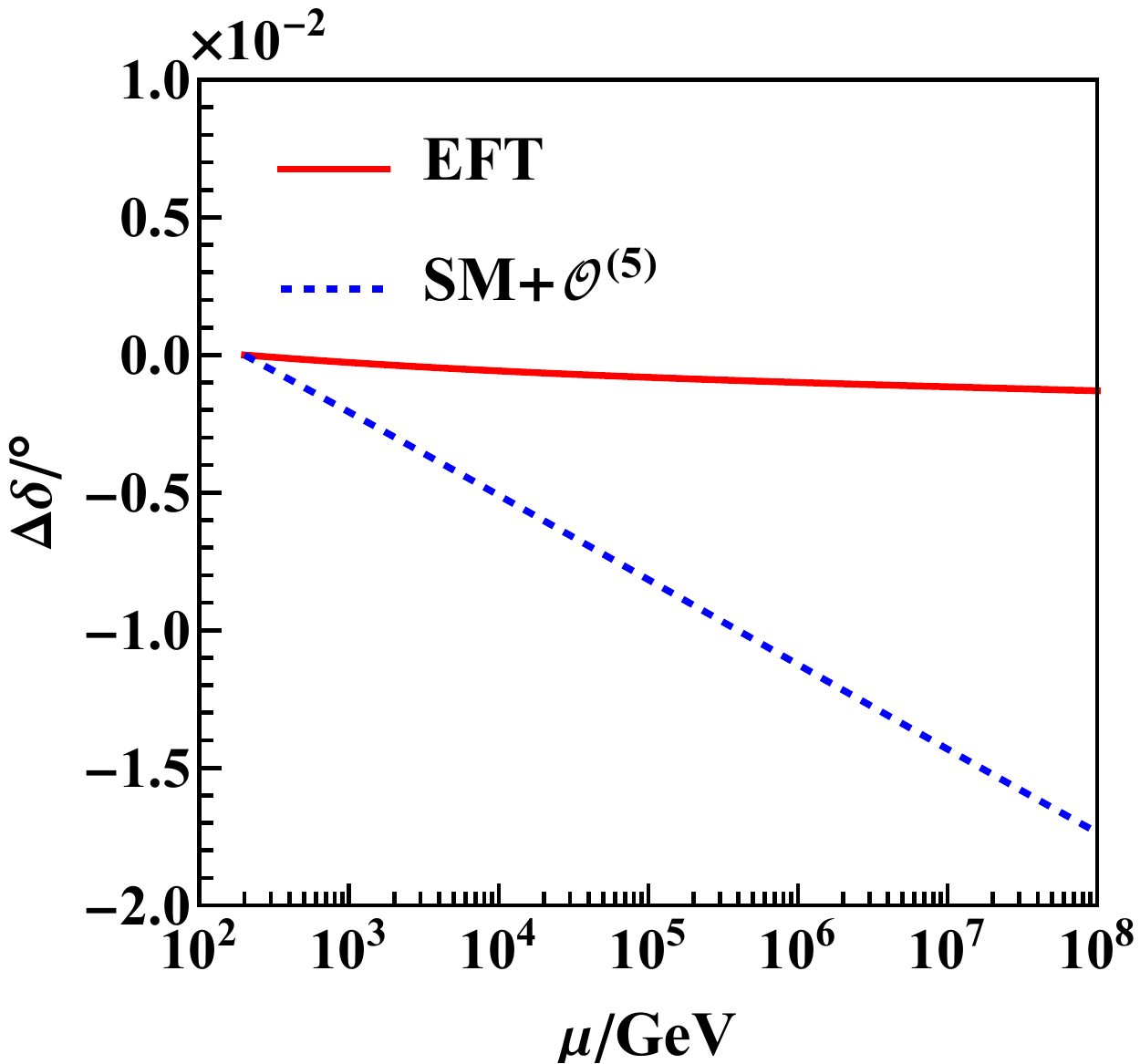}\\
	\includegraphics[scale=0.62]{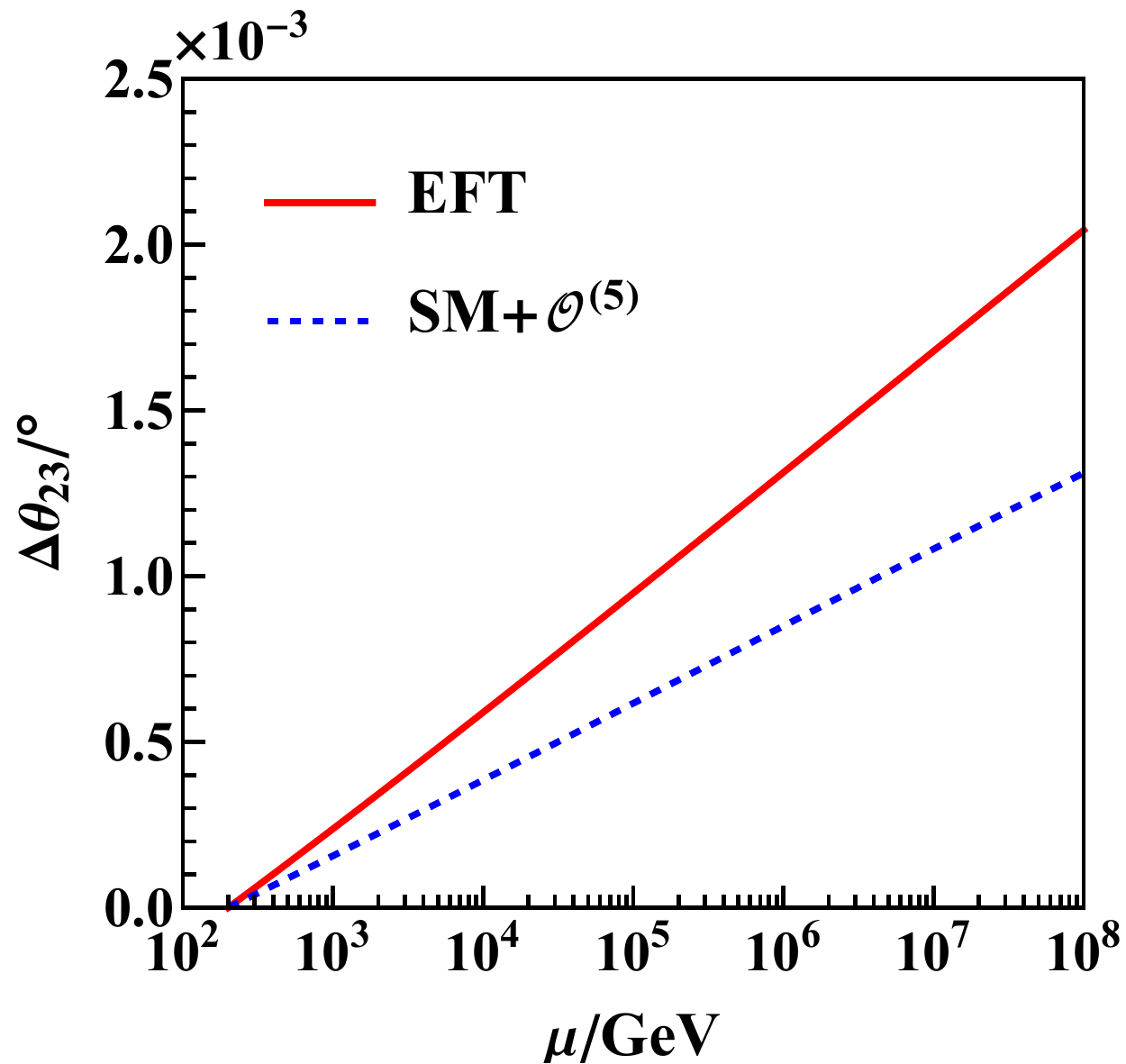}\quad
    \includegraphics[scale=0.62]{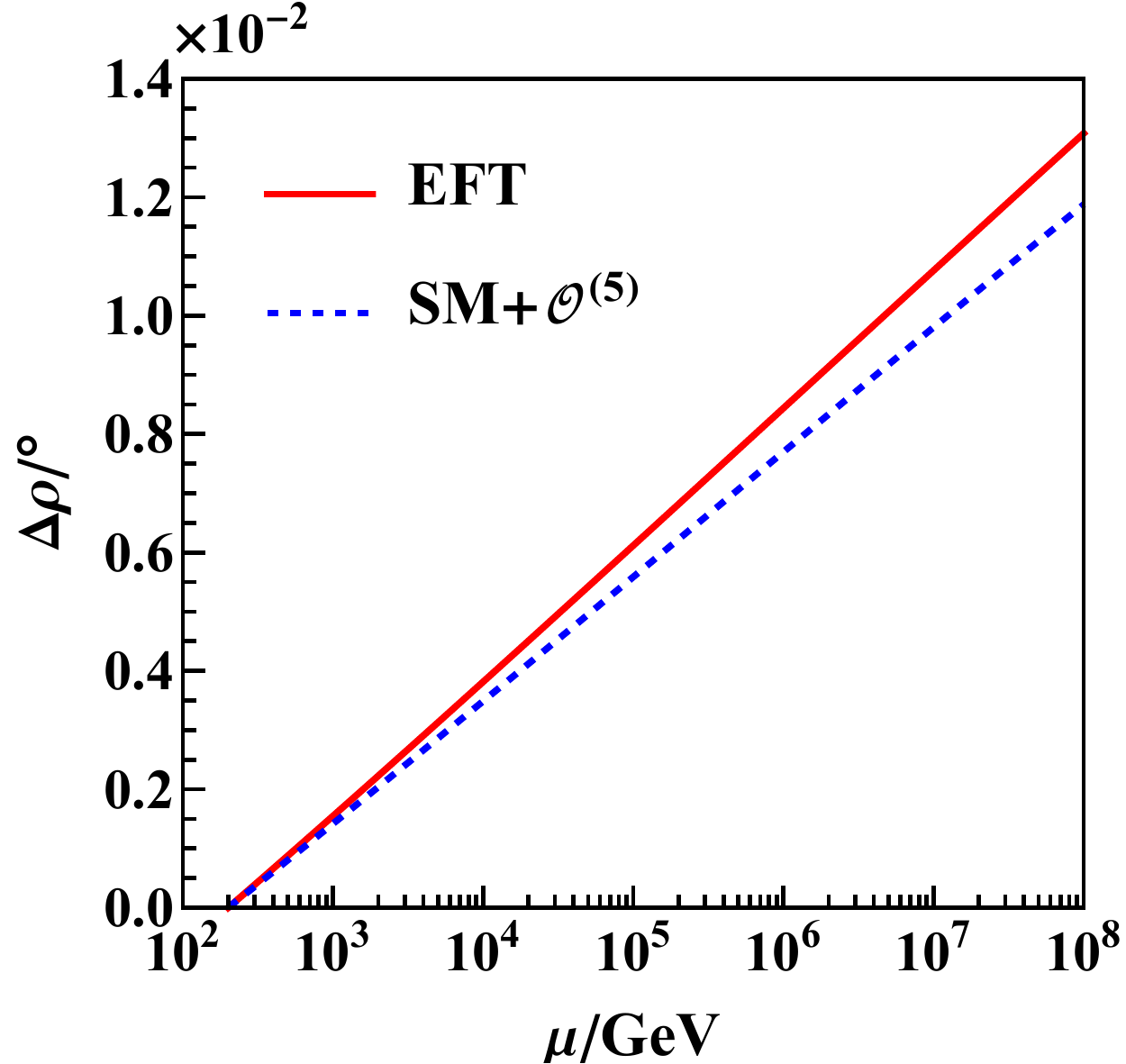}\\
	\includegraphics[scale=0.62]{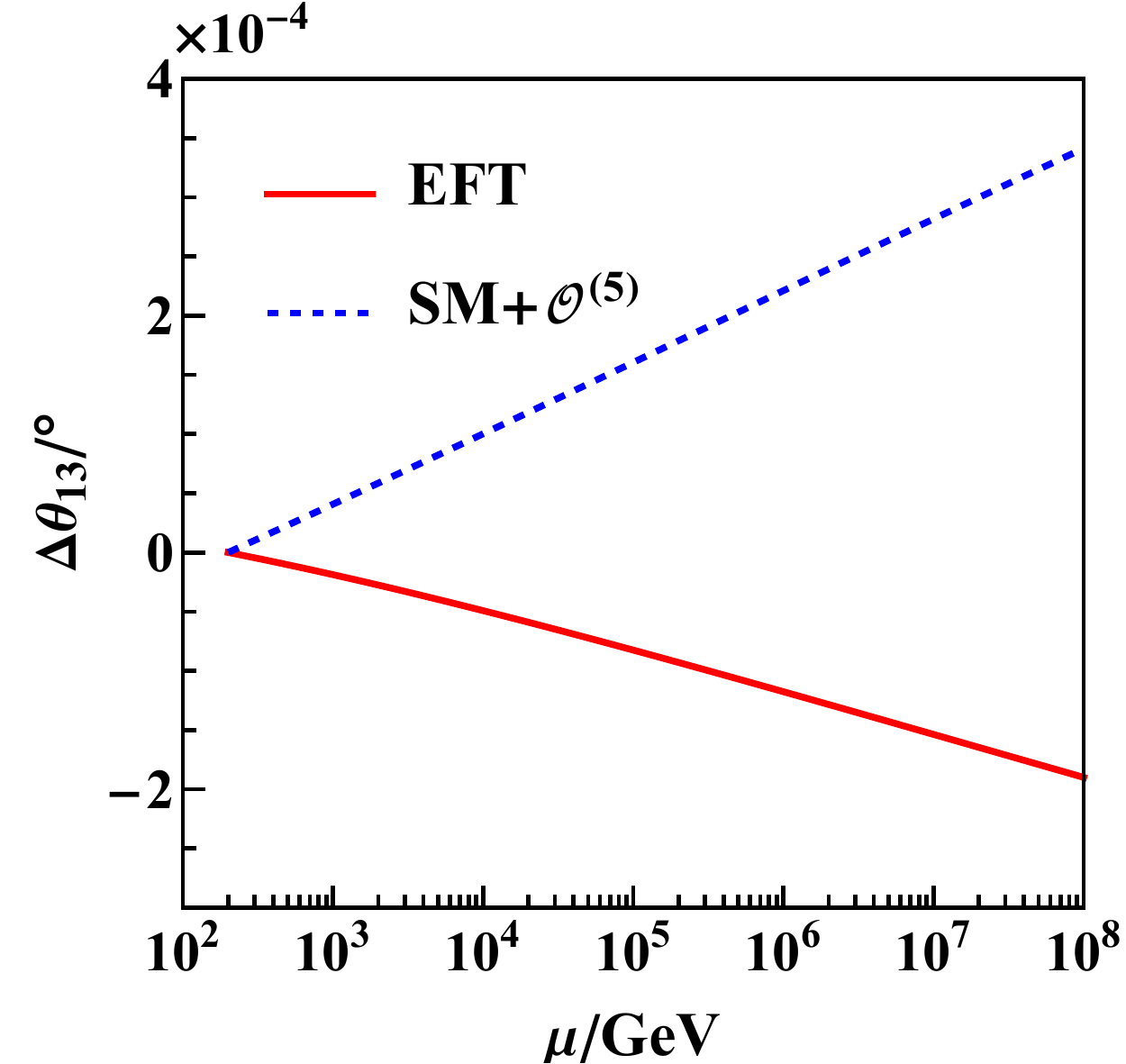}\quad
	\includegraphics[scale=0.62]{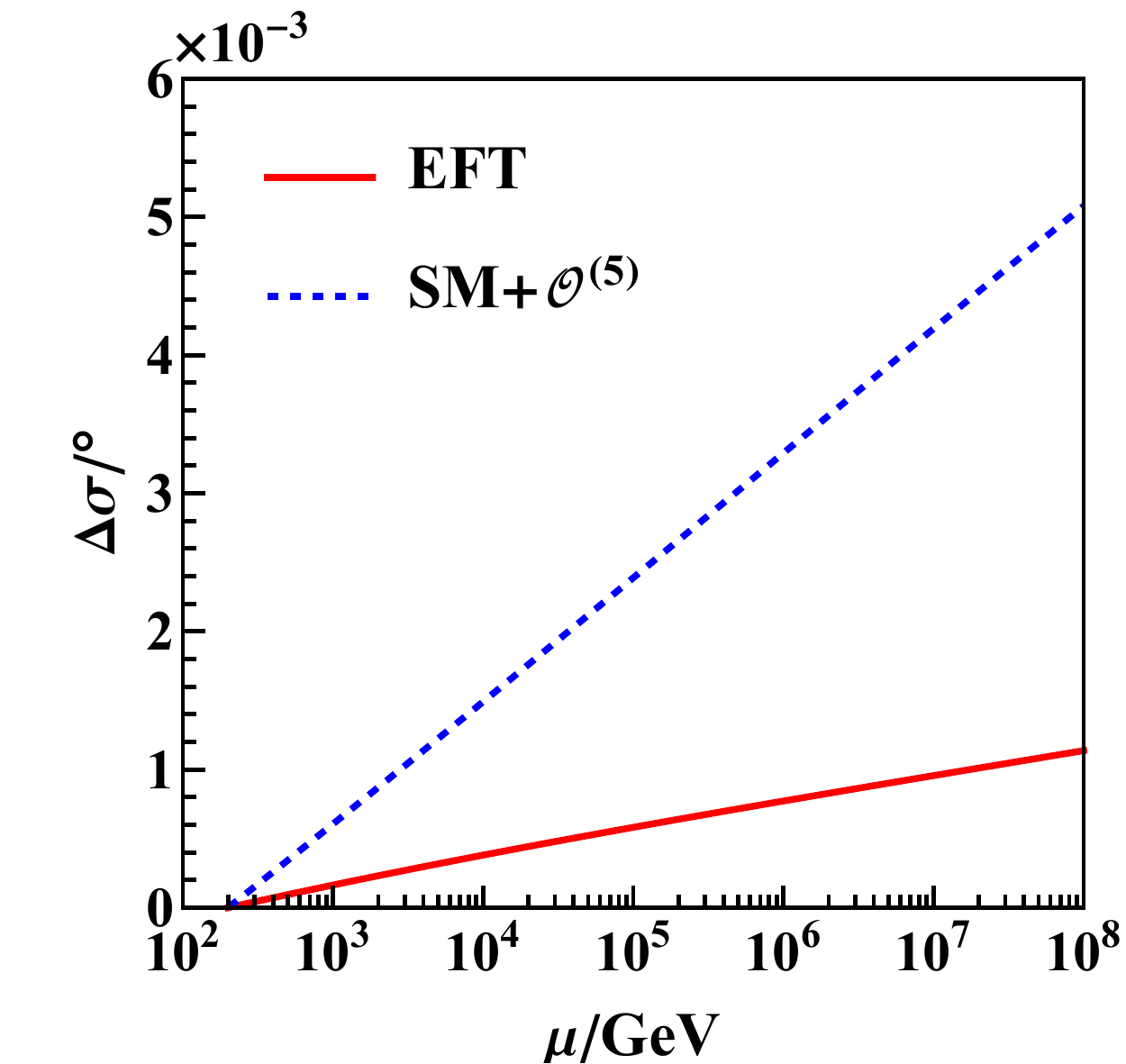}
	\vspace{-0.3cm}
	\caption{Running behaviors of three mixing angles $\{\theta_{12}^{},\theta_{23}^{},\theta_{13}^{}\}$ (left panel) and Dirac and Majorana CP-violating phases $\{\delta,\rho,\sigma\}$ (right panel) in a general EFT from $\mu_{\rm B}^{} = 200~{\rm GeV}$ to the cutoff scale $\mu^{}_\Lambda = 10^8~{\rm GeV}$. The values of $\Delta P \equiv P(\mu) - P(\mu^{}_{\rm B})$ for all parameters in the EFT are shown as red lines, while those in the case with only the dim-5 operator as blue dashed lines. }
	\label{fig:mixing}
\end{figure}

Then, we take the matching scale in the SEFT scenario to be $\mu^{}_{\rm M} = {\cal O}(M^{}_{\rm R}) = 10^4~{\rm GeV}$ and ${\cal O}(Y^{}_\nu) \sim 1$ in order that the absolute values of the matrix elements of $\eta$ and $\eta^\prime$ could be sizable. This can be achieved in the case where some underlying symmetry is introduced to guarantee reasonable values of neutrino masses, i.e., ${\cal O}(Y^{}_\nu M^{-1}_{\rm R} Y^{\rm T}_\nu v^2) \sim {\cal O}(0.1~{\rm eV})$~\cite{Kersten:2007vk,Abada:2007ux}. With the initial values summarized in Table~\ref{table: initial values}, one can utilize the RGEs of relevant parameters to obtain the ultimate values at $\mu^{}_{\rm M} = 10^4~{\rm GeV}$. As has been mentioned before, these values are then adjusted to satisfy the matching condition $\eta^\prime(\mu^{}_{\rm M}) = 2\eta(\mu^{}_{\rm M})$ and evolved from $\mu^{}_{\rm M}$ to the benchmark energy scale $\mu^{}_{\rm B}$ via the RGEs. In our calculations, the adjustment of the parameters at $\mu^{}_{\rm M}$ exactly reproduces the values at $\mu^{}_{\rm B}$ in Table~\ref{table: initial values} except for that of $\eta^\prime$, which is now different from $2\eta$. In the EFT scenario, through the same set of RGEs, all parameters with their initial values in Table~\ref{table: initial values} are evolved from $\mu^{}_{\rm B} = 200~{\rm GeV}$ up to the cutoff scale $\mu^{}_{\Lambda} = 10^8~{\rm GeV}$. The final numerical results are shown in Figs.~\ref{fig:SEFT-eta}-\ref{fig:mixing}. Some comments on the numerical results are as follows.

To quantify the running effects in both the SEFT and the general EFT scenarios, we define the ratios $R_{\alpha\beta}^{(\prime)}(\mu) \equiv [|\eta_{\alpha\beta}^{(\prime)}(\mu)| - |\eta_{\alpha\beta}^{(\prime)}(\mu^{}_{\rm init})|] / |\eta_{\alpha\beta}^{(\prime)}(\mu^{}_{\rm init})| \times 100\%$ (for $\alpha, \beta=e,\mu,\tau$) and the absolute differences $\Delta P(\mu) = P(\mu) - P(\mu^{}_{\rm init}) $ (for $P = \phi^{(\prime)}_{\alpha\beta}, \theta^{}_{ij}, \delta, \rho,\sigma$)  with the initial energy scale $\mu^{}_{\rm init} = \mu^{}_{\rm M}$ or $\mu^{}_{\rm B}$ for the type-I SEFT or the general EFT. The evolution of $R_{\alpha\beta}^{(\prime)}(\mu)$ and $\Delta P(\mu)$ with respect to the energy scale $\mu$ in the SEFT has been shown in Fig.~\ref{fig:SEFT-eta} and Fig.~\ref{fig:SEFT-mixing}, respectively. Note that the corresponding results of the mixing parameters in the ``${\rm SM}+{\cal O}^{(5)}$" scenario, which actually refers to the type-I SEFT with only the dim-5 operator, have also been given in Fig.~\ref{fig:SEFT-mixing} for comparison. The running behaviors of $R_{\alpha\beta}^{(\prime)}(\mu)$ and $\Delta P(\mu)$ can be understood in an approximate and analytical way. For the results in Fig.~\ref{fig:SEFT-eta}, we consider the beta functions in Eqs.~\eqref{eq:diaeta}-\eqref{eq:nondiaetap} and take account of the strong hierarchy among fermion Yukawa couplings, which can be neglected except for the top-quark Yukawa coupling $y^{}_t \approx 1$. Furthermore, with the input values in Table~\ref{table: initial values} and the smallness of $\eta^{(\prime)}$ and $\eta^\prime \simeq 2\eta$, one can approximately get
\begin{eqnarray}\label{eq:ele-aprox}
R^{}_{\alpha\alpha} &\sim& \Delta t\left[ \frac{2}{3}g^2_2 \frac{\tr{\eta}}{\eta^{}_{\alpha\alpha}} - \frac{17}{3} g^2_2 + 6y^2_t \right] \;,
\nonumber
\\
R^{\prime}_{\alpha\alpha} &\sim& \Delta t \left[ \frac{1}{3}\left( g^2_1 + g^2_2 \right) \frac{\tr{\eta}}{\eta^{}_{\alpha\alpha}} + \frac{1}{6} \left( g^2_1 - 17 g^2_2 \right) + 6 y^2_t \right] \;,
\nonumber
\\
R^{}_{\alpha\beta} &\sim& \Delta t \left( - \frac{17}{3} g^2_2 + 6y^2_t \right) \;,
\nonumber
\\
R^\prime_{\alpha\beta} &\sim& \Delta t \left[ \frac{1}{6} \left( g^2_1 - 17g^2_2 \right) + 6y^2_t \right] \;
\end{eqnarray}
with $\Delta t = \ln{(\mu^{}_{\rm B}/\mu^{}_{\rm M})}/(16\pi^2) < 0$.  From the first two lines of Eq.~\eqref{eq:ele-aprox} together with the input values, it is easy to see that $R^{(\prime)}_{\mu\mu} < R^{(\prime)}_{e e} < R^{(\prime)}_{\tau\tau} < 0$, which mainly arises from the factor $\eta^{-1}_{\alpha\alpha}$ (for $\alpha =e, \mu, \tau$) and its initial value (i.e., $\eta^{}_{\mu\mu} < \eta^{}_{ee} < \eta^{}_{\tau\tau}$). Moreover, we have 
\begin{eqnarray}
R^{\prime}_{\alpha \alpha} - R^{}_{\alpha \alpha} \sim \frac{\Delta t}{6} \left[ \left(g^2_1 + 17 g^2_2 \right) - 2 \left( g^2_2 - g^2_1 \right) \frac{\tr{\eta}}{\eta^{}_{\alpha\alpha}}\right] \; ,
\label{eq:RRpalfa}
\end{eqnarray}
from which one can observe $R^\prime_{\alpha\alpha} <  R^{}_{\alpha\alpha}$ (for $\alpha=e, \tau$) but $R^\prime_{\mu\mu} > R^{}_{\mu\mu}$ because of the competition between those two terms in the square brackets on the right-hand side. On the other hand, from the last two lines of Eq.~(\ref{eq:ele-aprox}), one can conclude that $R^{}_{\alpha \beta}$ ($R^\prime_{\alpha \beta}$) is roughly the same for $\alpha\beta=e\mu,e\tau,\mu\tau$. In addition, one can easily see $R^\prime_{\alpha\beta} - R^{}_{\alpha\beta} \sim (g^2_1 + 17 g^2_2) \Delta t/6 < 0$. All these observations are well consistent with the numerical results shown in Fig.~\ref{fig:SEFT-eta}. Similarly, with the help of Eqs.~\eqref{eq:phiemu}-\eqref{eq:phimutaup}, we obtain
\begin{eqnarray} \label{eq:pha-aprox}
\Delta \phi^{(\prime)}_{e\mu} &\sim& \left[ 8 \frac{m^2}{v^2} \left| \frac{\eta^{}_{e\tau}\eta^{}_{\mu\tau}}{\eta^{}_{e\mu}} \right| s_{e\mu}^{} + \frac{3}{4} \zeta^{-1}_{12} y^2_\tau s^2_{23} s^{}_{2(\rho-\sigma)} \right] \Delta t \;,
\nonumber
\\
\Delta \phi^{(\prime)}_{e\tau} &\sim&  \frac{3}{4} \zeta^{-1}_{12} y^2_\tau s^2_{23} s^{}_{2(\rho-\sigma)} \Delta t \;,
\nonumber
\\
\left| \Delta \phi^{(\prime)}_{\mu\tau} \right| &\sim& \mathcal{O} \left(|\eta|, \zeta^{-1}_{23} y^2_\tau \right) |\Delta t| \sim 10^{-4} |\Delta t| \;,
\end{eqnarray}
where the smallness of $\sin\theta^{}_{13}$ and $|\zeta^{-1}_{12}|\gg |\zeta^{-1}_{13}| \sim |\zeta^{-1}_{23}|$ are taken into account. As can be seen from Eq.~\eqref{eq:pha-aprox}, $\Delta \phi^{}_{\alpha\beta}\approx \Delta \phi^\prime_{\alpha\beta}$ (for $\alpha\beta = e\mu,e\tau,\mu\tau$) are valid, so they evolve in a similar way. The last line of Eq.~\eqref{eq:pha-aprox} shows that $\Delta \phi^{(\prime)}_{\mu\tau}$ is highly suppressed. Noticing that $s^{}_{2(\rho - \sigma)} < 0$ for the chosen initial values of $\rho$ and $\sigma$, one can see that $\Delta \phi^{(\prime)}_{e\tau} < 0$ and it is enhanced by a factor of $\zeta^{-1}_{12}$ when compared to $\Delta \phi^{(\prime)}_{\mu\tau}$. In the first line of Eq.~\eqref{eq:pha-aprox}, the two terms in square brackets on the right-hand side have opposite signs and the absolute value of the first term is much larger due to $|\eta^{}_{e\mu}|^{-1} \gg 1$ than that of the second one, so we have $\Delta \phi^{(\prime)}_{e\mu}  > | \Delta \phi^{(\prime)}_{e\tau} |$. These evolving behaviors can be clearly observed from Fig.~\ref{fig:SEFT-eta}. On the  other hand, the above analysis and results manifested in Fig.~\ref{fig:SEFT-eta} demonstrate that the RG-running effects on non-unitary parameters can be quite large, and they may have a significant impact on extracting the UV parameters from low-energy observables or on predicting low-energy observables given the UV model.

It is worthwhile to point out that the non-unitary parameters can affect the running of the mixing angles and CP-violating phases in a significant way. For comparison, we plot the results with and without non-unitary parameters  in Fig.~\ref{fig:SEFT-mixing} as red solid and blue dashed lines, respectively. The running behaviors of the mixing angles and CP-violating phases can be well understood by means of Eqs.~\eqref{eq:theta12}-\eqref{eq:delta}. Taking $\theta^{}_{13}$ for example, with the help of Eq.~\eqref{eq:theta13}, we can get
\begin{eqnarray}\label{eq:th13-apro}
    \Delta \theta^{}_{13} \sim - \Delta t \left[ 4 \frac{m^2}{v^2} \left| \eta^{}_{e\tau} \right| c^{}_{23} c^{}_{{e\tau}+\delta} + \frac{3}{8} \zeta^{-1}_{23} y^2_\tau \sin2\theta^{}_{12} \sin2\theta^{}_{23} s^{}_{\rho - \sigma} s^{}_{\delta + \rho + \sigma} \right] \;,
\end{eqnarray}
where the first term in the square brackets comes from non-unitary effects. With our inputs or outputs in the SEFT given in Table \ref{table: initial values}, the two terms in the square brackets have opposite signs and the absolute value of the first one is slightly larger than that of the second one. Consequently, $\theta^{}_{13}$ runs in opposite directions in the ``${\rm SM}+{\cal O}^{(5)}$" and SEFT scenarios. However, such an observation largely depends on the inputs of CP-violating phases and the initial values of $|\eta^{}_{\alpha \beta}|$.

Second, the running of non-unitary parameters and that of the mixing parameters in a general EFT from $\mu^{}_{\rm B} = 200~{\rm GeV}$ to $\mu^{}_{\Lambda} = 10^{8}$ GeV have been shown in Fig.~\ref{fig:eta} and Fig.~\ref{fig:mixing}, respectively. The initial inputs at $\mu^{}_{\rm B} = 200$ GeV are given in Table.~\ref{table: initial values}. The running behaviors of all these parameters can also be understood with the help of the approximate results given in Eqs.~\eqref{eq:ele-aprox}-\eqref{eq:th13-apro} but now with $\Delta t = \ln\left( \mu^{}_{\rm \Lambda}/\mu^{}_{\rm B} \right) / (16\pi^2) > 0$. The analysis is quite similar to that in the SEFT, but the cutoff scale is much higher, implying more remarkable running effects.  

Finally, we briefly mention the FCNC interaction of charged leptons. As can be seen from Eq.~\eqref{eq:effLagmass}, the FCNC term of charged leptons is determined by $\eta^\prime - 2\eta$, which is vanishing at the matching scale $\mu^{}_{\rm M}$ in the SEFT. However, the RG runnings of $\eta^\prime$ and $\eta$ are slightly different, leading to a non-vanishing value of $\eta^\prime - 2\eta$ and non-trivial phases as well. The running of absolute values and CP-violating phases of $\eta^\prime - 2\eta$ can be figured out by solving the complete set of RGEs. Numerically, we find the absolute values of the matrix elements of $\eta^\prime - 2\eta$ are about $10^{-4}$ and their phases are not vanishing. This can result in the FCNC processes and also CP violation in those processes.

\section{Summary}
\label{sec:summary}

In this paper, we derive the complete set of one-loop RGEs for all SM couplings and Wilson coefficients of operators up to dim-6 in the type-I SEFT, including the contributions from both double insertions of the dim-5 operator and single insertions of dim-6 operators. We find that there are 19 dim-6 operators in total, which can be generated by the RGEs at the one-loop level in the type-I SEFT. Although the one-loop RGEs for dim-6 operators have been considered in Refs.~\cite{Jenkins:2013zja,Jenkins:2013wua,Alonso:2013hga,Alonso:2014zka,Broncano:2004tz,Davidson:2018zuo}, we notice that the results therein are still not complete and fully correct. The contributions from double insertions of the dim-5 operator to the beta function of quartic Higgs coupling are not included, while a factor of $1/2$ is missing in the flavor conserving contributions from double insertions of the dim-5 operator to the one-loop anomalous dimensions of the Wilson coefficients of $\Op^{}_{H\square}$, $\Op^{}_{HD}$, $\Op^{}_H$, $\Op^{}_{eH}$, $\Op^{}_{uH}$ and $\Op^{}_{dH}$. We confirm that the contributions of single insertions of dim-6 operators are compatible with the results in the SMEFT~\cite{Jenkins:2013zja,Jenkins:2013wua,Alonso:2013hga,Alonso:2014zka}. Besides the one-loop RGEs of type-I SEFT, we also present the complete one-loop RGEs for type-II and type-III SEFTs. 

After spontaneous gauge symmetry breaking, two tree-level dim-6 operators in the type-I SEFT result in a non-unitary leptonic flavor mixing matrix appearing in the charged-current interaction of leptons. In addition, the neutral-current interaction of neutrinos and that of charged leptons are also modified by the Wilson coefficients of these two dim-6 operators. Concentrating on leptonic flavor mixing parameters and weak interactions of leptons, we derive the explicit expressions of one-loop RGEs of non-unitary parameters and leptonic mixing parameters. As a by-product, the RGEs of the mixing parameters in the standard parametrization of a unitary leptonic mixing matrix are obtained for the first time. Numerical solutions to the one-loop RGEs are provided in two specific scenarios. We demonstrate that the non-unitary parameters may affect significantly the running of leptonic flavor mixing parameters. In an approximate and analytical way, the running behaviors of all these parameters can be understood. It should be emphasized that other dim-6 operators may come into play at the one-loop level via the RGEs of the Wilson coefficients. Although the numerical solutions to an enlarged set of one-loop RGEs are always possible, it is difficult to solve the RGEs in an analytical way. In this sense, our calculations provide an illustrative example for the impact of dim-6 operators on leptonic flavor mixing and CP violation. 

Together with the one-loop matching results at the seesaw scale $M = {\cal O}(M^{}_{\rm R})$, the one-loop RGEs obtained in this paper establish a self-consistent theoretical framework to investigate low-energy phenomena of the type-I seesaw model up to the order ${\cal O}(M^{-2})$ at the one-loop level. In the era of precision measurements, such a framework will be helpful for testing the seesaw models and exploring the origin of neutrino masses.

\section*{Acknowledgements}
This work was supported in part by the National Natural Science Foundation of China under grants No. 12075254 and No. 11835013, and by the Alexander von Humboldt Foundation (D.Z.).

\begin{appendices}

\section{Counterterms in the Warsaw Basis}\label{app:counterterms}

The counterterms in the Warsaw basis can be derived by applying the EOMs of relevant fields to the results in the Green's basis, which are obtained by calculating the selected set of Feynman diagrams. An example for calculating the counterterms in the Green's basis has been given in Sec.~\ref{sec:deriveRGEs} in detail. Following similar procedures and making use of the EOMs, one can achieve all the counterterms in the Warsaw basis needed to cancel out one-loop UV divergences in the type-I SEFT, namely, those given in Eqs.~\eqref{eq:yl-warsaw}-\eqref{eq:llrr-warsaw}. Those counterterms together with that for the Weinberg operator and the SM counterterms constitute the full counterterms in the type-I SEFT, but the SM ones are not explicitly shown here. With those counterterms in Eqs.~\eqref{eq:yl-warsaw}-\eqref{eq:llrr-warsaw}, one can derive the RGEs in Eqs.~\eqref{eq:YukawaRGE}-\eqref{eq:LLRRRGE} by means of the formulas in Ref.~\cite{Antusch:2001ck}.

\subsubsection*{Renormalizable terms}
\begin{itemize}
	\item  {\bf The charged lepton Yukawa coupling}
	\begin{eqnarray} \label{eq:yl-warsaw}
		\delta \Yl &=& - \loopfactore m^2 \left( \co + 3\ct \right) \Yl \;.
	\end{eqnarray}

	\item {\bf The Higgs quartic coupling}
	\begin{eqnarray}
		\delta \lambda &=& \loopfactore 4 m^2 \tr{ \frac{1}{4} C^{}_5 C^\dagger_5 - \frac{1}{3} g^2_2 \ct+\ct \Yl\DYl} \;.
	\end{eqnarray}
\end{itemize}

\subsubsection*{Dimension-six operators}

\begin{itemize}
	\item $\bm{H^6} $ {\bf and} $\bm{H^4D^2}$ 
	\begin{eqnarray}
		\dC{}{H\square} &=& - \loopfactore \trs{ \frac{1}{2} C^{}_5 C^\dagger_5 + \frac{1}{3} g^2_1 \co - g^2_2 \ct + (\co+3\ct)\Yl\DYl } \;,
		\nonumber
		\\
		\dC{}{HD} &=& - \loopfactore \tr{ C^{}_5 C^\dagger_5 +  \frac{4}{3} g^2_1 \co + 4 \co \Yl\DYl } \;,
		\nonumber
		\\
		\dC{}{H} &=& \loopfactore \tr{ -2 \lambda  C^{}_5 C^\dagger_5 + \frac{8}{3} \lambda g^2_2 \ct - 8\lambda \ct \Yl\DYl } \;.
	\end{eqnarray}
	
	\item $\bm{\psi^2 H^3}$
	\begin{eqnarray}
		\dC{}{eH} &=& \loopfactore \left[ \frac{3}{4}  C^{}_5 C^\dagger_5  \Yl + \tr{ - \frac{1}{2} C^{}_5 C^\dagger_5 + \frac{2}{3} g^2_2 \ct - 2\ct \Yl \DYl} \Yl + \co \Yl\DYl\Yl  \right.
		\nonumber
		\\
		&& + \left. \left( 2\lambda - 3g^2_1 \right) \co \Yl + 3 \left( 2\lambda - g^2_1 \right) \ct \Yl \right] \;,
		\nonumber
		\\
		\dC{}{uH} &=& \loopfactore \tr{ - \frac{1}{2} C^{}_5 C^\dagger_5 + \frac{2}{3} g^2_2 \ct - 2 \ct \Yl\DYl} \Yu \;,
		\nonumber
		\\
		\dC{}{dH} &=& \loopfactore \tr{ - \frac{1}{2} C^{}_5 C^\dagger_5 + \frac{2}{3} g^2_2 \ct - 2 \ct \Yl\DYl} \Yd \;.
	\end{eqnarray}
	
	\item $\bm{\psi^2 H^2 D}$
	\begin{eqnarray}
		\dC{(1)}{H\ell} &=& \loopfactore \left[ - \frac{3}{4} C^{}_5 C^\dagger_5 + \frac{1}{3} g^2_1 \tr{\co} \mathbbm{1} + \frac{3}{4} \left( \co + 3\ct \right) \Yl\DYl + \frac{3}{4} \Yl\DYl \left( \co + 3\ct \right) \right.
		\nonumber
		\\
		&& + \left. \frac{1}{6} g^2_1 \co + \frac{1}{4} \left( \left( 3 - 2\xi^{}_1 \right)g^2_1 + 3\left( 3 - 2\xi^{}_2 \right)g^2_2 \right) \co \right] \;,
		\nonumber
		\\
		\dC{(3)}{H\ell} &=& \loopfactore \left[ \frac{1}{2} C^{}_5 C^\dagger_5 + \frac{1}{3} g^2_2 \tr{\ct} \mathbbm{1} + \frac{1}{4} \left( 3\co + \ct \right) \Yl\DYl + \frac{1}{4} \Yl\DYl \left( 3\co + \ct \right) \right.
		\nonumber
		\\
		&& + \left. \frac{1}{6} g^2_2 \ct + \frac{1}{4} \left( \left( 3-2\xi^{}_1 \right)g^2_1 - 3\left( 1+ 2\xi^{}_2 \right)g^2_2 \right) \ct \right] \;,
		\nonumber
		\\
		\dC{}{He} &=& \loopfactore \left[ \frac{2}{3} g^2_1 \tr{\co} \mathbbm{1} - \DYl\co\Yl \right] \;,
		\nonumber
		\\
		\dC{(1)}{Hq} &=& - \loopfactore \frac{1}{9} g^2_1 \tr{\co} \mathbbm{1} \;,
		\nonumber
		\\
		\dC{(3)}{Hq} &=& \loopfactore \frac{1}{3} g^2_2 \tr{\ct} \mathbbm{1} \;,
		\nonumber
		\\
		\dC{}{Hu} &=& -\loopfactore \frac{4}{9} g^2_1 \tr{\co} \mathbbm{1} \;,
		\nonumber
		\\
		\dC{}{Hd} &=& \loopfactore \frac{2}{9} g^2_1 \tr{\co} \mathbbm{1} \;.
	\end{eqnarray}
	
	\item $\bf{\left( \overline{L}L \right) \left( \overline{L}L\right)}$
	\begin{eqnarray}
		\dC{\alpha\beta\gamma\lambda}{\ell\ell} &=& - \loopfactore \left[ \frac{1}{4} C^{\alpha\gamma}_5 C^{\beta\lambda \ast}_5 + \frac{1}{6} g^2_1 \delta^{\gamma\lambda} C^{(1)\alpha\beta}_{H\ell} - \frac{1}{6} g^2_2 \left( 2\delta^{\gamma\beta} C^{(3)\alpha\lambda}_{H\ell} - \delta^{\gamma\lambda} C^{(3) \alpha\beta}_{H\ell} \right) \right.
		\nonumber
		\\
		&& + \frac{1}{4} \left( \Yl \DYl \right)^{}_{\alpha\beta} \left( \co - \ct \right)^{\gamma\lambda} + \frac{1}{4} \left( \co - \ct \right)^{\alpha\beta} \left( \Yl\DYl \right)^{}_{\gamma\lambda} 
		\nonumber
		\\
		&& + \left. \frac{1}{2} \left( \Yl\DYl \right)^{}_{\alpha\lambda} C^{(3)\gamma\beta}_{H\ell} + \frac{1}{2} C^{(3)\alpha\lambda}_{H\ell} \left( \Yl \DYl \right)^{}_{\gamma\beta} \right] \;,
		\nonumber
		\\
		\dC{(1)\alpha\beta\gamma\lambda}{\ell q} &=&  \loopfactore \left[ \frac{1}{18} g^2_1 \delta^{\gamma\lambda} C^{(1)\alpha\beta}_{H\ell} + \frac{1}{2} C^{(1)\alpha\beta}_{H\ell} \left( \Yu\DYu - \Yd\DYd \right)^{}_{\gamma\lambda} \right] \;,
		\nonumber
		\\
		\dC{(3)\alpha\beta\gamma\lambda}{\ell q} &=&  \loopfactore \left[ \frac{1}{6} g^2_2 \delta^{\gamma\lambda} C^{(3)\alpha\beta}_{H\ell} - \frac{1}{2} C^{(3)\alpha\beta}_{H\ell} \left( \Yu\DYu + \Yd\DYd \right)^{}_{\gamma\lambda} \right] \;.
	\end{eqnarray}
	
	\item $\bf{\left( \overline{L}L \right) \left( \overline{R}R\right)}$
	\begin{eqnarray}\label{eq:llrr-warsaw}
		\dC{\alpha\beta\gamma\lambda}{\ell e} &=& \loopfactore \left[ - \frac{1}{3} g^2_1 \delta^{\gamma\lambda} C^{(1)\alpha\beta}_{H\ell} + C^{(1)\alpha\beta}_{H\ell} \left( \DYl\Yl \right)^{}_{\gamma\lambda} \right] \;,
		\nonumber
		\\
		\dC{\alpha\beta\lambda\gamma}{\ell u} &=& \loopfactore \left[ \frac{2}{9} g^2_1 \delta^{\gamma\lambda} C^{(1)\alpha\beta}_{H\ell} - C^{(1)\alpha\beta}_{H\ell} \left( \DYu\Yu \right)^{}_{\gamma\lambda} \right] \;,
		\nonumber
		\\
		\dC{\alpha\beta\lambda\gamma}{\ell d} &=& \loopfactore \left[ - \frac{1}{9} g^2_1 \delta^{\gamma\lambda} C^{(1)\alpha\beta}_{H\ell}  + C^{(1)\alpha\beta}_{H\ell} \left( \DYd \Yd \right)^{}_{\gamma\lambda} \right] \;.
	\end{eqnarray}
\end{itemize}

\section{Complete One-loop RGEs for the Type-II and Type-III SEFTs}
\label{app:SEFTs}

\subsection{Generic Contributions from the Weinberg Operator} \label{subapp:con-5}

\subsubsection*{Renormalizable terms}
\begin{itemize}
	\item {\bf The Higgs quartic coupling}
	\begin{eqnarray}
		\bda{\lambda} &=&   2 m^2   \tr{ C^{}_5 C^\dagger_5 } \;.
	\end{eqnarray}
\end{itemize}

\subsubsection*{Dimension-six operators}

\begin{itemize}
	\item {\bf $\bm{ H^6}$ and $\bm{ H^4 D^2}$}
	\begin{eqnarray}
		\bda {C^{}_{H\square}} &=& - \tr{ C^{}_5 C^\dagger_5 } \;,
		\nonumber
		\\
		\bda{C^{}_{HD}} &=& - 2 \tr{ C^{}_5 C^\dagger_5 } \;,
		\nonumber
		\\
		\bda{C^{}_{H}} &=& -4\lambda  \tr{ C^{}_5 C^\dagger_5} \;.
	\end{eqnarray}
	
	\item $\bm {\psi^2 H^3}$
	\begin{eqnarray}
		\bda{C^{}_{eH}} &=& \frac{3}{2}  C^{}_5 C^\dagger_5  \Yl - \tr{ C^{}_5 C^\dagger_5} \Yl \;,
		\nonumber
		\\
		\bda{C^{}_{uH}} &=& - \tr{ C^{}_5 C^\dagger_5 } \Yu \;,
		\nonumber
		\\
		\bda{C^{}_{dH}} &=& - \tr{  C^{}_5 C^\dagger_5 } \Yd \;.
	\end{eqnarray}
	
	\item $\bm{\psi^2H^2 D}$
	\begin{eqnarray}
		\bda{C^{(1)}_{H\ell}}&=& -\frac{3}{2} C^{}_5 C^\dagger_5 \;,
		\nonumber
		\\
		\bda{C^{(3)}_{H\ell}}&=& C^{}_5 C^\dagger_5  \;.
	\end{eqnarray}
	
	\item $\mathbf {\left( \overline{L} L \right) \left( \overline{L} L \right) }$
	\begin{eqnarray}
		\bda{C^{\alpha\beta\gamma\lambda}_{\ell\ell}} &=& -\frac{1}{2} C^{\alpha\gamma}_5 C^{\beta\lambda \ast}_5 \;.
	\end{eqnarray}
\end{itemize}

\subsection{Contributions from Dim-6 Operators in the Type-II SEFT}

\subsubsection*{Renormalizable terms}

\begin{itemize}
	\item {\bf The Yukawa couplings}
	\begin{eqnarray}
		\bda{Y^{}_l} &=& - m^2 \left( 6 C^{}_{eH} - 2\Yl C^{}_{H\square} + \Yl C^{}_{HD} \right) \;,
		\nonumber
		\\
		\bda{Y^{}_{\rm u}} &=& - m^2 \left( 6 C^{}_{uH} - 2 \Yu C^{}_{H\square} + \Yu C^{}_{HD} \right) \;,
		\nonumber
		\\
		\bda{Y^{}_{\rm d}} &=& - m^2 \left( 6 C^{}_{dH} - 2 \Yd C^{}_{H\square} + \Yd C^{}_{HD} \right) \;.
	\end{eqnarray}
	\item {\bf The Higgs quadratic and quartic couplings}
	\begin{eqnarray}\label{eq:t2higgs}
		\bda{m^2} &=& 4m^4 \left( 2C^{}_{H\square} - C^{}_{HD} \right) \;,
		\nonumber
		\\
		\bda{\lambda} &=& - 2m^2 \left[ 12 C^{}_H + 2 \left( -16\lambda + \frac{5}{3} g^2_2 \right) C^{}_{H\square} + 3 \left( 4 \lambda + \frac{1}{2} g^2_1 - \frac{1}{2} g^2_2  \right) C^{}_{HD} \right.
		\nonumber
		\\
		&& + \left.  \tr{ 3 C^{}_{dH} \DYd + 3 \Yd C^\dagger_{dH} + 3C^{}_{uH} \DYu + 3\Yu C^\dagger_{uH} + C^{}_{eH} \DYl + \Yl C^\dagger_{eH} } \right] \;.
	\end{eqnarray}
\end{itemize}

\subsubsection*{Dimension-six operators}

\begin{itemize}
	\item {\bf $\bm{ H^6}$ and $\bm{ H^4 D^2}$}
	\begin{eqnarray}\label{eq:type-II-rge1}
		\bda{C^{}_H} &=& \left[ 108 \lambda - \frac{9}{2} \left( g^2_1 + 3g^2_2 \right) + 6T \right] C^{}_H + 40 \lambda \left( \frac{1}{3} g^2_2 - 4\lambda \right) C^{}_{H\square} + \left[ 48 \lambda^2  \right.
		\nonumber
		\\
		&&+ \left. 6 \lambda \left( g^2_1 - g^2_2 \right) -  \frac{3}{4} \left( g^2_1 + g^2_2 \right)^2 \right]  C^{}_{HD} + 4{\rm tr}\left[ 3C^{}_{dH} \DYd \left( \lambda - \Yd\DYd \right) \right.
		\nonumber
		\\
		&& + 3 \left( \lambda - \Yd\DYd \right) \Yd C^\dagger_{dH} + 3C^{}_{uH} \DYu \left( \lambda - \Yu\DYu \right) + 3 \left( \lambda - \Yu\DYu \right) \Yu C^\dagger_{uH} 
		\nonumber
		\\
		&& + \left. C^{}_{eH} \DYl \left( \lambda - \Yl\DYl \right) + \left( \lambda - \Yl\DYl \right) \Yl C^\dagger_{eH} \right] \;,
		\nonumber
		\\
		\bda{C^{}_{H\square}} &=& 4 \left[ 6 \lambda - \left( \frac{1}{3} g^2_1 + g^2_2 \right) + T \right]C^{}_{H\square} + \frac{5}{3} g^2_1 C^{}_{HD} \;,
		\nonumber
		\\
		\bda{C^{}_{HD}} &=&  \left( 12\lambda + 4T + \frac{9}{2} g^2_2 - \frac{5}{6} g^2_1 \right) C^{}_{HD} + \frac{20}{3} g^2_1 C^{}_{H\square} \;.
	\end{eqnarray}

	\item $\bm{\psi^2H^3}$
	\begin{eqnarray}\label{eq:type-II-rge2}
		\bda{C^{}_{eH}} &=& \left( -4\lambda \Yl + \frac{10}{3} g^2_2 \Yl - 6 \Yl\DYl\Yl \right) C^{}_{H\square} + \left[ 2\lambda \Yl + \frac{3}{2} \left( g^2_1 - g^2_2 \right) \Yl + \Yl\DYl\Yl \right] C^{}_{HD}
		\nonumber
		\\
		&& + \left[ 24\lambda - \frac{3}{4} \left( 7g^2_1 + 9 g^2_2 \right) + 3T \right] C^{}_{eH} + 2 \tr{ 3 C^{}_{dH} \DYd + 3C^\dagger_{uH} \Yu + C^{}_{eH} \DYl} \Yl 
		\nonumber
		\\
		&& + 5 C^{}_{eH} \DYl \Yl + \frac{11}{2} \Yl \DYl C^{}_{eH} \;,
		\nonumber
		\\
		\bda{C^{}_{uH}} &=&  \left( -4\lambda \Yu + \frac{10}{3} g^2_2 \Yu - 6\Yu\DYu\Yu \right) C^{}_{H\square} + \left[ 2 \lambda \Yu + \frac{3}{2} \left( g^2_1 - g^2_2 \right) \Yu + \Yu\DYu\Yu \right] C^{}_{HD}
		\nonumber
		\\
		&& + \left[ 24\lambda - \left( \frac{35}{12} g^2_1 + \frac{27}{4} g^2_2 + 8 g^2_s \right) + 3T \right] C^{}_{uH} - 2 \Yd C^\dagger_{dH} \Yu - C^{}_{dH} \DYd \Yu  
		\nonumber
		\\
		&& - \frac{3}{2} \Yd\DYd C^{}_{uH} + 2 \tr{ 3 C^\dagger_{dH} \Yd + 3 C^{}_{uH} \DYu + C^\dagger_{eH} \Yl } \Yu + 5 C^{}_{uH} \DYu\Yu 
        \nonumber
        \\
        && + \frac{11}{2} \Yu\DYu C^{}_{uH} \;,
		\nonumber
		\\
		\bda{C^{}_{dH}} &=&  \left( -4\lambda \Yd + \frac{10}{3} g^2_2 \Yd - 6\Yd\DYd\Yd \right) C^{}_{H\square} + \left[ 2 \lambda \Yd + \frac{3}{2} \left( g^2_1 - g^2_2 \right) \Yd + \Yd\DYd\Yd \right] C^{}_{HD}
		\nonumber
		\\
		&& + \left[ 24\lambda - \left( \frac{23}{12} g^2_1 + \frac{27}{4} g^2_2 + 8 g^2_s \right) + 3T \right] C^{}_{dH} - 2 \Yu C^\dagger_{uH} \Yd - C^{}_{uH} \DYu \Yd  
		\nonumber
		\\
		&& - \frac{3}{2} \Yu\DYu C^{}_{dH} + 2 \tr{ 3 C^{}_{dH} \DYd + 3 C^\dagger_{uH} \Yu + C^{}_{eH} \DYl } \Yd + 5 C^{}_{dH} \DYd\Yd 
        \nonumber
        \\
        && + \frac{11}{2} \Yd\DYd C^{}_{dH} \;.
	\end{eqnarray}
	
	\item $\bm{\psi^2 H^2 D}$
	\begin{eqnarray}\label{eq:type-II-rge}
		\bda{C^{(1)\alpha\beta}_{H\ell}} &=& - \frac{1}{2} \left( \frac{1}{3} g^2_1 + \Yl \DYl  \right)^{}_{\alpha\beta} \left( C^{}_{H\square} + C^{}_{HD} \right) - 2 \left( \frac{1}{3} g^2_1 + \Yl\DYl \right)^{}_{\lambda\gamma} \left( 2 C^{\alpha\beta\gamma\lambda}_{\ell\ell} + C^{\alpha\lambda\gamma\beta}_{\ell\ell} \right) \;,
		\nonumber
		\\
		\bda{C^{(3)\alpha\beta}_{H\ell}} &=& \frac{1}{2} \left( \frac{1}{3} g^2_2 - \Yl\DYl \right)^{}_{\alpha\beta} C^{}_{H\square} + 2 \left( \frac{1}{3} g^2_2 - \Yl\DYl \right)^{}_{\gamma\lambda} C^{\alpha\gamma\lambda\beta}_{\ell\ell} \;,
		\nonumber
		\\
		\bda{C^{\alpha\beta}_{He}} &=& \left( - \frac{1}{3} g^2_1 + \DYl \Yl \right)^{}_{\alpha\beta} \left( C^{}_{H \square} + C^{}_{HD} \right) \;,
		\nonumber
		\\
		\bda{C^{(1)\alpha\beta}_{Hq}} &=& \frac{1}{2} \left( \frac{1}{9} g^2_1 + \Yu\DYu - \Yd\DYd \right)^{}_{\alpha\beta} \left( C^{}_{H\square} + C^{}_{HD} \right) \;,
		\nonumber
		\\
		\bda{C^{(3)\alpha\beta}_{Hq}} &=& \frac{1}{2} \left( \frac{1}{3} g^2_2 - \Yu\DYu - \Yd\DYd \right)^{}_{\alpha\beta} C^{}_{H\square} \;,
		\nonumber
		\\
		\bda{C^{\alpha\beta}_{Hu}} &=& \left( \frac{2}{9} g^2_1 - \DYu \Yu \right)^{}_{\alpha\beta} \left( C^{}_{H\square} + C^{}_{HD} \right) \;,
		\nonumber
		\\
		\bda{C^{\alpha\beta}_{Hd}} &=& \left( - \frac{1}{9} g^2_1 + \DYd \Yd \right)^{}_{\alpha\beta} \left( C^{}_{H\square} + C^{}_{HD} \right) \;,
		\nonumber
		\\
		\bda{C^{\alpha\beta}_{Hud}} &=& \left( \DYu \Yd \right)^{}_{\alpha\beta} \left( 2C^{}_{H\square} - C^{}_{HD} \right) \;.
	\end{eqnarray}

	\item $\bf{\left( \overline{L}L \right) \left( \overline{L}L\right)}$
	\begin{eqnarray}
		\bda{C^{\alpha\beta\gamma\lambda}_{\ell\ell}} &=& \frac{1}{3} \left( g^2_1 - g^2_2 \right) \left( C^{\alpha\rho\rho\beta}_{\ell\ell} \delta^{}_{\gamma\lambda} + C^{\rho\lambda\gamma\rho}_{\ell\ell} \delta^{}_{\alpha\beta} \right) + \frac{2}{3} g^2_1 \left( C^{\alpha\beta\rho\rho}_{\ell\ell} \delta^{}_{\gamma\lambda} + C^{\rho\rho\gamma\lambda}_{\ell\ell} \delta^{}_{\alpha\beta} \right)
		\nonumber
		\\
		&& + \frac{2}{3} g^2_2 \left( C^{\rho\beta\gamma\rho}_{\ell\ell} \delta^{}_{\alpha\lambda} + C^{\alpha\rho\rho\lambda}_{\ell\ell} \delta^{}_{\beta\gamma} \right) + 6 g^2_2 C^{\alpha\lambda\gamma\beta}_{\ell\ell} - 3 \left( g^2_2 - g^2_1 \right) C^{\alpha\beta\gamma\lambda}_{\ell\ell}
		\nonumber
		\\
		&& + \frac{1}{2} \left( \Yl \DYl \right)^{}_{\alpha\rho} C^{\rho\beta\gamma\lambda}_{\ell\ell} + \frac{1}{2} \left( \Yl\DYl \right)^{}_{\gamma\rho} C^{\alpha\beta\rho\lambda}_{\ell\ell} + \frac{1}{2} C^{\alpha\rho\gamma\lambda}_{\ell\ell} \left( \Yl\DYl\right)^{}_{\rho\beta} 
		\nonumber
		\\
		&& + \frac{1}{2} C^{\alpha\beta\gamma\rho}_{\ell\ell} \left( \Yl\DYl \right)^{}_{\rho\lambda} \;,
		\nonumber
		\\
		\bda{C^{(1)\alpha\beta\gamma\lambda}_{\ell q}} &=& -\frac{2}{9} g^2_1 \delta^{}_{\gamma\lambda} \left( C^{\alpha\rho\rho\beta}_{\ell\ell} + 2 C^{\alpha\beta\rho\rho}_{\ell\ell} \right) \;,
		\nonumber
		\\
		\bda{C^{(3)\alpha\beta\gamma\lambda}_{\ell q}} &=& \frac{2}{3} g^2_2 \delta^{}_{\gamma\lambda} C^{\alpha\rho\rho\beta}_{\ell\ell} \;.
	\end{eqnarray}

	\item $\bf{\left( \overline{L}L \right) \left( \overline{R}R\right)}$
	\begin{eqnarray}
		\bda{C^{\alpha\beta\gamma\lambda}_{\ell e}} &=& \frac{4}{3} g^2_1 \delta^{}_{\gamma\lambda} \left( 2 C^{\alpha\beta\rho\rho}_{\ell\ell} + C^{\alpha\rho\rho\beta}_{\ell\ell} \right) - 2 \left( C^{\alpha\rho\sigma\beta}_{\ell\ell} + 2 C^{\alpha\beta\sigma\rho}_{\ell\ell} \right) \left( \Yl \right)^{}_{\rho\lambda} \left( \DYl \right)^{}_{\gamma\sigma} \;,
		\nonumber
		\\
		\bda{C^{\alpha\beta\gamma\lambda}_{\ell u}} &=& - \frac{8}{9} g^2_1 \delta^{}_{\gamma\lambda} \left( 2 C^{\alpha\beta\rho\rho}_{\ell\ell} + C^{\alpha\rho\rho\beta}_{\ell\ell} \right) \;,
		\nonumber
		\\
		\bda{C^{\alpha\beta\gamma\lambda}_{\ell d}} &=& \frac{4}{9} g^2_1 \delta^{}_{\gamma\lambda} \left( 2 C^{\alpha\beta\rho\rho}_{\ell} + C^{\alpha\rho\rho\beta}_{\ell\ell} \right) \;.
	\end{eqnarray}
\end{itemize}
Note that the quartic coupling $\lambda$ in Eqs.~\eqref{eq:t2higgs}-\eqref{eq:type-II-rge2} is the effective one at the tree level, e.g., $\lambda \to \lambda - 2\lambda^2_\Delta \left( 1 + 2m^2/M^2_\Delta \right)$.

\subsection{Contributions from Dim-6 Operators in the Type-III SEFT}

\subsubsection*{Renormalizable terms}

\begin{itemize}
	\item {\bf The Yukawa couplings}
	\begin{eqnarray}
		\bda{Y^{}_l} &=& - 2m^2 \left[\left( C^{(1)}_{H\ell} + 3C^{(3)}_{H\ell} \right) \Yl +  3 C^{}_{eH} \right] \;.
	\end{eqnarray}
	\item {\bf the Higgs quadratic and quartic couplings}
	\begin{eqnarray}
		\bda{\lambda} &=& - 2m^2 \trs{  4  C^{(3)}_{H\ell} \left( \frac{1}{3} g^2_2 - \Yl \DYl \right)+ C^{}_{eH} \DYl + \Yl C^\dagger_{eH} }\;.
	\end{eqnarray}
\end{itemize}

\subsubsection*{Dimension-six operators}

\begin{itemize}
	\item {\bf $\bm{ H^6}$ and $\bm{ H^4 D^2}$}
	\begin{eqnarray}
		\bda {C^{}_{H\square}} &=& - 2\trs{ \frac{1}{3} g^2_1 \co - g^2_2 \ct + (\co+3\ct)\Yl\DYl } \;,
		\nonumber
		\\
		\bda{C^{}_{HD}} &=& - 2 \tr{  \frac{4}{3} g^2_1 \co +  4\co \Yl\DYl } \;,
		\nonumber
		\\
		\bda{C^{}_{H}} &=& 4 \trs{ 4\lambda \ct \left( \frac{1}{3}  g^2_2  -  \Yl\DYl \right)  + C^{}_{eH} \DYl \left( \lambda - \Yl\DYl \right) + \left( \lambda - \Yl\DYl \right) \Yl C^\dagger_{eH} } \;.
		\nonumber
		\\
	\end{eqnarray}
	
	\item $\bm {\psi^2 H^3}$
	\begin{eqnarray}
		\bda{C^{}_{eH}} &=& 2 \left[ \tr{  \frac{2}{3} g^2_2 \ct - 2\ct \Yl \DYl} \Yl + \co \Yl\DYl\Yl  + \left( 2\lambda - 3g^2_1 \right) \co \Yl  \right.
		\nonumber
		\\
		&& + \left. 3 \left( 2\lambda - g^2_1 \right) \ct \Yl \right] + \left[ 24\lambda - \frac{3}{4} \left( 7g^2_1 + 9 g^2_2 \right) + 3T \right] C^{}_{eH} 
		\nonumber
		\\
		&& + 2 \tr{ C^{}_{eH} \DYl} \Yl + 5 C^{}_{eH} \DYl \Yl + \frac{11}{2} \Yl \DYl C^{}_{eH} \;,
		\nonumber
		\\
		\bda{C^{}_{uH}} &=& \tr{  \frac{4}{3} g^2_2 \ct -  4\ct \Yl\DYl + 2 \Yl C^\dagger_{eH}} \Yu \;,
		\nonumber
		\\
		\bda{C^{}_{dH}} &=& \tr{  \frac{4}{3} g^2_2 \ct - 4\ct \Yl\DYl + 2 C^{}_{eH} \DYl} \Yd \;.
	\end{eqnarray}
	
	\item $\bm{\psi^2H^2 D}$
	\begin{eqnarray}
		\bda{C^{(1)}_{H\ell}}&=&  \frac{2}{3} g^2_1  \tr{\co} \mathbbm{1} + \left( \frac{1}{3} g^2_1 + 2T \right)\co 
		\nonumber
		\\
		&& + \frac{1}{2} \left[ \left( 4\co + 9\ct \right) \Yl\DYl + \Yl\DYl \left( 4\co + 9\ct \right) \right] \;,
		\nonumber
		\\
		\bda{C^{(3)}_{H\ell}}&=&  \frac{2}{3} g^2_2  \tr{\ct} \mathbbm{1} + \left( -\frac{17}{3}  g^2_2 + 2T \right)\ct 
		\nonumber
		\\
		&& + \frac{1}{2} \left[ \left( 3\co + 2\ct \right) \Yl\DYl + \Yl\DYl \left( 3\co + 2\ct \right) \right] \;,
		\nonumber
		\\
		\bda{C^{}_{He}} &=&  \frac{4}{3} g^2_1 \tr{\co} \mathbbm{1} - 2\DYl\co\Yl  \;,
		\nonumber
		\\
		\bda{C^{(1)}_{Hq}} &=& - \frac{2}{9} g^2_1 \tr{\co} \mathbbm{1} \;,
		\nonumber
		\\
		\bda{C^{(3)}_{Hq}} &=& \frac{2}{3} g^2_2 \tr{\ct} \mathbbm{1} \;,
		\nonumber
		\\
		\bda{C^{}_{Hu}} &=& - \frac{8}{9} g^2_1 \tr{\co} \mathbbm{1} \;,
		\nonumber
		\\
		\bda{C^{}_{Hd}} &=& \frac{4}{9} g^2_1 \tr{\co} \mathbbm{1} \;,
	\end{eqnarray}
	
	\item $\mathbf {\left( \overline{L} L \right) \left( \overline{L} L \right) }$
	\begin{eqnarray}
		\bda{C^{\alpha\beta\gamma\lambda}_{\ell\ell}} &=& -  \frac{1}{3} g^2_1 \delta^{\gamma\lambda} C^{(1)\alpha\beta}_{H\ell} + \frac{1}{3} g^2_2 \left( 2\delta^{\gamma\beta} C^{(3)\alpha\lambda}_{H\ell} - \delta^{\gamma\lambda} C^{(3) \alpha\beta}_{H\ell} \right) 
		\nonumber
		\\
		&& - \frac{1}{2} \left( \Yl \DYl \right)^{}_{\alpha\beta} \left( \co - \ct \right)^{\gamma\lambda} - \frac{1}{2} \left( \co - \ct \right)^{\alpha\beta} \left( \Yl\DYl \right)^{}_{\gamma\lambda} 
		\nonumber
		\\
		&& -  \left( \Yl\DYl \right)^{}_{\alpha\lambda} C^{(3)\gamma\beta}_{H\ell} - C^{(3)\alpha\lambda}_{H\ell} \left( \Yl \DYl \right)^{}_{\gamma\beta} \;,
		\nonumber
		\\
		\bda{C^{(1)\alpha\beta\gamma\lambda}_{\ell q}} &=&   \frac{1}{9} g^2_1 \delta^{\gamma\lambda} C^{(1)\alpha\beta}_{H\ell} +  C^{(1)\alpha\beta}_{H\ell} \left( \Yu\DYu - \Yd\DYd \right)^{}_{\gamma\lambda} \;,
		\nonumber
		\\
		\bda{C^{(3)\alpha\beta\gamma\lambda}_{\ell q}} &=&  \frac{1}{3} g^2_2 \delta^{\gamma\lambda} C^{(3)\alpha\beta}_{H\ell} -  C^{(3)\alpha\beta}_{H\ell} \left( \Yu\DYu + \Yd\DYd \right)^{}_{\gamma\lambda} \;.
	\end{eqnarray}
	
	\item $\mathbf {\left( \overline{L} L \right) \left( \overline{R} R \right) }$
	\begin{eqnarray}
		\bda{C^{\alpha\beta\gamma\lambda}_{\ell e}} &=& - \frac{2}{3} g^2_1 \delta^{\gamma\lambda} C^{(1)\alpha\beta}_{H\ell} + 2C^{(1)\alpha\beta}_{H\ell} \left( \DYl\Yl \right)^{}_{\gamma\lambda} \;,
		\nonumber
		\\
		\bda{C^{\alpha\beta\lambda\gamma}_{\ell u}} &=& \frac{4}{9} g^2_1 \delta^{\gamma\lambda} C^{(1)\alpha\beta}_{H\ell} - 2C^{(1)\alpha\beta}_{H\ell} \left( \DYu\Yu \right)^{}_{\gamma\lambda} \;,
		\nonumber
		\\
		\bda{C^{\alpha\beta\lambda\gamma}_{\ell d}} &=& - \frac{2}{9} g^2_1 \delta^{\gamma\lambda} C^{(1)\alpha\beta}_{H\ell}  + 2C^{(1)\alpha\beta}_{H\ell} \left( \DYd \Yd \right)^{}_{\gamma\lambda} \;.
	\end{eqnarray}
\end{itemize}

\section{Explicit Expressions for Mixing Parameters}
\label{app:rge}
In this appendix, we outline the strategy to diagonalize $Y_l^{}$ and $\kappa$ and to derive the explicit expressions of the RGEs of three mixing angles, three CP-violating phases, the moduli and phases of the $\eta$ and $\eta^\prime$ elements. According to the RGE of $Y_l^{}$ in Eq.~(\ref{eq:RGEYlnew}), by inserting $Y^{}_l = U^{}_l \widehat{Y}^{}_l U^{\prime \dagger}_l$ into the left-handed side of Eq.~(\ref{eq:RGEYlnew}), we have
\begin{eqnarray}
	\dot{U}_l^{} \, \widehat{Y}_l^{} \,  U_l^{\prime\dagger} + U_l^{} \, \dot{\widehat{Y}}_l^{} \,  U_l^{\prime\dagger} + U_l^{} \, \widehat{Y}_l^{} \,  \dot{U}_l^{\prime\dagger} = \left[ \alpha_l^{} + C_l^l \left( Y_l^{} Y_l^\dagger \right) - 2 \, \frac{m^2_{}}{v^2_{}} U_l^{} P \left(\eta^\prime_{} - 4 \eta \right) P^\dagger U_l^\dagger \right] Y_l^{} \;.
	\label{eq:ylder}
\end{eqnarray}
Then multiplying $U_l^\dagger$ from left and $U_l^\prime \, \widehat{Y_l^{}}$ from right on both sides of Eq.~(\ref{eq:ylder}) and then adding it to its Hermitian conjugate~\cite{Xing:2020ijf}, one arrives at
\begin{eqnarray}
	&& U_l^\dagger \, \dot{U_l^{}} \, \widehat{Y}^2_l - \widehat{Y}^2_l U_l^\dagger \, \dot{U}_l^{} + 2 \, \dot{\widehat{Y}}_l^{} \, \widehat{Y}_l^{} \nonumber\\
	= && 2 \, \alpha_l^{} \, \widehat{Y}_l^2 + 2 \, C_l^l \, \widehat{Y}^4_l - 2 \, \frac{m^2_{}}{v^2_{}} P \left(\eta^\prime_{} - 4 \eta \right) P^\dagger_{} \widehat{Y}^2_l - 2 \, \frac{m^2_{}}{v^2_{}} \widehat{Y}^2_l P \left(\eta^\prime_{} - 4 \eta \right) P^\dagger_{} \; ,
	\label{eq:rgeyl}
\end{eqnarray}
where $U^\dagger_l \dot{U}^{}_l = - \dot{U}^\dagger_l U^{}_l$ and $U^{\prime \dagger}_l \dot{U}^{\prime}_l = - \dot{U}^{\prime \dagger}_l U^{\prime}_l$ from the unitarity conditions $U^\dagger_l U^{}_l = \mathbbm{1}$ and $U^{\prime \dagger}_l U^\prime_l = \mathbbm{1}$ have been used. Taking the diagonal elements on both sides of Eq.~(\ref{eq:rgeyl}) leads to the RGEs of the eigenvalues of $Y_l^{}$, namely,
\begin{eqnarray} \label{eq:rge-lep1}
	\dot{y}_\alpha = \left[ \alpha_l^{} + C_l^l \, y_\alpha^2 - 2 \, \frac{m^2_{}}{v^2_{}} \left(\eta^\prime_{} - 4 \eta \right)_{\alpha \alpha}^{} \right] y_\alpha \;, 
\end{eqnarray}
whereas the off-diagonal elements can be rewritten as
\begin{eqnarray}
	\left(U_l^\dagger \dot{U}_l^{}\right)_{\alpha\beta}^{} = -\left(\dot{U}_l^\dagger U_l^{} \right)_{\alpha\beta}^{}= - 2 \, \frac{m^2_{}}{v^2_{}} y_{\alpha\beta}^{} \left(\eta^\prime_{} - 4 \eta \right)_{\alpha \beta}^{} {\rm e}^{{\rm i} \left(\phi_\alpha^{}-\phi_\beta^{} \right)}_{} \;,
	\label{eq:Uloffd}
\end{eqnarray}
for $\alpha \neq \beta$, where $y_{\alpha\beta}^{} \equiv (y_\beta^2 + y_\alpha^2)/(y_\beta^2 - y_\alpha^2)$ for $\alpha,\beta = e,\mu,\tau$ have been defined. Notice that $\dot{U}_l^\dagger U_l^{}$ is anti-Hermitian, so ${\rm Re} [(\dot{U}_l^\dagger U^{}_l)_{\alpha\alpha}^{}] = 0$ holds. However, no simple information about ${\rm Im} [(\dot{U}_l^\dagger U^{}_l)_{\alpha\alpha}^{}]$ can be acquired. In the neutrino sector, we insert $\kappa = U_\nu^\ast \widehat{\kappa} U^{\rm T}_\nu$ into the left-hand side of Eq.~(\ref{eq:RGEkappa}) and obtain~\cite{Chankowski:1993tx,Babu:1993qv,Antusch:2001ck,Antusch:2001vn,Xing:2005fw}
\begin{eqnarray}
	\dot{U}_\nu^{} \widehat{\kappa} \, U_\nu^{\rm T} + U_\nu^{} \dot{\widehat{\kappa}} \, U_\nu^{\rm T} + U_\nu^{} \widehat{\kappa} \, \dot{U}_\nu^{\rm T} = \alpha_\kappa^{} \kappa + C_\kappa^{} \left[ \left( Y_l^{} Y_l^\dagger \right) \kappa + \kappa  \left( Y_l^{} Y_l^\dagger \right)^{\rm T}_{} \right] \;.
	\label{eq:kprge}
\end{eqnarray}
By multiplying $U_\nu^\dagger$ from left and $U_\nu^\ast$ from right on both sides of Eq.~(\ref{eq:kprge}), we have~\cite{Xing:2020ijf}
\begin{eqnarray}
	\dot{\widehat{\kappa}} = \alpha_\kappa^{} \widehat{\kappa} + C_\kappa^{} \left( V^{\prime \dagger}_{} \widehat{Y}_l^2 V^\prime \widehat{\kappa} +  \widehat{\kappa} \, V^{\prime \rm T}_{} \widehat{Y}_l^2 V^{\prime \ast}_{} \right) + \left(\dot{U}_\nu^\dagger U_\nu^{} \right) \widehat{\kappa} - \widehat{\kappa} \left(\dot{U}_\nu^\dagger U_\nu^{} \right)^\ast_{} \;,
	\label{eq:rgekp}
\end{eqnarray}
where $ V^\prime \equiv U_l^\dagger U_\nu^{}$ and $\dot{U}^\dagger_\nu U^{}_\nu = - U^\dagger_\nu \dot{U}^{}_\nu$ due to the unitarity condition $U^\dagger_\nu U^{}_\nu = \mathbbm{1}$ have been used. Since $\widehat{\kappa}_i^{}$ are real and positive, the diagonal elements on the right-hand side of Eq.~(\ref{eq:rgekp}) should be so as well, leading to ${\rm Im} [(\dot{U}_\nu^\dagger U_\nu^{})_{ii}^{}] = 0$. Moreover, as $\dot{U}^\dagger_\nu U^{}_\nu$ is anti-Hermitian with ${\rm Re} [(\dot{U}_\nu^\dagger U^{}_\nu)_{ii}^{}] = 0$, the diagonal elements of $\dot{U}_\nu^\dagger U_\nu^{}$ are actually vanishing. 

From the diagonal elements on both sides of Eq.~(\ref{eq:rgekp}), one can immediately extract the RGEs of eigenvalues of $\kappa$, i.e.,
\begin{eqnarray} \label{eq:rge-neu1}
	\dot{\kappa}_i^{} = \left( \alpha_\kappa^{} + 2 C_\kappa^{} {\rm Re} {\cal S}_{ii}^{} \right) \kappa_i^{} \;, 
\end{eqnarray}
and the off-diagonal elements give rise to
\begin{eqnarray}
	\left(U_\nu^\dagger \dot{U}_\nu^{}\right)_{ij}^{} = -\left(\dot{U}_\nu^\dagger U_\nu^{} \right)_{ij}^{}= \frac{ C_\kappa^{} }{\kappa_j^2 - \kappa_i^2} \left[ \left(\kappa_i^2 + \kappa_j^2\right) {\cal S}_{ij}^{} + 2 \kappa_i^{} \kappa_j^{} {\cal S}_{ij}^\ast \right] \;,
    \label{eq:Unuoffd}
\end{eqnarray}
for $i \neq j$. Then, with the help of Eqs.~\eqref{eq:Uloffd} and \eqref{eq:Unuoffd}, the RGEs in Eqs.~\eqref{eq:RGEYlnew}-\eqref{eq:RGEetaprime} can be utilized to derive the RGEs of eigenvalues of $Y^{}_l$ and $\kappa$ in Eqs.~\eqref{eq:rge-lep} and \eqref{eq:rge-neu}, as well as those of the unitary matrix $V^\prime$ and non-unitary parameters in $\eta$ and $\eta^\prime$. To get the RGEs of the mixing parameters in $V^\prime$, we introduce $\mathcal{T} \equiv V^{\prime \dagger} \dot{V}^\prime$ and $\mathcal{T}^\prime \equiv Q\cdot \mathcal{T} \cdot Q^\dagger$, where the latter is explicitly given by
\begin{eqnarray}
	\mathcal{T}^\prime_{} = \dot{Q} Q^{\dagger} + U^\dagger_{} \dot{U} + U^\dagger_{} P^\dagger_{} \dot{P} \, U \;.
	\label{eq:Tprimeexp}
\end{eqnarray}
On the one hand, the matrix elements of $\mathcal{T}$ can be calculated by using Eq.~(\ref{eq:pmnsrge}), i.e.,
\begin{eqnarray}
	\mathcal{T}^{}_{ii} &=& \sum_\alpha \sum_{\beta \neq \alpha} 2 \, \frac{m^2_{}}{v^2_{}} y_{\alpha\beta}^{} \, U_{\alpha i}^\ast \, U_{\beta i}^{} \left(\eta^\prime_{} - 4 \eta\right)_{\alpha\beta}^{} \;, \label{eq:Tii}\\
	\mathcal{T}^{}_{ij} &=& \sum_\alpha \sum_{\beta\neq \alpha} 2 \, \frac{m^2_{}}{v^2_{}} y_{\alpha\beta}^{} \left(U Q\right)_{\alpha i}^\ast \left(U Q\right)_{\beta j}^{} \left(\eta^\prime_{} - 4 \eta\right)_{\alpha\beta}^{} \nonumber \\
	&~& + \frac{C_\kappa^{}}{\kappa_j^2 - \kappa_i^2} \left[\left(\kappa_i^2 + \kappa_j^2\right) {\cal S}_{ij}^{} + 2 \kappa_i^{} \kappa_j^{} {\cal S}_{ij}^\ast \right] \;, \label{eq:Tij}
\end{eqnarray}
where $i = 1, 2, 3$ in Eq. (\ref{eq:Tii}) and $ij = 12, 13, 23$ in Eq.~(\ref{eq:Tij}) are implied. Via the definitions of ${\cal T}$ and ${\cal T}^\prime$ and the explicit expression of ${\cal T}^\prime$ in Eq.~(\ref{eq:Tprimeexp}), one can find
\begin{eqnarray}
	\mathcal{T}^\prime_{11} &=& \mathcal{T}^{}_{11} = {\rm i} \dot{\rho} + \sum_\alpha \left[ U_{\alpha 1}^\ast \left(\dot{U}_{\alpha 1}^{} + {\rm i} U_{\alpha 1}^{} \dot{\phi}_\alpha^{} \right) \right] \;,
	\label{eq:t11}\\
	\mathcal{T}^\prime_{22} &=& \mathcal{T}^{}_{22} = {\rm i} \dot{\sigma} + \sum_\alpha \left[ U_{\alpha 2}^\ast \left(\dot{U}_{\alpha 2}^{} + {\rm i} U_{\alpha 2}^{} \dot{\phi}_\alpha^{} \right) \right] \;,
	\label{eq:t22} \\
	\mathcal{T}^\prime_{33} &=& \mathcal{T}^{}_{33} = \sum_\alpha \left[ U_{\alpha 3}^\ast \left(\dot{U}_{\alpha 3}^{} + {\rm i} U_{\alpha 3}^{} \dot{\phi}_\alpha^{} \right) \right] \;, 
	\label{eq:t33} \\
	\mathcal{T}^\prime_{12} &=& \mathcal{T}^{}_{12} {\rm e}^{{\rm i} (\rho - \sigma)}_{} = \sum_\alpha \left[ U_{\alpha 1}^\ast \left(\dot{U}_{\alpha 2}^{} + {\rm i} U_{\alpha 2}^{} \dot{\phi}_\alpha^{} \right) \right] \;, \label{eq:t12}\\
	\mathcal{T}^\prime_{13} &=& \mathcal{T}^{}_{13} {\rm e}^{{\rm i} \rho}_{} = \sum_\alpha \left[ U_{\alpha 1}^\ast \left(\dot{U}_{\alpha 3}^{} + {\rm i} U_{\alpha 3}^{} \dot{\phi}_\alpha^{} \right) \right] \;, \label{eq:t13}\\
	\mathcal{T}^\prime_{23} &=& \mathcal{T}^{}_{23} {\rm e}^{{\rm i} \sigma}_{} = \sum_\alpha \left[ U_{\alpha 2}^\ast \left(\dot{U}_{\alpha 3}^{} + {\rm i} U_{\alpha 3}^{} \dot{\phi}_\alpha^{} \right) \right] \;. \label{eq:t23}
\end{eqnarray}
By identifying the matrix elements in Eqs.~(\ref{eq:Tii}) and (\ref{eq:Tij}) with the corresponding ones in Eqs.~(\ref{eq:t11})-(\ref{eq:t23}) and solving them, one can obtain the RGEs for three mixing angles and the Dirac phase in $U$, the Majorana phases in $Q$, and the unphysical phases in $P$. Then, with those results and Eqs.~\eqref{eq:rge-eta} and \eqref{eq:rge-etap}, the RGEs for the moduli and phases of elements in $\eta$ and $\eta^\prime$ under the parametrizations shown in Eqs.~\eqref{eq:etapara} and \eqref{eq:etaprimepara} can be easily achieved. The exact analytical expressions of the one-loop RGEs for all leptonic flavor mixing parameters can be found below. 

\subsubsection*{Three lepton flavor mixing angles}

\begin{eqnarray}
	\dot{\theta}_{12}^{} &=& 2 \, \frac{m^2_{}}{v^2_{}} \left[y_{e \mu}^{} \, c_{13}^{-1}  c_{23}^{} \left( \left| \eta_{e\mu}^\prime \right| c_{e\mu}^\prime - 4 \left| \eta_{e\mu}^{} \right| c_{e\mu}^{} \right) - y_{e \tau}^{} c_{13}^{-1} s_{23}^{} \left( \left| \eta_{e\tau}^\prime \right| c_{e\tau}^\prime - 4 \left| \eta_{e\tau}^{} \right| c_{e\tau}^{} \right)  \right] \nonumber\\
	&& - C_\kappa^{} \, \zeta_{12}^{-1} \left\{ s_{12}^{} c_{12}^{} \left[ c_{13}^2 \, y_e^2 - \left( c_{23}^2 - s_{13}^2 \, s_{23}^2 \right) y_\mu^2 + \left( s_{13}^2 \, c_{23}^2 - s_{23}^2 \right) y_\tau^2 \right] c_{\rho-\sigma}^{} \right.\nonumber\\
	&& \left.\;\;\quad \qquad- s_{13}^{} s_{23}^{} c_{23}^{} \left(y_\mu^2- y_\tau^2 \right) \left(c_{12}^2 \, c_{\delta+\rho-\sigma}^{} - s_{12}^2 \, c_{\delta-\rho+\sigma}^{} \right) \right\} c_{\rho-\sigma}^{} \nonumber\\
	&& - C_\kappa^{} \; \zeta_{12}^{} \, \left\{ s_{12}^{} c_{12}^{} \left[ c_{13}^2 \, y_e^2 - \left( c_{23}^2 - s_{13}^2 \, s_{23}^2 \right) y_\mu^2 + \left( s_{13}^2 \, c_{23}^2 - s_{23}^2 \right) y_\tau^2 \right] s_{\rho-\sigma}^{} \right.\nonumber\\
	&& \left.\;\;\quad \qquad- s_{13}^{}  s_{23}^{} c_{23}^{} \left(y_\mu^2- y_\tau^2 \right) \left(c_{12}^2 \, s_{\delta+\rho-\sigma}^{} + s_{12}^2 \, s_{\delta-\rho+\sigma}^{} \right) \right\} s_{\rho-\sigma}^{} \nonumber\\
	&& + C_\kappa^{} \, \zeta_{13}^{-1} s_{12}^{} s_{13}^{} \left[ s_{12}^{} s_{23}^{} c_{23}^{} \left(y_\mu^2- y_\tau^2 \right) c_\rho^{} - c_{12}^{} s_{13}^{} \left(y_e^2 - s_{23}^2 \, y_\mu^2 - c_{23}^2 \, y_\tau^2\right) c_{\rho+\delta}^{}\right] c_{\rho+\delta}^{} \nonumber\\
	&& + C_\kappa^{} \, \zeta_{13}^{} \; s_{12}^{} s_{13}^{} \left[ s_{12}^{} s_{23}^{} c_{23}^{} \left(y_\mu^2- y_\tau^2 \right) s_\rho^{} - c_{12}^{} s_{13}^{} \left(y_e^2 - s_{23}^2 \, y_\mu^2 - c_{23}^2 \, y_\tau^2\right) s_{\rho+\delta}^{}\right] s_{\rho+\delta}^{} \nonumber\\
	&& + C_\kappa^{} \, \zeta_{23}^{-1} c_{12}^{} s_{13}^{} \left[ c_{12}^{} s_{23}^{} c_{23}^{} \left(y_\mu^2- y_\tau^2 \right) c_\sigma^{} + s_{12}^{} s_{13}^{} \left(y_e^2 - s_{23}^2 \, y_\mu^2 - c_{23}^2 \, y_\tau^2\right) c_{\sigma+\delta}^{}\right] c_{\sigma+\delta}^{} \nonumber\\
	&& + C_\kappa^{} \, \zeta_{23}^{} \; c_{12}^{} s_{13}^{} \left[ c_{12}^{} s_{23}^{} c_{23}^{} \left(y_\mu^2- y_\tau^2 \right) s_\sigma^{} + s_{12}^{} s_{13}^{} \left(y_e^2 - s_{23}^2 \, y_\mu^2 - c_{23}^2 \, y_\tau^2\right) s_{\sigma+\delta}^{}\right] s_{\sigma+\delta}^{} \;,
    \label{eq:theta12}\\
	\dot{\theta}_{13}^{} &=& 2 \, \frac{m^2_{}}{v^2_{}} \left[ y_{e \mu}^{} s_{23}^{} \left( \left| \eta_{e\mu}^\prime \right| c_{e\mu + \delta}^\prime - 4 \left| \eta_{e\mu}^{} \right| c_{e\mu + \delta}^{} \right) + y_{e \tau}^{} \,  c_{23}^{} \left( \left| \eta_{e\tau}^\prime \right| c_{e\tau + \delta}^\prime - 4 \left| \eta_{e\tau}^{} \right| c_{e\tau + \delta}^{} \right) \right] \nonumber\\
	&& + C_\kappa^{} \, \zeta^{-1}_{13} c_{12}^{} \, c_{13}^{} \left[ s_{12}^{} s_{23}^{} c_{23}^{} \left(y_\mu^2 - y_\tau^2\right) c_\rho^{} - c_{12}^{} s_{13}^{} \left(y_e^2 - s_{23}^2 \, y_\mu^2 - c_{23}^2 \, y_\tau^2\right) c_{\rho + \delta}\right] c_{\rho + \delta}^{}
	  \nonumber\\
	&& + C_\kappa^{} \, \zeta^{}_{13} \;c_{12}^{} \, c_{13}^{} \left[ s_{12}^{} s_{23}^{} c_{23}^{} \left(y_\mu^2 - y_\tau^2\right) s_\rho^{} - c_{12}^{} s_{13}^{} \left(y_e^2 - s_{23}^2 \, y_\mu^2 - c_{23}^2 \, y_\tau^2\right) s_{\rho + \delta}\right] s_{\rho + \delta}^{} \nonumber\\
	&& - C_\kappa^{} \, \zeta^{-1}_{23} s_{12}^{} \, c_{13}^{} \left[ c_{12}^{} s_{23}^{} c_{23}^{} \left(y_\mu^2 - y_\tau^2\right) c_\sigma^{} + s_{12}^{} s_{13}^{} \left(y_e^2 - s_{23}^2 \, y_\mu^2 - c_{23}^2 \, y_\tau^2\right) c_{\sigma + \delta}\right] c_{\sigma + \delta}^{}
	  \nonumber\\\
	&& - C_\kappa^{} \, \zeta^{}_{23} \; s_{12}^{} \, c_{13}^{} \left[ c_{12}^{} s_{23}^{} c_{23}^{} \left(y_\mu^2 - y_\tau^2\right) s_\sigma^{} + s_{12}^{} s_{13}^{} \left(y_e^2 - s_{23}^2 \, y_\mu^2 - c_{23}^2 \, y_\tau^2\right) s_{\sigma + \delta}\right] s_{\sigma + \delta}^{} \;,
    \label{eq:theta13}\\
    \dot{\theta}_{23}^{} &=& 2 \, \frac{m^2_{}}{v^2_{}} \left[ -y_{e \mu}^{} s_{13}^{} c_{13}^{-1} c_{23}^{} \left( \left| \eta_{e\mu}^\prime \right| c_{e\mu + \delta}^\prime - 4 \left| \eta_{e\mu}^{} \right| c_{e\mu + \delta}^{} \right)  + y_{\mu \tau}^{} \left( \left| \eta_{\mu\tau}^\prime \right| c_{\mu\tau}^\prime - 4 \left| \eta_{\mu\tau}^{} \right| c_{\mu\tau}^{} \right) \right.\nonumber\\
    && \left.\qquad\;\;\, + y_{e \tau}^{} s_{13}^{} c_{13}^{-1} s_{23}^{} \left( \left| \eta_{e\tau}^\prime \right| c_{e\tau + \delta}^\prime - 4 \left| \eta_{e\tau}^{} \right| c_{e\tau + \delta}^{} \right) \right] \nonumber\\
	&& - C_\kappa^{} \, \zeta^{-1}_{13} s_{12}^{} \left[ s_{12}^{} s_{23}^{} c_{23}^{} \left(y_\mu^2 - y_\tau^2\right) c_\rho^{} - c_{12}^{} s_{13}^{} \left(y_e^2 - s_{23}^2 \, y_\mu^2 - c_{23}^2 \, y_\tau^2\right) c_{\rho + \delta}\right] c_\rho^{}
	  \nonumber\\
	&& - C_\kappa^{} \, \zeta^{}_{13} \; s_{12}^{} \left[ s_{12}^{} s_{23}^{} c_{23}^{} \left(y_\mu^2 - y_\tau^2\right) s_\rho^{} - c_{12}^{} s_{13}^{} \left(y_e^2 - s_{23}^2 \, y_\mu^2 - c_{23}^2 \, y_\tau^2\right) s_{\rho + \delta}\right] s_\rho^{} \nonumber\\
	&& - C_\kappa^{} \, \zeta^{-1}_{23} c_{12}^{} \left[ c_{12}^{} s_{23}^{} c_{23}^{} \left(y_\mu^2 - y_\tau^2\right) c_\sigma^{} + s_{12}^{} s_{13}^{} \left(y_e^2 - s_{23}^2 \, y_\mu^2 - c_{23}^2 \, y_\tau^2\right) c_{\sigma + \delta}\right] c_\sigma^{}
	 \nonumber\\
	&& - C_\kappa^{} \, \zeta^{}_{23} \; c_{12}^{} \left[ c_{12}^{} s_{23}^{} c_{23}^{} \left(y_\mu^2 - y_\tau^2\right) s_\sigma^{} + s_{12}^{} s_{13}^{} \left(y_e^2 - s_{23}^2 \, y_\mu^2 - c_{23}^2 \, y_\tau^2\right) s_{\sigma + \delta}\right] s_\sigma^{} \;, \label{eq:theta23}
\end{eqnarray}
where $\zeta_{ij}^{} \equiv (\kappa_i^{} - \kappa_j^{})/(\kappa_i^{} + \kappa_j^{})$ for $ij = 12,13,23$. In addition, we have written $\{\sin \theta_{ij}^{}, \cos \theta_{ij}^{}\}$ as $\{s_{ij}^{}, c_{ij}^{}\}$ for $ij = 12,13,23$, and likewise for the CP-violating phases $\{\sin \varphi, \cos \varphi,\sin (\varphi + \varphi^\prime), \cos (\varphi + \varphi^\prime)\}$ as $\{s_\varphi, c_\varphi, s_{\varphi + \varphi^\prime}^{}, c_{\varphi + \varphi^\prime}^{} \}$ for $\varphi,\varphi^\prime = \rho,\sigma,\delta$, $\{ \sin \phi_{\alpha\beta}^{}, \cos \phi_{\alpha\beta}^{}, \sin \phi_{\alpha\beta}^\prime, \cos \phi_{\alpha\beta}^\prime, \sin (\phi_{\alpha\beta}^{} + \delta), \cos (\phi_{\alpha\beta}^{} + \delta), \sin (\phi_{\alpha\beta}^\prime + \delta), \cos (\phi_{\alpha\beta}^\prime + \delta)\}$ as $\{ s_{\alpha\beta}^{}, c_{\alpha\beta}^{}, s_{\alpha\beta}^\prime, c_{\alpha\beta}^\prime, s_{\alpha\beta + \delta}^{}, c_{\alpha\beta + \delta}^{}, s_{\alpha\beta + \delta}^\prime, c_{\alpha\beta + \delta}^\prime\}$ for $\alpha\beta = e\mu,e\tau,\mu\tau$.

\subsubsection*{Majorana and Dirac CP-violating phases}
\begin{eqnarray}
    \dot{\rho} &=& - 2 \, \frac{m^2_{}}{v^2_{}}  y_{e\mu}^{} \frac{c_{23}^{}}{c_{13}^{}} \left[ \frac{c_{12}^{}}{s_{12}^{}} \left( \left| \eta_{e\mu}^\prime \right| s_{e\mu}^\prime - 4 \left| \eta_{e\mu}^{} \right| s_{e\mu}^{} \right) + \frac{s_{13}^{} \, c_{23}^{} }{s_{23}^{}} \left( \left| \eta_{e\mu}^\prime \right| s_{e\mu + \delta}^\prime - 4 \left| \eta_{e\mu}^{} \right| s_{e\mu + \delta}^{} \right) \right] \nonumber\\
    && + 2 \, \frac{m^2_{}}{v^2_{}} y_{e\tau}^{} \frac{s_{23}^{}}{c_{13}^{}} \left[ \frac{c_{12}^{}}{s_{12}^{}} \left( \left| \eta_{e\tau}^\prime \right| s_{e\tau}^\prime - 4 \left| \eta_{e\tau}^{} \right| s_{e\tau}^{} \right) - \frac{s_{13}^{} \, s_{23}^{}}{c_{23}^{}} \left( \left| \eta_{e\tau}^\prime \right| s_{e\tau + \delta}^\prime - 4 \left| \eta_{e\tau}^{} \right| s_{e\tau + \delta}^{} \right) \right]\nonumber\\
    && - 2 \, \frac{m^2_{}}{v^2_{}} y_{\mu\tau}^{} s_{23}^{-1} c_{23}^{-1} \left( \left| \eta_{\mu\tau}^\prime \right| s_{\mu\tau}^\prime - 4 \left| \eta_{\mu\tau}^{} \right| s_{\mu\tau}^{} \right) \nonumber\\
    && + C_\kappa^{} \, \zeta^{-1}_{12}  c_{12}^2 \left\{ \left[ c_{13}^2 \, y_e^2 - \left( c_{23}^2 - s_{13}^2 \, s_{23}^2 \right) y_\mu^2 + \left( s_{13}^2 \, c_{23}^2 - s_{23}^2 \right) y_\tau^2 \right] c_{\rho-\sigma}^{} \right.\nonumber\\
    && \left.\qquad\;\; - s_{12}^{-1} c _{12}^{-1} s_{13}^{} s_{23}^{} c_{23}^{} \left(y_\mu^2 - y_\tau^2\right) \left( c_{12}^2 \, c_{\delta+\rho-\sigma}^{} - s_{12}^2 \, c_{\delta-\rho+\sigma}^{} \right)\right\} s_{\rho-\sigma}^{} \nonumber\\
    && - C_\kappa^{} \, \zeta^{}_{12} \, c_{12}^2 \left\{ \left[ c_{13}^2 \, y_e^2 - \left( c_{23}^2 - s_{13}^2 \, s_{23}^2 \right) y_\mu^2 + \left( s_{13}^2 \, c_{23}^2 - s_{23}^2 \right) y_\tau^2 \right] s_{\rho-\sigma}^{} \right.\nonumber\\
    && \left. \qquad\;\; - s_{12}^{-1} c_{12}^{-1} s_{13}^{} s_{23}^{} c_{23}^{} \left(y_\mu^2 - y_\tau^2\right) \left( c_{12}^2 \, s_{\delta+\rho-\sigma}^{} + s_{12}^2 \, s_{\delta-\rho+\sigma}^{} \right)\right\} c_{\rho-\sigma}^{} \nonumber\\
    && + C_\kappa^{} \, \zeta^{-1}_{13} \left[ s_{12}^{} \left(y_\mu^2 - y_\tau^2\right) c_\rho^{} - c_{12}^{} s_{13}^{} s_{23}^{-1} c_{23}^{-1} \left(y_e^2 - s_{23}^2 \, y_\mu^2 - c_{23}^2 \, y_\tau^2\right)  c_{\rho+\delta}^{} \right] \nonumber\\
    && \qquad \;\;\; \times \left[ s_{12}^{} \left(c_{23}^2 - s_{23}^2\right) s_\rho + 2 \, c_{12}^{} s_{13}^{} s_{23}^{} c_{23}^{} s_{\rho+\delta}^{} \right] 
    \nonumber\\
    && - C_\kappa^{} \, \zeta^{}_{13} \; \left[ s_{12}^{} \left(y_\mu^2 - y_\tau^2\right) s_\rho^{} - c_{12}^{} s_{13}^{} s_{23}^{-1} c_{23}^{-1} \left(y_e^2 - s_{23}^2 \, y_\mu^2 - c_{23}^2 \, y_\tau^2\right)  s_{\rho+\delta}^{} \right] \nonumber\\
    && \qquad \;\;\; \times \left[ s_{12}^{} \left(c_{23}^2 - s_{23}^2\right) c_\rho + 2 \, c_{12}^{} s_{13}^{} s_{23}^{} c_{23}^{} c_{\rho+\delta}^{} \right] \nonumber\\
    && + C_\kappa^{} \, \zeta^{-1}_{23} \left[ s_{12}^{-1} c_{12}^{} \left(y_\mu^2 - y_\tau^2\right) c_\sigma^{} +  s_{13}^{} s_{23}^{-1} c_{23}^{-1} \left(y_e^2 - s_{23}^2 \, y_\mu^2 - c_{23}^2 \, y_\tau^2\right)  c_{\sigma+\delta}^{} \right] \nonumber\\
    && \qquad \;\;\; \times \left[ s_{12}^{} c_{12}^{} \left(c_{23}^2 - s_{23}^2\right) s_\sigma + \left(c_{12}^2 - s_{12}^2\right) s_{13}^{} s_{23}^{} c_{23}^{} s_{\delta+\sigma}^{} \right]\nonumber\\
    && - C_\kappa^{} \, \zeta^{}_{23} \; \left[ s_{12}^{-1} c_{12}^{} \left(y_\mu^2 - y_\tau^2\right) s_\sigma^{} + s_{13}^{} s_{23}^{-1} c_{23}^{-1}  \left(y_e^2 - s_{23}^2 \, y_\mu^2 - c_{23}^2 \, y_\tau^2\right)  s_{\sigma+\delta}^{} \right] \nonumber\\
    && \qquad \;\;\; \times \left[ s_{12}^{} c_{12}^{} \left(c_{23}^2 - s_{23}^2\right) c_\sigma + \left(c_{12}^2 - s_{12}^2\right) s_{13}^{}s_{23}^{} c_{23}^{} c_{\delta+\sigma}^{} \right] \;, \label{eq:rho}\\
    \dot{\sigma} &=& + 2 \, \frac{m^2_{}}{v^2_{}} y_{e\mu}^{} \frac{c_{23}^{}}{c_{13}^{}} \left[ \frac{s_{12}^{}}{c_{12}^{}} \left( \left| \eta_{e\mu}^\prime \right| s_{e\mu}^\prime - 4 \left| \eta_{e\mu}^{} \right| s_{e\mu}^{} \right) - \frac{s_{13}^{} \, c_{23}^{}}{s_{23}^{}} \left( \left| \eta_{e\mu}^\prime \right| s_{e\mu+\delta}^\prime - 4 \left| \eta_{e\mu}^{} \right| s_{e\mu+\delta}^{} \right) \right]\nonumber\\ 
    && - 2 \, \frac{m^2_{}}{v^2_{}} y_{e\tau}^{} \frac{s_{23}^{}}{c_{13}^{}} \left[ \frac{s_{12}^{}}{c_{12}^{}} \left( \left| \eta_{e\tau}^\prime \right| s_{e\tau}^\prime - 4 \left| \eta_{e\tau}^{} \right| s_{e\tau}^{} \right) + \frac{ s_{13}^{} \, s_{23}^{}}{c_{23}^{}} \left( \left| \eta_{e\tau}^\prime \right| s_{e\tau + \delta}^\prime - 4 \left| \eta_{e\tau}^{} \right| s_{e\tau + \delta}^{} \right) \right] \nonumber\\
    && - 2 \, \frac{m^2_{}}{v^2_{}} y_{\mu\tau}^{} s_{23}^{-1} c_{23}^{-1} \left( \left| \eta_{\mu\tau}^\prime \right| s_{\mu\tau}^\prime - 4 \left| \eta_{\mu\tau}^{} \right| s_{\mu\tau}^{} \right) \nonumber\\
    && + C_\kappa^{} \, \zeta^{-1}_{12}  s_{12}^2 \left\{ \left[ c_{13}^2 \, y_e^2 - \left( c_{23}^2 - s_{13}^2 \, s_{23}^2 \right) y_\mu^2 + \left( s_{13}^2 \, c_{23}^2 - s_{23}^2 \right) y_\tau^2 \right] c_{\rho-\sigma}^{} \right.\nonumber\\
    && \left. \qquad\;\; - s_{12}^{-1} c_{12}^{-1} s_{13}^{} s_{23}^{} c_{23}^{} \left(y_\mu^2 - y_\tau^2\right) \left( c_{12}^2 \, c_{\delta+\rho-\sigma}^{} - s_{12}^2 \, c_{\delta-\rho+\sigma}^{} \right)\right\} s_{\rho-\sigma}^{} \nonumber\\
    && - C_\kappa^{} \, \zeta^{}_{12} \, s_{12}^2 \left\{ \left[ c_{13}^2 \, y_e^2 - \left( c_{23}^2 - s_{13}^2 \, s_{23}^2 \right) y_\mu^2 + \left( s_{13}^2 \, c_{23}^2 - s_{23}^2 \right) y_\tau^2 \right] s_{\rho-\sigma}^{} \right.\nonumber\\
    && \left.\qquad\;\; - s_{12}^{-1} c_{12}^{-1} s_{13}^{} s_{23}^{} c_{23}^{} \left(y_\mu^2 - y_\tau^2\right) \left( c_{12}^2 \, s_{\delta+\rho-\sigma}^{} + s_{12}^2 \, s_{\delta-\rho+\sigma}^{} \right)\right\} c_{\rho-\sigma}^{} \nonumber\\
    && + C_\kappa^{} \, \zeta^{-1}_{13} \left[ s_{12}^{} c_{12}^{-1}\left(y_\mu^2 - y_\tau^2\right) c_\rho^{} - s_{13}^{} s_{23}^{-1} c_{23}^{-1} \left(y_e^2 - s_{23}^2 \, y_\mu^2 - c_{23}^2 \, y_\tau^2\right)  c_{\rho+\delta}^{} \right] \nonumber\\
    && \qquad \;\;\; \times \left[ s_{12}^{} c_{12}^{} \left(c_{23}^2 - s_{23}^2\right) s_\rho + \left(c_{12}^2 - s_{12}^2\right) s_{13}^{} s_{23}^{} c_{23}^{} s_{\rho+\delta}^{} \right] 
    \nonumber\\
    && - C_\kappa^{} \, \zeta^{}_{13} \; \left[ s_{12}^{} c_{12}^{-1} \left(y_\mu^2 - y_\tau^2\right) s_\rho^{} - s_{13}^{} s_{23}^{-1} c_{23}^{-1} \left(y_e^2 - s_{23}^2 \, y_\mu^2 - c_{23}^2 \, y_\tau^2\right)  s_{\rho+\delta}^{} \right] \nonumber\\
    && \qquad \;\;\; \times \left[ s_{12}^{} c_{12}^{} \left(c_{23}^2 - s_{23}^2\right) c_\rho + \left(c_{12}^2 - s_{12}^2\right) s_{13}^{} s_{23}^{} c_{23}^{} c_{\rho+\delta}^{} \right]\nonumber\\
    && + C_\kappa^{} \, \zeta^{-1}_{23} \left[ c_{12}^{} \left(y_\mu^2 - y_\tau^2\right) c_\sigma^{} + s_{12}^{} s_{13}^{}s_{23}^{-1} c_{23}^{-1} \left(y_e^2 - s_{23}^2 \, y_\mu^2 - c_{23}^2 \, y_\tau^2\right)  c_{\sigma+\delta}^{} \right] \nonumber\\
    && \qquad \;\;\; \times \left[ c_{12}^{} \left(c_{23}^2 - s_{23}^2\right) s_\sigma - 2 s_{12}^{} s_{13}^{} s_{23}^{} c_{23}^{} s_{\delta+\sigma}^{} \right]\nonumber\\
    && - C_\kappa^{} \, \zeta^{}_{23} \; \left[ c_{12}^{} \left(y_\mu^2 - y_\tau^2\right) s_\sigma^{} + s_{12}^{} s_{13}^{} s_{23}^{-1} c_{23}^{-1} \left(y_e^2 - s_{23}^2 \, y_\mu^2 - c_{23}^2 \, y_\tau^2\right) s_{\sigma+\delta}^{} \right] \nonumber\\
    && \qquad \;\;\; \times \left[ c_{12}^{} \left(c_{23}^2 - s_{23}^2\right) c_\sigma - 2 s_{12}^{} s_{13}^{} s_{23}^{} c_{23}^{} c_{\delta+\sigma}^{} \right]\;, \label{eq:sigma}\\
    \dot{\delta} &=& + 2 \, \frac{m^2_{}}{v^2_{}} y_{e\mu}^{} \left[ \frac{c_{23}^{} \cos 2 \theta_{12}^{}}{s_{12}^{} \, c_{12}^{} \, c_{13}^{}} \left( \left| \eta_{e\mu}^\prime \right| s_{e\mu}^\prime - 4 \left| \eta_{e\mu}^{} \right| s_{e\mu}^{} \right) + \frac{s_{13}^2 c_{23}^2 - s_{23}^2}{ s_{13}^{} \, c_{13}^{} \,s_{23}^{}}\left( \left| \eta_{e\mu}^\prime \right| s_{e\mu + \delta}^\prime - 4 \left| \eta_{e\mu}^{} \right| s_{e\mu + \delta}^{} \right) \right] \nonumber\\
    && - 2 \, \frac{m^2_{}}{v^2_{}} y_{e\tau}^{} \left[ \frac{s_{23}^{} \cos 2 \theta_{12}^{} }{s_{12}^{} \, c_{12}^{} \, c_{13}^{}} \left( \left| \eta_{e\tau}^\prime \right| s_{e\tau}^\prime - 4 \left| \eta_{e\tau}^{} \right| s_{e\tau}^{} \right) - \frac{ s_{13}^2 s_{23}^2 - c_{23}^2}{ s_{13}^{} \, c_{13}^{} \, c_{23}^{}} \left( \left| \eta_{e\tau}^\prime \right| s_{e\tau + \delta}^\prime - 4 \left| \eta_{e\tau}^{} \right| s_{e\tau + \delta}^{} \right) \right] \nonumber\\
    && + 2\, \frac{m^2_{}}{v^2_{}} y_{\mu\tau}^{} s_{23}^{-1} c_{23}^{-1} \left( \left| \eta_{\mu\tau}^\prime \right| s_{\mu\tau}^\prime - 4 \left| \eta_{\mu\tau}^{} \right| s_{\mu\tau}^{} \right) \nonumber\\
    && - C_\kappa^{} \zeta_{12}^{-1} \left\{ \left[c_{13}^2 y_e^2 - \left( c_{23}^2 - s_{13}^2 \, s_{23}^2 \right) y_\mu^2 + \left( s_{13}^2 \, c_{23}^2 - s_{23}^2 \right) y_\tau^2 \right] c_{\rho-\sigma}^{}\right.\nonumber\\
    &&\left. \qquad\quad - s_{12}^{-1} c_{12}^{-1} s_{13}^{} s_{23}^{} c_{23}^{} \left(y_\mu^2 - y_\tau^2\right) \left( c_{12}^2 \, c_{\delta+\rho-\sigma}^{} - s_{12}^2 \, c_{\delta-\rho+\sigma}^{} \right) \right\} s_{\rho-\sigma}^{} \nonumber\\
    && + C_\kappa^{} \zeta_{12}^{} \; \left\{ \left[c_{13}^2 y_e^2 - \left( c_{23}^2 - s_{13}^2 \, s_{23}^2 \right) y_\mu^2 + \left( s_{13}^2 \, c_{23}^2 - s_{23}^2 \right)  y_\tau^2 \right] s_{\rho-\sigma}^{}\right.\nonumber\\
    &&\left.\qquad\quad - s_{12}^{-1} c_{12}^{-1} s_{13}^{} s_{23}^{} c_{23}^{} \left(y_\mu^2 - y_\tau^2\right) \left( c_{12}^2 \, s_{\delta+\rho-\sigma}^{} + s_{12}^2 \, s_{\delta-\rho+\sigma}^{} \right) \right\} c_{\rho-\sigma}^{} \nonumber\\
    && - C_\kappa^{} \, \zeta^{-1}_{13} \left[ s_{12}^{} s_{23}^{} c_{23}^{} \left(y_\mu^2 - y_\tau^2\right) c_\rho^{} - c_{12}^{} s_{13}^{} \left(y_e^2 - s_{23}^2 \, y_\mu^2 - c_{23}^2 \, y_\tau^2\right)  c_{\rho+\delta}^{} \right] \nonumber\\
    && \qquad \;\;\; \times \left[ s_{12}^{} s_{23}^{-1} c_{23}^{-1} \left(c_{23}^2 - s_{23}^2\right) s_\rho + c_{12}^{-1}  s_{13}^{-1} \left(c_{12}^2 - s_{12}^2 s_{13}^2 \right) s_{\rho+\delta}^{} \right] \nonumber\\
    && + C_\kappa^{} \, \zeta^{}_{13} \; \left[ s_{12}^{} s_{23}^{} c_{23}^{} \left(y_\mu^2 - y_\tau^2\right) s_\rho^{} - c_{12}^{} s_{13}^{} \left(y_e^2 - s_{23}^2 \, y_\mu^2 - c_{23}^2 \, y_\tau^2\right)  s_{\rho+\delta}^{} \right] \nonumber\\
    && \qquad \;\;\; \times \left[ s_{12}^{} s_{23}^{-1} c_{23}^{-1} \left(c_{23}^2 - s_{23}^2\right) c_\rho + c_{12}^{-1}  s_{13}^{-1} \left(c_{12}^2 - s_{12}^2 s_{13}^2 \right) c_{\rho+\delta}^{} \right]  \nonumber\\
    && - C_\kappa^{} \, \zeta^{-1}_{23} \left[ c_{12}^{} s_{23}^{} c_{23}^{} \left(y_\mu^2 - y_\tau^2\right) c_\sigma^{} + s_{12}^{} s_{13}^{} \left(y_e^2 - s_{23}^2 \, y_\mu^2 - c_{23}^2 \, y_\tau^2\right)  c_{\sigma+\delta}^{} \right] \nonumber\\
    && \qquad \;\;\; \times \left[ c_{12}^{} s_{23}^{-1} c_{23}^{-1} \left(c_{23}^2 - s_{23}^2\right) s_\sigma - s_{12}^{-1}  s_{13}^{-1} \left(s_{12}^2 - c_{12}^2 s_{13}^2 \right) s_{\sigma+\delta}^{} \right] \nonumber\\
    && + C_\kappa^{} \, \zeta^{}_{23} \; \left[ c_{12}^{} s_{23}^{} c_{23}^{} \left(y_\mu^2 - y_\tau^2\right) s_\sigma^{} + s_{12}^{} s_{13}^{} \left(y_e^2 - s_{23}^2 \, y_\mu^2 - c_{23}^2 \, y_\tau^2\right)  s_{\sigma+\delta}^{} \right] \nonumber\\
    && \qquad \;\;\; \times \left[ c_{12}^{} s_{23}^{-1} c_{23}^{-1} \left(c_{23}^2 - s_{23}^2\right) c_\sigma - s_{12}^{-1}  s_{13}^{-1} \left(s_{12}^2 - c_{12}^2 s_{13}^2 \right) c_{\sigma+\delta}^{} \right]\;.
    \label{eq:delta}
    \end{eqnarray}

\subsubsection*{Moduli of $\eta$ and $\eta^\prime$ elements }
\begin{eqnarray}
	\dot{\eta}_{\alpha\alpha}^{} &=& \sum_{\gamma \neq \alpha}^{} 4 \, \frac{m^2_{}}{v^2_{}} y_{\alpha\gamma}^{} \left[ \left|\eta_{\alpha \gamma}^\prime \right| \cos \left(\phi_{\alpha \gamma}^\prime - \phi_{\alpha \gamma}^{} \right) - 4 \left|\eta_{\alpha \gamma}^{} \right| \right] \left|\eta_{\alpha \gamma}^{} \right| + \frac{2}{3} \, g_2^2 \, {\rm tr} \left(\eta\right) \nonumber\\
	&&- \sum_i^{} \kappa_i^2 v^2_{} \left|U_{\alpha i}^{}\right|^2_{} + y_\alpha^2 \left(5 \eta_{\alpha\alpha}^{} - 3 \eta_{\alpha\alpha}^\prime\right) + \left(-\frac{17}{3} g_2^2 + 2 T \right) \eta_{\alpha\alpha}^{} \;, \label{eq:diaeta}\\
	\dot{\eta}_{\alpha\alpha}^\prime &=& \sum_{\gamma\neq \alpha}^{} 4 \, \frac{m^2_{}}{v^2_{}} y_{\alpha\gamma}^{} \left[ \left|\eta_{\alpha \gamma}^\prime \right| - 4 \left|\eta_{\alpha \gamma}^{} \right| \cos\left(\phi_{\alpha \gamma}^{} - \phi_{\alpha \gamma}^\prime \right) \right] \left|\eta_{\alpha \gamma}^\prime \right| + \frac{2}{3} g_1^2 {\rm tr} \left(\eta^\prime_{}\right) + \frac{2}{3} \left( g_2^2- g_1^2 \right) {\rm tr} \left(\eta\right) \nonumber\\
	&&- \frac{5}{2} \sum_i^{} \kappa_i^2 v^2_{} \left|U_{\alpha i}^{}\right|^2_{} + y_\alpha^2 \left( \eta_{\alpha\alpha}^\prime - 8 \eta_{\alpha\alpha}^{} \right) + \left(\frac{1}{3} g_1^2 + 2 T\right) \eta_{\alpha\alpha}^\prime - \frac{1}{3} \left( 17 g_2^2 + g_1^2 \right) \eta_{\alpha\alpha}^{}  \;,
    \label{eq:diaetap}
    \\
	\left| \dot{\eta}^{}_{\alpha\beta} \right| &=& +\sum_{\gamma\neq \alpha}^{} 2 \, \frac{m^2_{}}{v^2_{}} y_{\alpha\gamma}^{} \left[ \left|\eta_{\alpha \gamma}^\prime \right| \cos\left(\phi_{\alpha\gamma}^\prime + \phi_{\gamma\beta}^{} - \phi_{\alpha\beta}^{} \right) - 4 \left|\eta_{\alpha\gamma}^{} \right| \cos\left(\phi_{\alpha\gamma}^{} + \phi_{\gamma\beta}^{} - \phi_{\alpha\beta}^{} \right) \right] \left|\eta_{\beta\gamma}^{} \right| \nonumber\\
	&&+ \sum_{\varrho \neq \beta}^{} 2\, \frac{m^2_{}}{v^2_{}} y_{\beta\varrho}^{} \left[ \left|\eta_{\varrho\beta}^\prime \right| \cos\left(\phi_{\varrho\beta}^\prime + \phi_{\alpha\varrho}^{} - \phi_{\alpha\beta}^{} \right) - 4 \left|\eta_{\varrho\beta}^{} \right| \cos\left(\phi_{\varrho\beta}^{} + \phi_{\alpha\varrho}^{} - \phi_{\alpha\beta}^{} \right) \right] \left|\eta_{\alpha\varrho}^{} \right| \nonumber\\
	&&- \sum_i \kappa_i^2 v^2_{} \left[ \cos \phi^{}_{\alpha\beta} \, {\rm Re} \left(U_{\alpha i}^{} U_{\beta i}^\ast \right) + \sin \phi{}_{\alpha\beta} \, {\rm Im} \left(U_{\alpha i}^{} U_{\beta i}^\ast \right) \right] +\left(-\frac{17}{3} g_2^2 + 2 T \right) \left| \eta^{}_{\alpha\beta} \right| \nonumber\\ 
	&&+ \frac{5}{2} \left( y_\alpha^2 +  y_\beta^2 \right) \left| \eta^{}_{\alpha\beta} \right|-\frac{3}{2} \left( y_\alpha^2 + y_\beta^2\right) \left| \eta^\prime_{\alpha\beta} \right| \cos \left( \phi^{}_{\alpha\beta} - \phi^\prime_{\alpha\beta} \right)  \;, \label{eq:nondiaeta}\\
	\left| \dot{\eta}^\prime_{\alpha\beta} \right| &=& +\sum_{\gamma\neq \alpha}^{} 2 \, \frac{m^2_{}}{v^2_{}} y_{\alpha\gamma}^{} \left[ \left|\eta_{\alpha \gamma}^\prime \right| \cos\left(\phi_{\alpha\gamma}^\prime + \phi_{\gamma\beta}^\prime - \phi_{\alpha\beta}^\prime \right) - 4 \left|\eta_{\alpha\gamma}^{} \right| \cos\left(\phi_{\alpha\gamma}^{} + \phi_{\gamma\beta}^\prime - \phi_{\alpha\beta}^\prime \right) \right] \left|\eta_{\beta\gamma}^\prime \right| \nonumber\\
	&& + \sum_{\varrho\neq\beta}^{} 2 \, \frac{m^2_{}}{v^2_{}} y_{\beta\varrho}^{} \left[ \left|\eta_{\varrho\beta}^\prime \right| \cos\left(\phi_{\varrho\beta}^\prime + \phi_{\alpha\varrho}^\prime - \phi_{\alpha\beta}^\prime \right) - 4 \left|\eta_{\varrho\beta}^{} \right| \cos\left(\phi_{\varrho\beta}^{} + \phi_{\alpha\varrho}^\prime - \phi_{\alpha\beta}^\prime \right) \right] \left|\eta_{\alpha\varrho}^\prime \right| \nonumber\\
	&&- \frac{5}{2} \sum_i \kappa_i^2 v^2_{} \left[ \cos \phi^\prime_{\alpha\beta} {\rm Re} \left(U_{\alpha i}^{} U_{\beta i}^\ast \right) + \sin \phi^\prime_{\alpha\beta} {\rm Im} \left(U_{\alpha i}^{} U_{\beta i}^\ast \right) \right] +\left(\frac{1}{3} g_1^2 + 2 T + \frac{1}{2} y_\alpha^2 + \frac{1}{2} y_\beta^2\right) \left|\eta_{\alpha\beta}^\prime \right|\nonumber\\ 
	&&-  \left( \frac{17}{3} g_2^2 + \frac{1}{3} g_1^2 + 4 y_\alpha^2 + 4 y_\beta^2 \right) \left| \eta^{}_{\alpha\beta} \right| \cos \left( \phi^\prime_{\alpha\beta} - \phi^{}_{\alpha\beta} \right)  \;,
 \label{eq:nondiaetap}
\end{eqnarray}
where we do not separately write out the result for each element for brevity but one can simply substitute the expressions of elements of $U$ to get the results for specific elements of $\eta$ and $\eta^\prime$.

\subsubsection*{Phases of $\eta$ and $\eta^\prime$ elements}
\begin{eqnarray}
	\dot{\phi}_{e\mu}^{} &=& + v^2_{} c_{13}^{} \left|\eta_{e\mu}^{}\right|_{}^{-1} \left[ - \kappa_1^2 \, c_{12}^{}  \left( s_{12}^{} c_{23}^{} s_{e\mu}^{} + c_{12}^{} s_{13}^{} s_{23}^{} s_{e\mu+ \delta}^{} \right)  + \kappa_3^2 \, s_{13}^{} s_{23}^{} s_{e\mu+ \delta}^{} \right] \nonumber\\
	&&+ v^2_{} c_{13}^{} \left|\eta_{e\mu}^{}\right|_{}^{-1} \left[ 
	+ \kappa_2^2 s_{12}^{} \left( c_{12}^{} c_{23}^{} s_{e\mu}^{} - s_{12}^{} s_{13}^{} s_{23}^{} s_{e\mu + \delta}^{} \right) \right] \nonumber\\
	&&+ \frac{3}{2} \left(y_e^2 + y_\mu^2\right) \left|\eta_{e\mu}^\prime\right| \left|\eta_{e\mu}^{}\right|_{}^{-1} \sin \left(\phi_{e\mu}^{} - \phi_{e\mu}^\prime \right) \nonumber\\
    &&+ 2 \, \frac{m^2_{}}{v^2_{}} \left|\eta_{e\mu}^{}\right|_{}^{-1}  \left\{ y_{e\mu}^{} \left(\eta_{\mu\mu}^{} - \eta_{ee}^{} \right) \left|\eta_{e\mu}^\prime\right| \sin \left( \phi_{e\mu}^\prime - \phi_{e\mu}^
	{} \right) \right.\nonumber\\
	&&+ \, y_{e\tau}^{} \left|\eta_{\mu\tau}^{} \right| \left[ \left|\eta_{e\tau}^\prime\right| \sin \left( \phi_{e\tau}^\prime - \phi_{\mu\tau}^{} - \phi_{e\mu}^{} \right) - 4 \left|\eta_{e\tau}^{}\right| \sin \left( \phi_{e\tau}^{} - \phi_{\mu\tau}^{} - \phi_{e\mu}^{} \right) \right] \nonumber\\
	&&+\left. y_{\mu\tau}^{} \left|\eta_{e\tau}^{} \right| \left[ \left|\eta_{\mu\tau}^\prime\right| \sin \left( \phi_{e\tau}^{} - \phi_{e\mu}^{} - \phi_{\mu\tau}^\prime \right) - 4 \left|\eta_{\mu\tau}^{}\right| \sin \left( \phi_{e\tau}^{} - \phi_{e\mu}^{} - \phi_{\mu\tau}^{} \right) \right] \right\} \nonumber\\
	&&+ 2 \, \frac{m^2_{}}{v^2_{}} y_{e\mu}^{} \left[ - \frac{c_{23}^{} \cos 2 \theta_{12}^{}}{s_{12}^{} \, c_{12}^{} \, c_{13}^{}} \left( \left| \eta_{e\mu}^\prime \right| s_{e\mu}^\prime - 4 \left| \eta_{e\mu}^{} \right| s_{e\mu}^{} \right) + \frac{s_{13}^{} \sin 2 \theta_{23}^{}}{c_{13}^{} \, c_{23}^{}} \left( \left| \eta_{e\mu}^\prime \right| s_{e\mu + \delta}^\prime - 4 \left| \eta_{e\mu}^{} \right| s_{e\mu + \delta}^{} \right) \right] \nonumber\\
	&&+ 2 \, \frac{m^2_{}}{v^2_{}} y_{e\tau}^{} \left[ + \frac{s_{23}^{} \cos 2 \theta_{12}^{}}{s_{12}^{} \, c_{12}^{} \, c_{13}^{}} \left( \left| \eta_{e\tau}^\prime \right| s_{e\tau}^\prime - 4 \left| \eta_{e\tau}^{} \right| s_{e\tau}^{} \right) + \frac{s_{13}^{} \cos 2 \theta_{23}^{} }{ c_{13}^{} \, c_{23}^{}} \left( \left| \eta_{e\tau}^\prime \right| s_{e\tau + \delta}^\prime - 4 \left| \eta_{e\tau}^{} \right| s_{e\tau + \delta}^{} \right) \right] \nonumber\\
	&&- 2 \, \frac{m^2_{}}{v^2_{}} y_{\mu\tau}^{} s_{23}^{} \, c_{23}^{-1} \left( \left| \eta_{\mu\tau}^\prime \right| s_{\mu\tau}^\prime - 4 \left| \eta_{\mu\tau}^{} \right| s_{\mu\tau}^{} \right)  \nonumber\\
	&&+ C_\kappa^{} \zeta_{12}^{-1} \left\{ \left[c_{13}^2 \, y_e^2 - \left(c_{23}^2 - s_{13}^2 \, s_{23}^2 \right) y_\mu^2+ \left( s_{13}^2 \, c_{23}^2 - s_{23}^2 \right) y_\tau^2 \right] c_{\rho-\sigma}^{} \right.\nonumber\\
	&& \left.\qquad\quad\;\;\; - s_{12}^{-1} c_{12}^{-1} s_{13}^{} s_{23}^{} c_{23}^{} \left(y_\mu^2 - y_\tau^2\right) \left( c_{12}^2 \, c_{\delta+\rho-\sigma}^{} - s_{12}^2 \, c_{\delta-\rho+\sigma}^{} \right) \right\} s_{\rho-\sigma}^{} \nonumber\\
	&&- C_\kappa^{} \zeta_{12}^{} \; \left\{ \left[c_{13}^2 \, y_e^2 - \left(c_{23}^2 - s_{13}^2 \, s_{23}^2 \right) y_\mu^2+ \left( s_{13}^2  \, c_{23}^2 - s_{23}^2 \right) y_\tau^2 \right] s_{\rho-\sigma}^{} \right.\nonumber\\
	&& \left. \qquad\quad\;\;\; - s_{12}^{-1} c_{12}^{-1} s_{13}^{} s_{23}^{} c_{23}^{} \left(y_\mu^2 - y_\tau^2\right) \left( c_{12}^2 \, s_{\delta+\rho-\sigma}^{} + s_{12}^2 \, s_{\delta-\rho+\sigma}^{} \right) \right\} c_{\rho-\sigma}^{} \nonumber\\
	&&- C_\kappa^{} \zeta_{13}^{-1} s_{12}^{} \left[ s_{12}^{} c_{12}^{-1} s_{23}^{} \left(y_\mu^2 - y_\tau^2\right) c_\rho^{} - s_{13}^{} c_{23}^{-1} \left(y_e^2 - s_{23}^2 y_\mu^2 - c_{23}^2 y_\tau^2 \right) c_{\rho+\delta}^{} \right] \nonumber\\
	&& \qquad\qquad \times \left( c_{12}^{} s_{23}^{} s_{\rho}^{} + s_{12}^{} s_{13}^{} c_{23}^{} s_{\rho+\delta} \right) \nonumber\\
	&&+ C_\kappa^{} \zeta_{13}^{} \; s_{12}^{} \left[ s_{12}^{} c_{12}^{-1} s_{23}^{} \left(y_\mu^2 - y_\tau^2\right) s_\rho^{} - s_{13}^{} c_{23}^{-1} \left(y_e^2 - s_{23}^2 y_\mu^2 - c_{23}^2 y_\tau^2 \right) s_{\rho+\delta}^{} \right]\nonumber\\
	&& \qquad\qquad \times \left( c_{12}^{} s_{23}^{} c_{\rho}^{} + s_{12}^{} s_{13}^{} c_{23}^{} c_{\rho+\delta} \right)\nonumber\\
	&&- C_\kappa^{} \zeta_{23}^{-1} c_{12}^{} \left[  s_{12}^{-1} c_{12}^{} s_{23}^{} \left(y_\mu^2 - y_\tau^2\right) c_\sigma^{} + s_{13}^{} c_{23}^{-1} \left(y_e^2 - s_{23}^2 y_\mu^2 - c_{23}^2 y_\tau^2 \right) c_{\sigma+\delta}^{} \right]\nonumber\\
	&& \qquad\qquad \times \left( s_{12}^{} s_{23}^{} s_{\sigma}^{} - c_{12}^{} s_{13}^{} c_{23}^{} s_{\sigma+\delta} \right) \nonumber\\
	&&+ C_\kappa^{} \zeta_{23}^{} \; c_{12}^{} \left[  s_{12}^{-1} c_{12}^{} s_{23}^{} \left(y_\mu^2 - y_\tau^2\right) s_\sigma^{} + s_{13}^{} c_{23}^{-1} \left(y_e^2 - s_{23}^2 y_\mu^2 - c_{23}^2 y_\tau^2 \right) s_{\sigma+\delta}^{} \right]\nonumber\\
	&& \qquad\qquad \times \left( s_{12}^{} s_{23}^{} c_{\sigma}^{} - c_{12}^{} s_{13}^{} c_{23}^{} c_{\sigma+\delta} \right)\;, \label{eq:phiemu}\\
	\dot{\phi}_{e\tau}^{} &=& + v^2_{} c_{13}^{} \left|\eta_{e\tau}^{}\right|_{}^{-1} \left[ + \kappa_1^2 c_{12}^{} \left( s_{12}^{} s_{23}^{} s_{e\tau}^{} - c_{12}^{} s_{13}^{} c_{23}^{} s_{e\tau + \delta}^{} \right) + \kappa_3^2 s_{13}^{} c_{23}^{} s_{e\tau + \delta}^{} \right] \nonumber\\
	&&+ v^2_{} c_{13}^{} \left|\eta_{e\tau}^{}\right|_{}^{-1} \left[ - \kappa_2^2  s_{12}^{} \left( c_{12}^{} s_{23}^{} s_{e\tau}^{} + s_{12}^{} s_{13}^{} c_{23}^{} s_{e\tau + \delta}^{} \right) \right] \nonumber\\
	&&+ \frac{3}{2} \left(y_e^2 + y_\tau^2 \right) \left|\eta_{e\tau}^\prime\right| \left|\eta_{e\tau}^{}\right|_{}^{-1} \sin \left(\phi_{e\tau}^{} - \phi_{e\tau}^\prime \right) + 2 \, \frac{m^2_{}}{v^2_{}} \left|\eta_{e\tau}^{}\right|_{}^{-1} \left\{ y_{e\tau}^{} \left(\eta_{\tau\tau}^{} - \eta_{ee}^{} \right) \left|\eta_{e\tau}^\prime\right| \sin \left( \phi_{e\tau}^\prime - \phi_{e\tau}^{} \right) \right. \nonumber\\
	&&+ \, y_{e\mu}^{} \left|\eta_{\mu\tau}^{} \right| \left[ \left|\eta_{e\mu}^\prime\right| \sin \left( \phi_{e\mu}^\prime + \phi_{\mu\tau}^{} - \phi_{e\tau}^{} \right) - 4 \left|\eta_{e\mu}^{}\right| \sin \left( \phi_{e\mu}^{} + \phi_{\mu\tau}^{} - \phi_{e\tau}^{} \right) \right] \nonumber\\
	&&- \left.y_{\mu\tau}^{} \left|\eta_{e\mu}^{} \right| \left[ \left|\eta_{\mu\tau}^\prime\right| \sin \left( \phi_{\mu\tau}^\prime + \phi_{e\mu}^
	{} - \phi_{e\tau}^{} \right) - 4 \left|\eta_{\mu\tau}^{}\right|\sin \left( \phi_{\mu\tau}^{} + \phi_{e\mu}^{} - \phi_{e\tau}^{} \right) \right] \right\}\nonumber\\
	&&- 2 \, \frac{m^2_{}}{v^2_{}} y_{e\mu}^{} \left[  \frac{c_{23}^{} \cos 2\theta_{12}^{} }{s_{12}^{} \, c_{12}^{} \, c_{13}^{}} \left( \left| \eta_{e\mu}^\prime \right| s_{e\mu}^\prime - 4 \left| \eta_{e\mu}^{} \right| s_{e\mu}^{} \right) + \frac{s_{13}^{} \cos 2 \theta_{23}^{}}{ c_{13}^{} \, s_{23}^{}} \left( \left| \eta_{e\mu}^\prime \right| s_{e\mu + \delta }^\prime  - 4 \left| \eta_{e\mu}^{} \right| s_{e\mu + \delta}^{} \right) \right] \nonumber\\
	&&+ 2 \, \frac{m^2_{}}{v^2_{}} y_{e\tau}^{} \left[  \frac{s_{23}^{} \cos 2\theta_{12}^{}}{s_{12}^{} \, c_{12}^{} \, c_{13}^{}} \left( \left| \eta_{e\tau}^\prime \right| s_{e\tau}^\prime - 4 \left| \eta_{e\tau}^{} \right| s_{e\tau}^{} \right) +  \frac{s_{13}^{} \sin 2\theta_{23}^{}}{c_{13}^{} \, s_{23}^{}} \left( \left| \eta_{e\tau}^\prime \right| s_{e\tau + \delta }^\prime - 4 \left| \eta_{e\tau}^{} \right| s_{e\tau + \delta}^{} \right) \right] \nonumber\\
	&&- 2 \, \frac{m^2_{}}{v^2_{}} y_{\mu\tau}^{} s_{23}^{-1} c_{23}^{} \left( \left| \eta_{\mu\tau}^\prime \right| s_{\mu\tau}^\prime - 4 \left| \eta_{\mu\tau}^{} \right| s_{\mu\tau}^{} \right)  \nonumber\\
	&&+ C_\kappa^{} \zeta_{12}^{-1} \left\{ \left[c_{13}^2 \, y_e^2 - \left(c_{23}^2 - s_{13}^2 \, s_{23}^2 \right) y_\mu^2+ \left(s_{13}^2 \, c_{23}^2 - s_{23}^2 \right) y_\tau^2 \right] c_{\rho-\sigma}^{} \right.\nonumber\\
	&& \left.\qquad\quad\;\;\; - s_{12}^{-1} c_{12}^{-1} s_{13}^{} s_{23}^{} c_{23}^{} \left(y_\mu^2 - y_\tau^2\right) \left( c_{12}^2 \, c_{\delta+\rho-\sigma}^{} - s_{12}^2 \, c_{\delta-\rho+\sigma}^{} \right) \right\} s_{\rho-\sigma}^{} \nonumber\\
	&&- C_\kappa^{} \zeta_{12}^{} \; \left\{ \left[c_{13}^2 \, y_e^2 - \left(c_{23}^2 - s_{13}^2 \, s_{23}^2 \right) y_\mu^2+ \left(s_{13}^2 \, c_{23}^2 - s_{23}^2 \right) y_\tau^2 \right] s_{\rho-\sigma}^{} \right.\nonumber\\
	&& \left.\qquad\quad\;\;\; -s_{12}^{-1} c_{12}^{-1} s_{13}^{} s_{23}^{} c_{23}^{} \left(y_\mu^2 - y_\tau^2\right) \left( c_{12}^2 \, s_{\delta+\rho-\sigma}^{} + s_{12}^2 \, s_{\delta-\rho+\sigma}^{} \right) \right\} c_{\rho-\sigma}^{} \nonumber\\
    &&+ C_\kappa^{} \zeta_{13}^{-1} s_{12}^{} \left[s_{12}^{} c_{12}^{-1} c_{23}^{} \left(y_\mu^2 - y_\tau^2 \right) c_\rho^{} - s_{13}^{} s_{23}^{-1} \left(y_e^2 - s_{23}^2 y_\mu^2 - c_{23}^2 y_\tau^2 \right) c_{\rho+\delta}^{} \right] \nonumber\\
    && \qquad \qquad \times \left(c_{12}^{} c_{23}^{} s_\rho^{} - s_{12}^{} s_{13}^{} s_{23}^{}  s_{\rho+\delta}^{} \right) \nonumber\\
    &&- C_\kappa^{} \zeta_{13}^{} \; s_{12}^{} \left[s_{12}^{} c_{12}^{-1} c_{23}^{} \left(y_\mu^2 - y_\tau^2 \right) s_\rho^{} - s_{13}^{} s_{23}^{-1} \left(y_e^2 - s_{23}^2 y_\mu^2 - c_{23}^2 y_\tau^2 \right) s_{\rho+\delta}^{} \right] \nonumber\\
    && \qquad \qquad \times \left(c_{12}^{} c_{23}^{} c_\rho^{} - s_{12}^{} s_{13}^{} s_{23}^{} c_{\rho+\delta}^{} \right)  \nonumber\\
    &&+ C_\kappa^{} \zeta_{23}^{-1} c_{12}^{} \left[s_{12}^{-1} c_{12}^{} c_{23}^{} \left(y_\mu^2 - y_\tau^2 \right) c_\sigma^{} + s_{13}^{} s_{23}^{-1} \left(y_e^2 - s_{23}^2 y_\mu^2 - c_{23}^2 y_\tau^2 \right) c_{\sigma+\delta}^{} \right] \nonumber\\
    && \qquad \qquad \times \left(s_{12}^{} c_{23}^{} s_\sigma^{} + c_{12}^{} s_{13}^{} s_{23}^{} s_{\sigma+\delta}^{} \right) \nonumber\\
    &&- C_\kappa^{} \zeta_{23}^{} \; c_{12}^{} \left[s_{12}^{-1} c_{12}^{} c_{23}^{} \left(y_\mu^2 - y_\tau^2 \right) s_\sigma^{} +s_{13}^{}  s_{23}^{-1} \left(y_e^2 - s_{23}^2 y_\mu^2 - c_{23}^2 y_\tau^2 \right) s_{\sigma+\delta}^{} \right] \nonumber\\
    && \qquad \qquad \times \left(s_{12}^{} c_{23}^{} c_\sigma^{} + c_{12}^{} s_{13}^{} s_{23}^{} c_{\sigma+\delta}^{} \right)\;, \label{eq:phietau}\\
	\dot{\phi}_{\mu\tau}^{} &=& + \kappa_1^2 v^2_{} \left|\eta_{\mu\tau}^{}\right|_{}^{-1} \left\{ \left[s_{23}^{} c_{23}^{} \left(c_{12}^2 s_{13}^2 - s_{12}^2 \right) + s_{12}^{} c_{12}^{} s_{13}^{} \left(c_{23}^2 - s_{23}^2 \right) c_\delta^{} \right] s_{\mu\tau}^{}  + s_{12}^{} c_{12}^{} s_{13}^{} s_\delta^{} c_{\mu\tau}^{} \right\} \nonumber\\
	&&+  \kappa_2^2 v^2_{} \left|\eta_{\mu\tau}^{}\right|_{}^{-1}  \left\{ \left[s_{23}^{} c_{23}^{} \left(s_{12}^2 s_{13}^2 - c_{12}^2 \right) - s_{12}^{} c_{12}^{} s_{13}^{} \left(c_{23}^2 - s_{23}^2 \right) c_\delta^{} \right] s_{\mu\tau}^{} - s_{12}^{} c_{12}^{} s_{13}^{} s_\delta^{} c_{\mu\tau}^{} \right\}\nonumber\\
	&&+  \kappa_3^2 \, v^2_{} \left|\eta_{\mu\tau}^{}\right|_{}^{-1} c_{13}^2 s_{23}^{} c_{23}^{} s_{\mu\tau}^{} + \frac{3}{2} \left(y_\mu^2 + y_\tau^2\right) \left|\eta_{\mu\tau}^\prime\right| \left|\eta_{\mu\tau}^{}\right|_{}^{-1} \sin \left(\phi_{\mu\tau}^{} - \phi_{\mu\tau}^\prime \right) \nonumber\\
	&& + 2 \, \frac{m^2_{}}{v^2_{}} \left|\eta_{\mu\tau}^{}\right|_{}^{-1} \left\{ y_{\mu\tau}^{} \left(\eta_{\tau\tau}^{} - \eta_{\mu\mu}^{} \right) \left|\eta_{\mu\tau}^\prime\right| \sin \left( \phi_{\mu\tau}^\prime - \phi_{\mu\tau}^{} \right) \right.\nonumber\\
	&&- \, y_{e\mu}^{} \left|\eta_{e\tau}^{} \right| \left[ \left|\eta_{e\mu}^\prime\right| \sin \left( \phi_{e\tau}^{} - \phi_{e\mu}^\prime - \phi_{\mu\tau}^{} \right) - 4 \left|\eta_{e\mu}^{}\right| \sin \left( \phi_{e\tau}^{} - \phi_{e\mu}^{} - \phi_{\mu\tau}^{} \right) \right] \nonumber\\
	&&- \left.y_{e\tau}^{} \left|\eta_{e\mu}^{} \right| \left[ \left|\eta_{e\tau}^\prime\right| \sin \left( \phi_{e\tau}^\prime - \phi_{e\mu}^{} - \phi_{\mu\tau}^{} \right) - 4 \left|\eta_{e\tau}^{}\right| \sin \left( \phi_{e\tau}^{} - \phi_{e\mu}^{} - \phi_{\mu\tau}^{} \right) \right] \right\}\nonumber\\
	&&- 2 \, \frac{m^2_{}}{v^2_{}} s_{13}^{} c_{13}^{-1}  \left[ y_{e\mu}^{} s_{23}^{-1} \left(\left| \eta_{e\mu}^\prime \right| s_{e\mu + \delta}^\prime - 4 \left| \eta_{e\mu}^{} \right| s_{e\mu + \delta}^{}\right) - y_{e\tau}^{} c_{23}^{-1} \left(\left| \eta_{e\tau}^\prime \right| s_{e\tau+ \delta}^\prime- 4 \left| \eta_{e\tau}^{} \right| s_{e\tau+ \delta}^{}\right) \right] \nonumber\\
	&&-  2 \, \frac{m^2_{}}{v^2_{}} y_{\mu\tau}^{} s_{23}^{-1} c_{23}^{-1} \left(c_{23}^2 - s_{23}^2\right) \left( \left| \eta_{\mu\tau}^\prime \right| s_{\mu\tau}^\prime - 4 \left| \eta_{\mu\tau}^{} \right| s_{\mu\tau}^{} \right) \nonumber\\
    &&+ C_\kappa^{} \zeta_{13}^{-1} s_{12}^{} \left[s_{12}^{} \left(y_\mu^2 - y_\tau^2 \right) c_\rho^{} - c_{12}^{} s_{13}^{} s_{23}^{-1} c_{23}^{-1} \left(y_e^2 - s_{23}^2 y_\mu^2 - c_{23}^2 y_\tau^2 \right) c_{\rho+\delta}^{} \right] s_\rho^{} \nonumber\\
    &&- C_\kappa^{} \zeta_{13}^{} \; s_{12}^{} \left[s_{12}^{} \left(y_\mu^2 - y_\tau^2 \right) s_\rho^{} - c_{12}^{} s_{13}^{} s_{23}^{-1} c_{23}^{-1} \left(y_e^2 - s_{23}^2 y_\mu^2 - c_{23}^2 y_\tau^2 \right) s_{\rho+\delta}^{} \right] c_\rho^{} \nonumber\\
    &&+ C_\kappa^{} \zeta_{23}^{-1} c_{12}^{} \left[c_{12}^{} \left(y_\mu^2 - y_\tau^2 \right) c_\sigma^{} + s_{12}^{} s_{13}^{} s_{23}^{-1} c_{23}^{-1} \left(y_e^2 - s_{23}^2 y_\mu^2 - c_{23}^2 y_\tau^2 \right) c_{\sigma+\delta}^{} \right] s_\sigma^{} \nonumber\\
    &&- C_\kappa^{} \zeta_{23}^{} \; c_{12}^{} \left[c_{12}^{} \left(y_\mu^2 - y_\tau^2 \right) s_\sigma^{} + s_{12}^{} s_{13}^{} s_{23}^{-1} c_{23}^{-1} \left(y_e^2 - s_{23}^2 y_\mu^2 - c_{23}^2 y_\tau^2 \right) s_{\sigma+\delta}^{} \right] c_\sigma^{}\;,
    \label{eq:phimutau}
\\
	\dot{\phi}_{e\mu}^\prime &=& \frac{5}{2} \frac{v^2_{} c_{13}^{} }{\left|\eta_{e\mu}^\prime\right|}  \left[ \kappa_2^2 \, s_{12}^{} \left( c_{12}^{} c_{23}^{} s_{e\mu}^\prime - s_{12}^{} s_{13}^{} s_{23}^{}  s_{e\mu+\delta}^\prime \right) -\kappa_1^2 \, c_{12}^{} \left( s_{12}^{} c_{23}^{} s_{e\mu}^\prime + c_{12}^{} s_{13}^{} s_{23}^{} s_{e\mu+\delta}^\prime \right) \right. \nonumber\\
	&&+ \left. \kappa_3^2 \, s_{13}^{} s_{23}^{} s_{e\mu+\delta}^\prime  \right] + \left( \frac{17}{3} g_2^2 + \frac{1}{3} g_1^2 + 4 y_e^2 + 4 y_\mu^2 \right)  \left|\eta_{e\mu}^{}\right| \left|\eta_{e\mu}^\prime\right|_{}^{-1} \sin \left(\phi_{e\mu}^\prime - \phi_{e\mu}^{} \right) \nonumber\\
	&&+ 2\, \frac{m^2_{}}{v^2_{}} \left|\eta_{e\mu}^\prime\right|_{}^{-1} \left\{ 4 \, y_{e\mu}^{} \left(\eta_{\mu\mu}^\prime - \eta_{ee}^\prime \right) \left|\eta_{e\mu}^{}\right| \sin \left( \phi_{e\mu}^\prime - \phi_{e\mu}^{} \right) \right.\nonumber\\
	&&+ \, y_{e\tau}^{} \left|\eta_{\mu\tau}^\prime \right| \left[ \left|\eta_{e\tau}^\prime \right| \sin \left( \phi_{e\tau}^\prime - \phi_{\mu\tau}^
	\prime - \phi_{e\mu}^\prime \right) - 4 \left|\eta_{e\tau}^{}\right| \sin \left( \phi_{e\tau}^{} - \phi_{\mu\tau}^\prime - \phi_{e\mu}^\prime \right) \right] \nonumber\\
	&&+\left. y_{\mu\tau}^{} \left|\eta_{e\tau}^\prime \right| \left[ \left|\eta_{\mu\tau}^\prime\right| \sin \left( \phi_{e\tau}^\prime - \phi_{\mu\tau}^\prime - \phi_{e\mu}^\prime \right) - 4 \left|\eta_{\mu\tau}^{}\right| \sin \left( \phi_{e\tau}^\prime - \phi_{\mu\tau}^{} - \phi_{e\mu}^\prime  \right) \right] \right\} \nonumber\\
	&&- 2\, \frac{m^2_{}}{v^2_{}} y_{e\mu}^{} \left[ \frac{c_{23}^{} \cos 2\theta_{12}^{} }{s_{12}^{} \, c_{12}^{} \, c_{13}^{}} \left( \left| \eta_{e\mu}^\prime \right| s_{e\mu}^\prime - 4 \left| \eta_{e\mu}^{} \right| s_{e\mu}^{} \right) - \frac{s_{13}^{} \sin 2\theta_{23}^{}}{c_{13}^{} \, c_{23}^{}}  \left( \left| \eta_{e\mu}^\prime \right| s_{e\mu+ \delta}^\prime - 4 \left| \eta_{e\mu}^{} \right| s_{e\mu + \delta}^{} \right) \right] \nonumber\\
	&&+ 2 \, \frac{m^2_{}}{v^2_{}}  y_{e\tau}^{} \left[ \frac{s_{23}^{} \cos 2\theta_{12}^{} }{s_{12}^{} \, c_{12}^{} \, c_{13}^{}} \left( \left| \eta_{e\tau}^\prime \right| s_{e\tau}^\prime - 4 \left| \eta_{e\tau}^{} \right| s_{e\tau}^{} \right) + \frac{s_{13}^{} \cos 2\theta_{23}^{}}{c_{13}^{} \, c_{23}^{}} \left( \left| \eta_{e\tau}^\prime \right| s_{e\tau + \delta}^\prime - 4 \left| \eta_{e\tau}^{} \right| s_{e\tau + \delta}^{} \right) \right] \nonumber\\
	&&- 2 \, \frac{m^2_{}}{v^2_{}} y_{\mu\tau}^{} s_{23}^{} c_{23}^{-1} \left( \left| \eta_{\mu\tau}^\prime \right| s_{\mu\tau}^\prime - 4 \left| \eta_{\mu\tau}^{} \right| s_{\mu\tau}^{} \right) \nonumber\\
	&&+ C_\kappa^{} \zeta_{12}^{-1} \left\{ \left[c_{13}^2 \, y_e^2 - \left(c_{23}^2 - s_{13}^2 \, s_{23}^2 \right) y_\mu^2+ \left(s_{13}^2 \, c_{23}^2 - s_{23}^2 \right) y_\tau^2 \right] c_{\rho-\sigma}^{} \right.\nonumber\\
	&& \left.\qquad\quad\;\;\; - s_{12}^{-1} c_{12}^{-1} s_{13}^{} s_{23}^{} c_{23}^{} \left(y_\mu^2 - y_\tau^2\right) \left( c_{12}^2 \, c_{\delta+\rho-\sigma}^{} - s_{12}^2 \, c_{\delta-\rho+\sigma}^{} \right) \right\} s_{\rho-\sigma}^{} \nonumber\\
	&&- C_\kappa^{} \zeta_{12}^{} \; \left\{ \left[c_{13}^2 \, y_e^2 - \left(c_{23}^2 - s_{13}^2 \, s_{23}^2 \right) y_\mu^2+ \left(s_{13}^2 \, c_{23}^2 - s_{23}^2 \right) y_\tau^2 \right] s_{\rho-\sigma}^{} \right.\nonumber\\
	&& \left.\qquad\quad\;\;\; - s_{12}^{-1} c_{12}^{-1} s_{13}^{} s_{23}^{} c_{23}^{} \left(y_\mu^2 - y_\tau^2\right) \left( c_{12}^2 s_{\delta
	+\rho-\sigma}^{} + s_{12}^2 s_{\delta
	-\rho+\sigma}^{} \right) \right\} c_{\rho-\sigma}^{} \nonumber\\
	&&- C_\kappa^{} \zeta_{13}^{-1} s_{12}^{} \left[ s_{12}^{} c_{12}^{-1} s_{23}^{} \left(y_\mu^2 - y_\tau^2\right) c_\rho^{} - s_{13}^{} c_{23}^{-1} \left(y_e^2 - s_{23}^2 y_\mu^2 - c_{23}^2 y_\tau^2 \right) c_{\rho+\delta}^{} \right] \nonumber\\
	&& \qquad \qquad \times \left( c_{12}^{} s_{23}^{} s_{\rho}^{} + s_{12}^{} s_{13}^{} c_{23}^{} s_{\rho+\delta} \right) \nonumber\\
	&&+ C_\kappa^{} \zeta_{13}^{} \; s_{12}^{} \left[ s_{12}^{} c_{12}^{-1} s_{23}^{} \left(y_\mu^2 - y_\tau^2\right) s_\rho^{} - s_{13}^{} c_{23}^{-1} \left(y_e^2 - s_{23}^2 y_\mu^2 - c_{23}^2 y_\tau^2 \right) s_{\rho+\delta}^{} \right]\nonumber\\
	&& \qquad \qquad \times \left( c_{12}^{} s_{23}^{} c_{\rho}^{} + s_{12}^{} s_{13}^{} c_{23}^{} c_{\rho+\delta} \right)\nonumber\\
	&&- C_\kappa^{} \zeta_{23}^{-1} c_{12}^{} \left[ s_{12}^{-1} c_{12}^{} s_{23}^{} \left(y_\mu^2 - y_\tau^2\right) c_\sigma^{} + s_{13}^{} c_{23}^{-1} \left(y_e^2 - s_{23}^2 y_\mu^2 - c_{23}^2 y_\tau^2 \right) c_{\sigma+\delta}^{} \right]\nonumber\\
	&& \qquad \qquad \times \left( s_{12}^{} s_{23}^{} s_{\sigma}^{} - c_{12}^{} s_{13}^{} c_{23}^{} s_{\sigma+\delta} \right) \nonumber\\
	&&+ C_\kappa^{} \zeta_{23}^{} \; c_{12}^{} \left[ s_{12}^{-1} c_{12}^{} s_{23}^{} \left(y_\mu^2 - y_\tau^2\right) s_\sigma^{} + s_{13}^{} c_{23}^{-1} \left(y_e^2 - s_{23}^2 y_\mu^2 - c_{23}^2 y_\tau^2 \right) s_{\sigma+\delta}^{} \right]\nonumber\\
	&& \qquad \qquad \times \left( s_{12}^{} s_{23}^{} c_{\sigma}^{} - c_{12}^{} s_{13}^{} c_{23}^{} c_{\sigma+\delta} \right) \;, \label{eq:phiemup}\\
	\dot{\phi}_{e\tau}^\prime &=& \frac{5}{2} \frac{v^2_{} c_{13}^{}}{ \left|\eta_{e\tau}^\prime\right|} \left[ \kappa_1^2 \, c_{12}^{} \left(s_{12}^{} s_{23}^{} s_{e\tau}^\prime - c_{12}^{} s_{13}^{} c_{23}^{} s_{e\tau+\delta}^\prime \right) -\kappa_2^2 s_{12}^{} \left( c_{12}^{} s_{23}^{} s_{e\tau}^\prime + s_{12}^{} s_{13}^{} c_{23}^{} s_{e\tau+\delta}^\prime \right) \right.\nonumber\\
	&&+ \left. \kappa_3^2 \, s_{13}^{} c_{23}^{} s_{e\tau+\delta}^\prime  \right] + \left(\frac{17}{3} g_2^2 + \frac{1}{3} g_1^2 + 4 y_e^2 + 4 y_\tau^2\right) \left|\eta_{e\tau}^{}\right| \left|\eta_{e\tau}^\prime\right|_{}^{-1} \sin \left(\phi_{e\tau}^\prime - \phi_{e\tau}^{} \right) \nonumber\\
	&&+ 2 \, \frac{m^2_{}}{v^2_{}} \left|\eta_{e\tau}^\prime\right|_{}^{-1} \left\{ 4 \, y_{e\tau}^{} \left(\eta_{\tau\tau}^\prime - \eta_{ee}^\prime \right) \left|\eta_{e\tau}^{}\right| \sin \left( \phi_{e\tau}^\prime - \phi_{e\tau}^{} \right) \right.\nonumber\\
	&&+ \, y_{e\mu}^{} \left|\eta_{\mu\tau}^\prime \right| \left[ \left|\eta_{e\mu}^\prime\right| \sin \left( \phi_{e\mu}^\prime + \phi_{\mu\tau}^\prime - \phi_{e\tau}^\prime \right) - 4 \left|\eta_{e\mu}^{}\right| \left( \phi_{e\mu}^{} + \phi_{\mu\tau}^\prime - \phi_{e\tau}^\prime \right) \right] \nonumber\\
	&&- \left.y_{\mu\tau}^{} \left|\eta_{e\mu}^\prime \right| \left[ \left|\eta_{\mu\tau}^\prime\right| \sin \left( \phi_{\mu\tau}^\prime + \phi_{e\mu}^\prime - \phi_{e\tau}^\prime \right) - 4 \left|\eta_{\mu\tau}^{}\right| \left(  \phi_{\mu\tau}^{} + \phi_{e\mu}^\prime - \phi_{e\tau}^\prime \right) \right] \right\}\nonumber\\
	&&- 2 \, \frac{m^2_{}}{v^2_{}} y_{e\mu}^{} \left[ \frac{c_{23}^{} \cos 2\theta_{12}^{}}{s_{12}^{} \, c_{12}^{} \,  c_{13}^{}} \left( \left| \eta_{e\mu}^\prime \right| s_{e\mu}^\prime - 4 \left| \eta_{e\mu}^{} \right| s_{e\mu}^{} \right) + \frac{s_{13}^{} \cos 2\theta_{23}^{}}{c_{13}^{} \, s_{23}^{} } \left( \left| \eta_{e\mu}^\prime \right| s_{e\mu+\delta}^\prime - 4 \left| \eta_{e\mu}^{} \right| s_{e\mu+\delta}^{} \right) \right] \nonumber\\
	&&+ 2 \, \frac{m^2_{}}{v^2_{}} y_{e\tau}^{} \left[ \frac{s_{23}^{} \cos 2\theta_{12}^{}}{ s_{12}^{} \, c_{12}^{} \, c_{13}^{}} \left( \left| \eta_{e\tau}^\prime \right| s_{e\tau}^\prime - 4 \left| \eta_{e\tau}^{} \right| s_{e\tau}^{} \right) +  \frac{s_{13}^{} \sin 2 \theta_{23}^{} }{c_{13}^{} \, s_{23}^{}} \left( \left| \eta_{e\tau}^\prime \right| s_{e\tau + \delta}^\prime - 4 \left| \eta_{e\tau}^{} \right| s_{e\tau + \delta}^{} \right) \right] \nonumber\\
	&&-  2 \, \frac{m^2_{}}{v^2_{}} y_{\mu\tau}^{} s_{23}^{-1} c_{23}^{} \left( \left| \eta_{\mu\tau}^\prime \right| s_{\mu\tau}^\prime - 4 \left| \eta_{\mu\tau}^{} \right| s_{\mu\tau}^{} \right) \nonumber\\
	&&+ C_\kappa^{} \zeta_{12}^{-1} \left\{ \left[c_{13}^2 \, y_e^2 - \left(c_{23}^2 - s_{23}^2 s_{13}^2 \right) y_\mu^2+ \left(c_{23}^2 s_{13}^2 - s_{23}^2 \right) y_\tau^2 \right] c_{\rho-\sigma}^{} \right.\nonumber\\
	&& \left. \qquad\quad - s_{12}^{-1} c_{12}^{-1} s_{13}^{} s_{23}^{} c_{23}^{} \left(y_\mu^2 - y_\tau^2\right) \left( c_{12}^2 \, c_{\delta+\rho-\sigma}^{} - s_{12}^2 \, c_{\delta-\rho+\sigma}^{} \right) \right\} s_{\rho-\sigma}^{} \nonumber\\
	&&- C_\kappa^{} \zeta_{12}^{} \; \left\{ \left[c_{13}^2 \, y_e^2 - \left(c_{23}^2 - s_{13}^2 \, s_{23}^2 \right) y_\mu^2 + \left(s_{13}^2 \, c_{23}^2 - s_{23}^2 \right) y_\tau^2 \right] s_{\rho-\sigma}^{} \right.\nonumber\\
	&& \left.\qquad\quad\;\;\; - s_{12}^{-1} c_{12}^{-1} s_{13}^{} s_{23}^{} c_{23}^{} \left(y_\mu^2 - y_\tau^2\right) \left( c_{12}^2 \, s_{\delta+\rho-\sigma}^{} + s_{12}^2 \, s_{\delta-\rho+\sigma}^{} \right) \right\} c_{\rho-\sigma}^{} \nonumber\\
    &&+ C_\kappa^{} \zeta_{13}^{-1} s_{12}^{} \left[s_{12}^{} c_{12}^{-1} c_{23}^{} \left(y_\mu^2 - y_\tau^2 \right) c_\rho^{} - s_{13}^{} s_{23}^{-1} \left(y_e^2 - s_{23}^2 y_\mu^2 - c_{23}^2 y_\tau^2 \right) c_{\rho+\sigma}^{} \right] \nonumber\\
    && \qquad\quad\;\;\; \times \left(c_{12}^{} c_{23}^{} s_\rho^{} - s_{12}^{} s_{13}^{} s_{23}^{} s_{\rho+\delta}^{} \right) \nonumber\\
    &&- C_\kappa^{} \zeta_{13}^{} \; s_{12}^{} \left[s_{12}^{} c_{12}^{-1} c_{23}^{} \left(y_\mu^2 - y_\tau^2 \right) s_\rho^{} - s_{13}^{} s_{23}^{-1} \left(y_e^2 - s_{23}^2 y_\mu^2 - c_{23}^2 y_\tau^2 \right) s_{\rho+\sigma}^{} \right] \nonumber\\
    && \qquad \qquad \times \left(c_{12}^{} c_{23}^{} c_\rho^{} - s_{12}^{} s_{13}^{} s_{23}^{} c_{\rho+\delta}^{} \right)\nonumber\\
    &&+ C_\kappa^{} \zeta_{23}^{-1} c_{12}^{} \left[s_{12}^{-1} c_{12}^{} c_{23}^{} \left(y_\mu^2 - y_\tau^2 \right) c_\sigma^{} + s_{13}^{}  s_{23}^{-1} \left(y_e^2 - s_{23}^2 y_\mu^2 - c_{23}^2 y_\tau^2 \right) c_{\sigma+\delta}^{} \right] \nonumber\\
    && \qquad \qquad \times \left(s_{12}^{} c_{23}^{} s_\sigma^{} + c_{12}^{} s_{13}^{} s_{23}^{} s_{\sigma+\delta}^{} \right) \nonumber\\
    &&- C_\kappa^{} \zeta_{23}^{} \; c_{12}^{} \left[s_{12}^{-1} c_{12}^{} c_{23}^{} \left(y_\mu^2 - y_\tau^2 \right) s_\sigma^{} + s_{13}^{}  s_{23}^{-1} \left(y_e^2 - s_{23}^2 y_\mu^2 - c_{23}^2 y_\tau^2 \right) s_{\sigma+\delta}^{} \right] \nonumber\\
    && \qquad \qquad \times \left(s_{12}^{} c_{23}^{} c_\sigma^{} + c_{12}^{} s_{13}^{} s_{23}^{} c_{\sigma+\delta}^{} \right) \;, \label{eq:phietaup}\\
	\dot{\phi}_{\mu\tau}^\prime &=& \frac{5 \, \kappa_1^2 v^2_{}}{ 2 \left|\eta_{\mu\tau}^\prime\right|}  \left\{ \left[s_{23}^{} c_{23}^{} \left(c_{12}^2 s_{13}^2 - s_{12}^2 \right) + s_{12}^{} c_{12}^{} s_{13}^{} \left(c_{23}^2 - s_{23}^2 \right) c_\delta^{} \right] s_{\mu\tau}^\prime  + s_{12}^{} c_{12}^{} s_{13}^{} s_\delta^{} c_{\mu\tau}^\prime \right\} \nonumber\\
	&&+ \frac{5 \, \kappa_2^2 v^2_{}}{2 \left|\eta_{\mu\tau}^\prime\right|}  \left\{ \left[s_{23}^{} c_{23}^{} \left(s_{12}^2 s_{13}^2 - c_{12}^2 \right) - s_{12}^{} c_{12}^{} s_{13}^{} \left(c_{23}^2 - s_{23}^2 \right) c_\delta^{} \right] s_{\mu\tau}^\prime - s_{12}^{} c_{12}^{} s_{13}^{} s_\delta^{} c_{\mu\tau}^\prime \right\}\nonumber\\
	&&+  \frac{5 \, \kappa_3^2 v^2_{}}{2 \left|\eta_{\mu\tau}^\prime\right|} c_{13}^2 s_{23}^{} c_{23}^{} s_{\mu\tau}^\prime + \left(\frac{17}{3} g_2^2 + \frac{1}{3} g_1^2 + 4 y_\mu^2 + 4 y_\tau^2\right) \frac{\left|\eta_{\mu\tau}^{}\right|}{  \left|\eta_{\mu\tau}^\prime\right|} \sin \left(\phi_{\mu\tau}^\prime - \phi_{\mu\tau}^{} \right) \nonumber\\
	&&+ 2 \, \frac{m^2_{}}{v^2_{}} \left|\eta_{\mu\tau}^\prime \right|_{}^{-1} \left\{ 4 \, y_{\mu\tau}^{} \left(\eta_{\tau\tau}^\prime - \eta_{\mu\mu}^\prime \right) \left|\eta_{\mu\tau}^{}\right| \sin \left( \phi_{\mu\tau}^\prime - \phi_{\mu\tau}^{} \right) \right.\nonumber\\
	&&- \, y_{e\mu}^{} \left|\eta_{e\tau}^\prime \right| \left[ \left|\eta_{e\mu}^\prime\right| \sin \left( \phi_{e\tau}^\prime - \phi_{e\mu}^\prime - \phi_{\mu\tau}^\prime \right) - 4 \left|\eta_{e\mu}^{}\right| \sin \left( \phi_{e\tau}^\prime - \phi_{e\mu}^{} - \phi_{\mu\tau}^\prime \right) \right] \nonumber\\
	&&- \left.y_{e\tau}^{} \left|\eta_{e\mu}^\prime \right| \left[ \left|\eta_{e\tau}^\prime\right| \sin \left( \phi_{e\tau}^\prime - \phi_{e\mu}^\prime - \phi_{\mu\tau}^\prime  \right) - 4 \left|\eta_{e\tau}^{}\right| \sin \left( \phi_{e\tau}^{} - \phi_{e\mu}^\prime - \phi_{\mu\tau}^\prime \right) \right] \right\}\nonumber\\
	&&- 2\, \frac{m^2_{}}{v^2_{}} s_{13}^{} c_{13}^{-1}  \left[ y_{e\mu}^{} s_{23}^{-1} \left(\left| \eta_{e\mu}^\prime \right| s_{e\mu + \delta}^\prime - 4 \left| \eta_{e\mu}^{} \right| s_{e\mu + \delta}^{}\right) - y_{e\tau}^{} c_{23}^{-1} \left(\left| \eta_{e\tau}^\prime \right| s_{e\tau+ \delta}^\prime- 4 \left| \eta_{e\tau}^{} \right| s_{e\tau+ \delta}^{}\right) \right] \nonumber\\
	&&- 2 \, \frac{m^2_{}}{v^2_{}} y_{\mu\tau}^{} s_{23}^{-1} c_{23}^{-1} \left(c_{23}^2 - s_{23}^2\right) \left( \left| \eta_{\mu\tau}^\prime \right| s_{\mu\tau}^\prime - 4 \left| \eta_{\mu\tau}^{} \right| s_{\mu\tau}^{} \right) \nonumber\\
    &&+ C_\kappa^{} \zeta_{13}^{-1} s_{12}^{} \left[s_{12}^{} \left(y_\mu^2 - y_\tau^2 \right) c_\rho^{} - c_{12}^{} s_{13}^{} s_{23}^{-1} c_{23}^{-1} \left(y_e^2 - s_{23}^2 y_\mu^2 - c_{23}^2 y_\tau^2 \right) c_{\rho+\delta}^{} \right] s_\rho^{} \nonumber\\
    &&- C_\kappa^{} \zeta_{13}^{} \; s_{12}^{} \left[s_{12}^{} \left(y_\mu^2 - y_\tau^2 \right) s_\rho^{} - c_{12}^{} s_{13}^{} s_{23}^{-1} c_{23}^{-1} \left(y_e^2 - s_{23}^2 y_\mu^2 - c_{23}^2 y_\tau^2 \right) s_{\rho+\delta}^{} \right] c_\rho^{} \nonumber\\
    &&+ C_\kappa^{} \zeta_{23}^{-1} c_{12}^{} \left[c_{12}^{} \left(y_\mu^2 - y_\tau^2 \right) c_\sigma^{} + s_{12}^{} s_{13}^{} s_{23}^{-1}  c_{23}^{-1} \left(y_e^2 - s_{23}^2 y_\mu^2 - c_{23}^2 y_\tau^2 \right) c_{\sigma+\delta}^{} \right] s_\sigma^{} \nonumber\\
    &&- C_\kappa^{} \zeta_{23}^{} \; c_{12}^{} \left[c_{12}^{} \left(y_\mu^2 - y_\tau^2 \right) s_\sigma^{} + s_{12}^{} s_{13}^{} s_{23}^{-1} c_{23}^{-1} \left(y_e^2 - s_{23}^2 y_\mu^2 - c_{23}^2 y_\tau^2 \right) s_{\sigma+\delta}^{} \right] c_\sigma^{}\;.
    \label{eq:phimutaup}
\end{eqnarray}

\end{appendices}

\end{document}